\documentclass[%
 reprint,
 amsmath,amssymb,
 aps,
prb,
]{revtex4-2}

\usepackage[utf8]{inputenc}
\usepackage{graphicx}
\usepackage{overpic}
 \usepackage{rotating}
 \usepackage{mwe, tikz}
 \usepackage{amsmath,amssymb}
\usepackage{pgfplots}
\usetikzlibrary{calc,arrows,decorations.markings,shapes.geometric} 
\usepackage{amssymb}

\usepackage[inline]{trackchanges}

\usepackage{lipsum}
\usepackage{romannum}
\usepackage{xfp}

\soulregister\ref7
\soulregister\cite7
\soulregister\citeA7
\soulregister\romannum7

\usepackage{graphicx}
\usepackage{amsmath}
\usepackage{mathtools} 
\usepackage{color}
\usepackage{overpic}
\usepackage[export]{adjustbox}
\usepackage{mwe, tikz}
\usepackage{rotating}
\usepackage{latexsym}
\usepackage{csquotes}
\usepackage{tikz} 
\usetikzlibrary{calc,arrows,decorations.markings} 
\usepackage{pict2e}
\usepackage{pgfplots}
\usepackage{float}
\usepackage{hyperref}
\usepackage{xr}
\usepackage{rotating}
\usepackage{amsmath,amssymb}
\usepackage{fp}
\usetikzlibrary{calc} 
\usepackage{booktabs, tabularx}
\usepackage{romannum}
\usepackage{bm}
\usepackage{float}
\usepackage{adjustbox}
\usepackage{booktabs, tabularx}
\usepackage{longtable}

\newcommand{\legLine}[5]{ 
			     \coordinate (center1) at (#1,#2); 
                \coordinate (b) at ($ (center1) - (.075,0) $);
                \coordinate (c) at ($ (b) - (.15,0) $); 
				\node at ($ (center1) + (#5,0.025) $) {{\color{#3}#4}}; 
                \draw[#3][line width=0.1mm] (b) -- (c); 
}

\definecolor{light-gray}{gray}{0.85}

\newcommand{\drawSlope}[6]{ 
				\coordinate (center1) at (#1,#2); 
				 \FPeval{\nx}{cos(#4*pi/180)}%
				 \FPeval{\ny}{sin(#4*pi/180)}%
				 \FPeval{\absNx}{abs(\nx)}%
				 \FPeval{\absNy}{abs(\ny)}%
				\coordinate (b) at ($ (center1) + #3*(-\ny,+\nx) $);
				\coordinate (c) at ($ (center1) - #3*(-\ny,+\nx) $); 
				 \FPeval{\xx}{(-\nx*\ny)}%
				\ifdim\xx pt < 0pt 
				\coordinate (d) at ($ (c) -2*#3*\ny*(1,0) $);
				\node at ($(d) +(0,#3*\nx)-0.2*(\nx/\absNx,0)$) {{\color{#5}#6}}; 
				\else
				\coordinate (d) at ($ (c) +2*#3*\nx*(0,1) $);
				\node at ($(d)-#3*\nx*(0,1)-0.2*\nx/\absNx*(1,0)$) {{\color{#5}#6}}; 
				\fi
				\draw[#5,line width=0.1mm] (d) -- (b); 
				\draw[#5,line width=0.1mm] (d) -- (c); 
				\draw[#5,line width=0.1mm] (b) -- (c); 
} 

\newcommand{\arrow}[6]{
			\coordinate (center2) at (#1,#2); 
            \coordinate (x1) at ($ (center2) + #3*(#4,#5) $); 
            \coordinate (x2) at ($ (center2) - .3*#3*(#4,#5) $); 
            \draw[->,>=stealth,black][line width=0.4mm] (x2) -- (x1);
			\node at ($(x1)+(0.2,0)$){#6}; 
}

\newcommand{\lline}[6]{
			\coordinate (center2) at (#1,#2); 
            \coordinate (x1) at ($ (center2) + #3*(#4,#5) $); 
            \coordinate (x2) at ($ (center2) - .3*#3*(#4,#5) $); 
            \draw[red][line width=0.4mm] (x2) -- (x1);
			\node at ($(x1)+(0.2,0)$){#6}; 
}
\newcommand*{\Labelxy}[4]{\put(#1,#2) {\setlength{\fboxsep}{0pt}{\strut\textcolor{black}{\begin{turn}{#3}{#4}\end{turn}}}}}
\newcommand*{\LabelFig}[3]{\put(#1,#2) {\setlength{\fboxsep}{0pt}\colorbox{white}{\textcolor{black}{#3}}} }

\newcommand{\glsix}{NiCoCr~}

\newcommand{\glfour}{NiCoCrFeMn~}

\newcommand{\rate}{\dot{\epsilon}_{zz}}
\newcommand{\ratezero}{\rate=10^8~\text{s}^{-1}}

\newcommand{\ratefour}{\rate=10^{10}~\text{s}^{-1}}

\newcommand{\sigmadot}{\partial_{t}\sigma_{zz}}
\definecolor{darkolivegreen}{rgb}{0.33, 0.42, 0.18}
\definecolor{darkspringgreen}{rgb}{0.09, 0.45, 0.27}
\definecolor{darkslategray}{rgb}{0.18, 0.31, 0.31}
\definecolor{darkred}{rgb}{0.55, 0.0, 0.0}
\definecolor{darkblue}{rgb}{0.12156862745098039, 0.4666666666666667, 0.7058823529411765}

\addeditor{kk}
\usepackage{xr}
\makeatletter
\newcommand*{\addFileDependency}[1]{
  \typeout{(#1)}
  \@addtofilelist{#1}
  \IfFileExists{#1}{}{\typeout{No file #1.}}
}
\makeatother
\newcommand*{\myexternaldocument}[1]{
    \externaldocument{#1}
    \addFileDependency{#1.tex}
    \addFileDependency{#1.aux}
}
\myexternaldocument{./sm}

\begin{document}

\title{Serrated plastic flow in slowly-deforming complex concentrated alloys: universal signatures of dislocation avalanches}

\author{Kamran Karimi$^1$}
\email{kamran.karimi@ncbj.gov.pl}
\author{Amin Esfandiarpour$^1$}%
\author{Stefanos Papanikolaou$^1$}
\email{Stefanos.Papanikolaou@ncbj.gov.pl}
\affiliation{%
$^1$ NOMATEN Centre of Excellence, National Center for Nuclear Research, ul. A. Sołtana 7, 05-400 Swierk/Otwock, Poland 
}%

\begin{abstract}
Under plastic flow, multi-element high/medium-entropy alloys (HEAs/MEAs) commonly exhibit complex intermittent and collective dislocation dynamics owing to inherent lattice distortion and atomic-level chemical complexities. Using atomistic simulations, we report on an avalanche study of slowly-driven model face-centered cubic (fcc) \glfour and \glsix chemically complex alloys aiming for microstructural/topological characterization of associated dislocation avalanches. The results of our avalanche simulations reveal a close correspondence between the observed serration features in the stress response of the deforming HEA/MEA and the incurred slip patterns within the bulk crystal. We show that such correlations become quite pronounced within the rate-independent (quasi-static) regime exhibiting scale-free statistics and critical scaling features as universal signatures of dislocation avalanches.
\end{abstract}

\maketitle

\section{Introduction}
The serrated response is a commonly observed phenomenon in a broad class of driven systems \cite{fisher1998collective}. 
Examples include crackling sounds due to plasticity \cite{miguel2001intermittent,sultan2022sheared,karimi2017inertia} or brittle fracture \cite{vu2019compressive,petri1994experimental,karimi2019plastic}, Barkhausen noise in ferromagnetism \cite{durin2000scaling} and even stick-slip dynamics of earthquakes at geological scales \cite{scholz2002mechanics}, just to name a few.
Under a sufficiently slow driving rate, serrations refer to highly-intermittent and irrecoverable dynamics that a quiescent (but driven) system undergoes, beyond its threshold, as a certain form of relaxation.
As for serrated plastic flow in deforming crystalline solids \cite{brechtl2020review}, the relaxation process typically occurs through slip bursts mainly due to the spontaneous nucleation/depinning of dislocations that exhibit a collective motion within the bulk leading to the so-called dislocation avalanches.
The scale-free nature of dislocation avalanches ---featuring a broad range of time, length, and energy scales \cite{zaiser2006scale}--- may indeed suggest some form of criticality/universality within the context of yielding transition \cite{friedman2012statistics}.
The notion of universality is not always strictly defined in light of close ties between avalanche statistics and dislocations' substructure as well as their complex mechanisms of nucleation, glide, and interactions which are believed to show certain non-universal features \cite{richeton2005breakdown,sparks2018shapes}, depending on various factors such as crystalline phase, lattice orientation, chemical composition, specimen size, temperature, and deformation rate sensitivity.
Given the above considerations, how could we infer such complex dislocation patterns and underlying interaction mechanisms by probing statistics of dislocation avalanches? 
The question posed here has very practical implications in the context of nanomechanical testing methods that, together with in-situ imaging techniques, can give us rich knowledge and insights about nanoscopic origins of plasticity.

The existing literature has a wealth of experimental information on serration features of driven crystalline metals and associated microstructural signatures across a broad range of laboratory settings (see \cite{papanikolaou2017avalanches} and references therein).
This includes a large suite of nano/micro scale mechanical tests (e.g. uniform tension, nano/micro pillars, nano-indentation) that are typically supplemented by acoustic emission (AE) measurements of intermittent bursts and/or in(ex)-situ microscopy.
A rather generic observation is the power-law distributed magnitude of the latter $P(S)\propto S^{-\tau}$ but with scaling exponent $\tau$ that gets typically affected by underlying mechanisms at play and shows deviations from the mean-filed estimate $3/2$ \cite{fisher1998collective}.
The experimental range mostly observed for pure metals (Ni, Al, Cu, Au, Mo, and Nb) is between $\tau=1.5-1.9$ \cite{papanikolaou2017avalanches,rizzardi2022mild}, bearing in mind that various metrics have been proposed as avalanche size based on types of experimentation and associated observables. 
In-situ electron-microscopy-based investigations have been mainly centered on establishing meaningful links between the occurrence of plastic avalanches and coinciding microstructural evolution.
In this context, significant size effects have been commonly identified in both stress serration features and dislocation morphologies, with the latter mainly arising from surface-induced limitations of deformation sources and augmented annhilation mechanisms \cite{shan2008mechanical,oh2009situ,kiener2011source,maass2007time}.
Rate effects \cite{zhang2013strain} have been consistently reported to influence dislocation glide mechanisms and associated relaxation processes which were shown to systematically alter statistics of slip avalanches \cite{papanikolaou2012quasi,ananthakrishna1999crossover}.

\begin{figure}[t]
    \begin{overpic}[width=0.49\columnwidth]{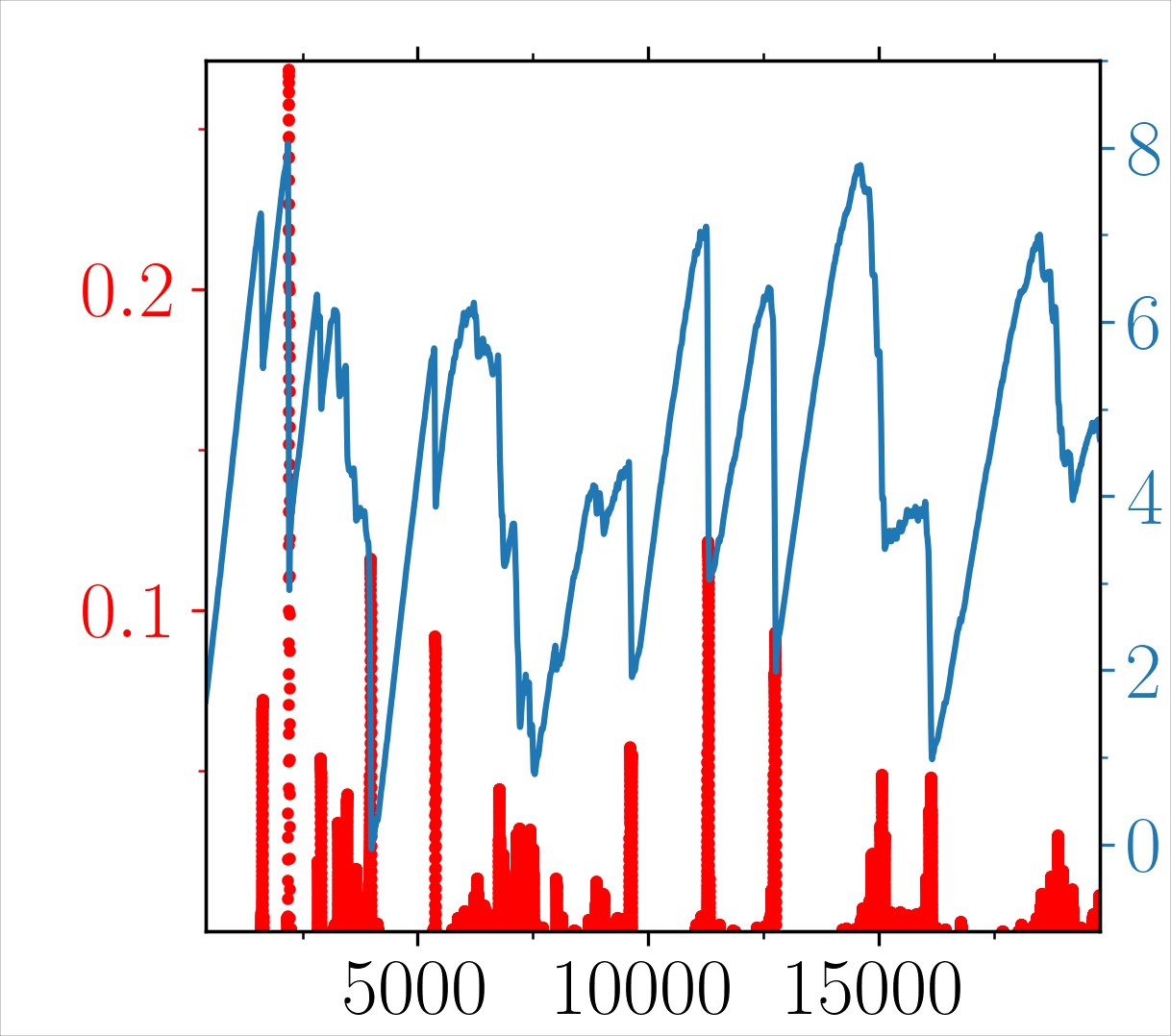}
        \LabelFig{18}{86}{$a)$}
        \Labelxy{50}{-8}{0}{$t$(ps)}
        \Labelxy{102}{35}{90}{\color{darkblue}$\sigma_{zz}$ \tiny(Gpa)}
        \Labelxy{-4}{35}{90}{\color{red} $-\sigmadot \scriptstyle ~(\text{Gpa.ps}^{-1})$}

     \end{overpic}
    \begin{overpic}[width=0.49\columnwidth]{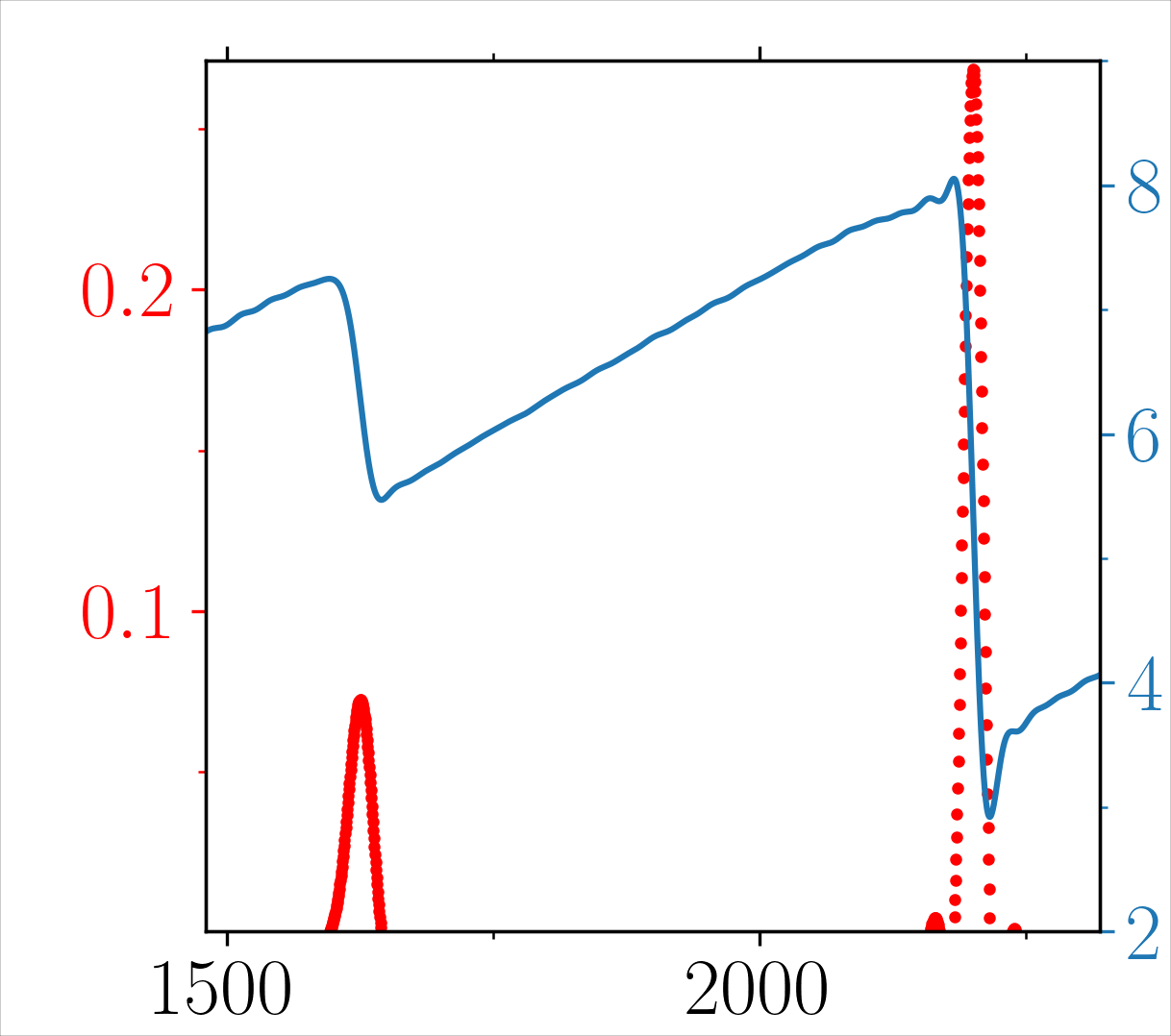}
        \LabelFig{18}{86}{$b)$}
        \Labelxy{50}{-8}{0}{$t$(ps)}
        \Labelxy{22}{12}{0}{\color{red} $t_i$}
        \Labelxy{31}{0}{0}{\color{red} $\mathcal{T}_i$}
       \begin{tikzpicture}
            \coordinate (a) at (0,0); 
            \node[white] at (a) {\tiny.};               %
             \lline{1.3}{0.2}{.1}{0}{1.0}{}
             \lline{1.0}{0.2}{.1}{0}{1.0}{}
 		    \end{tikzpicture}
     \end{overpic}
    \vspace{4pt}

    \begin{overpic}[width=0.49\columnwidth]{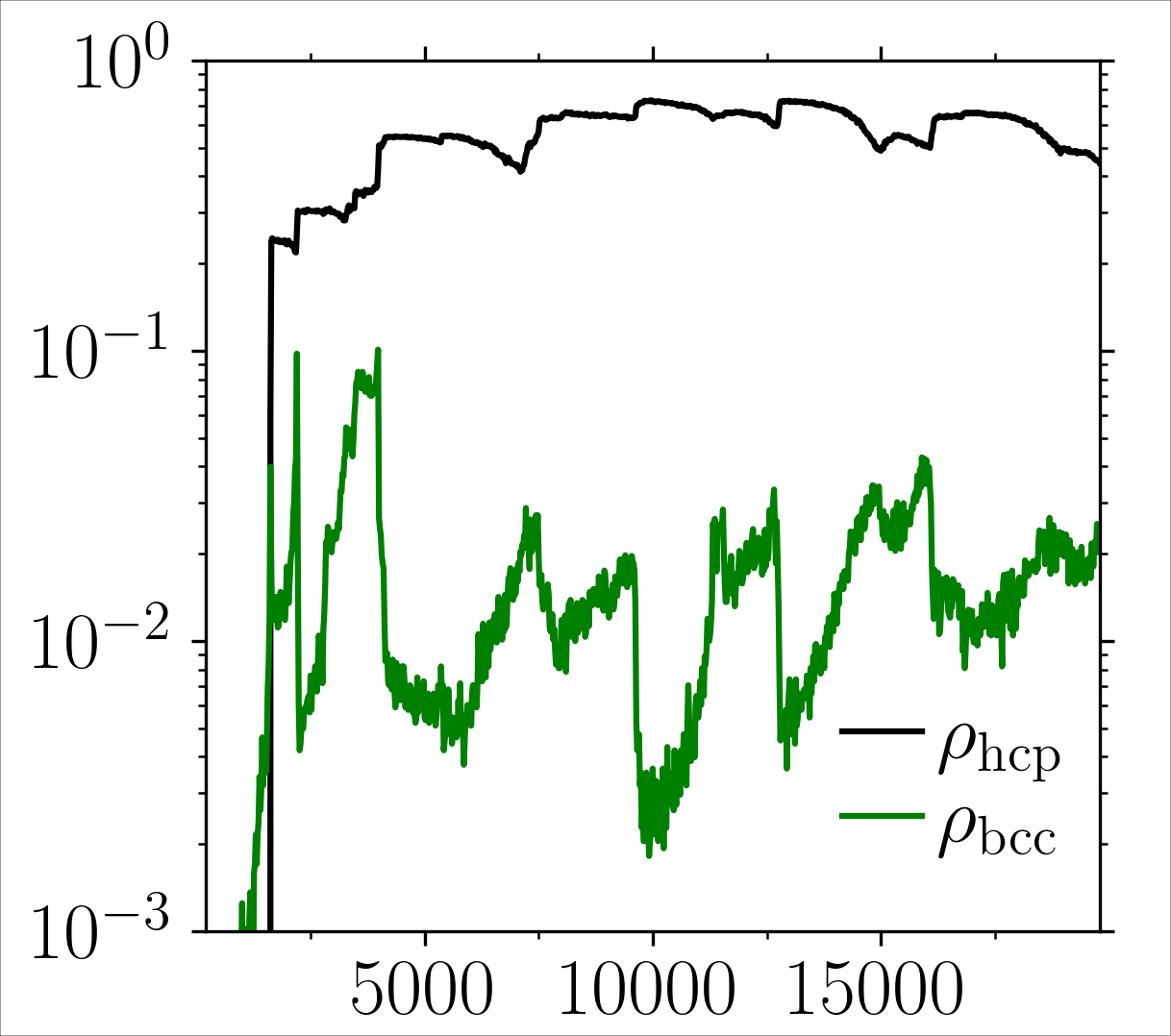}
        \LabelFig{18}{86}{$c)$}
        \Labelxy{50}{-8}{0}{$t$(ps)}
        \Labelxy{-4}{35}{90}{$\rho_\text{hcp},~\color{darkspringgreen}\rho_\text{bcc}$}

     \end{overpic}
    \begin{overpic}[width=0.49\columnwidth]{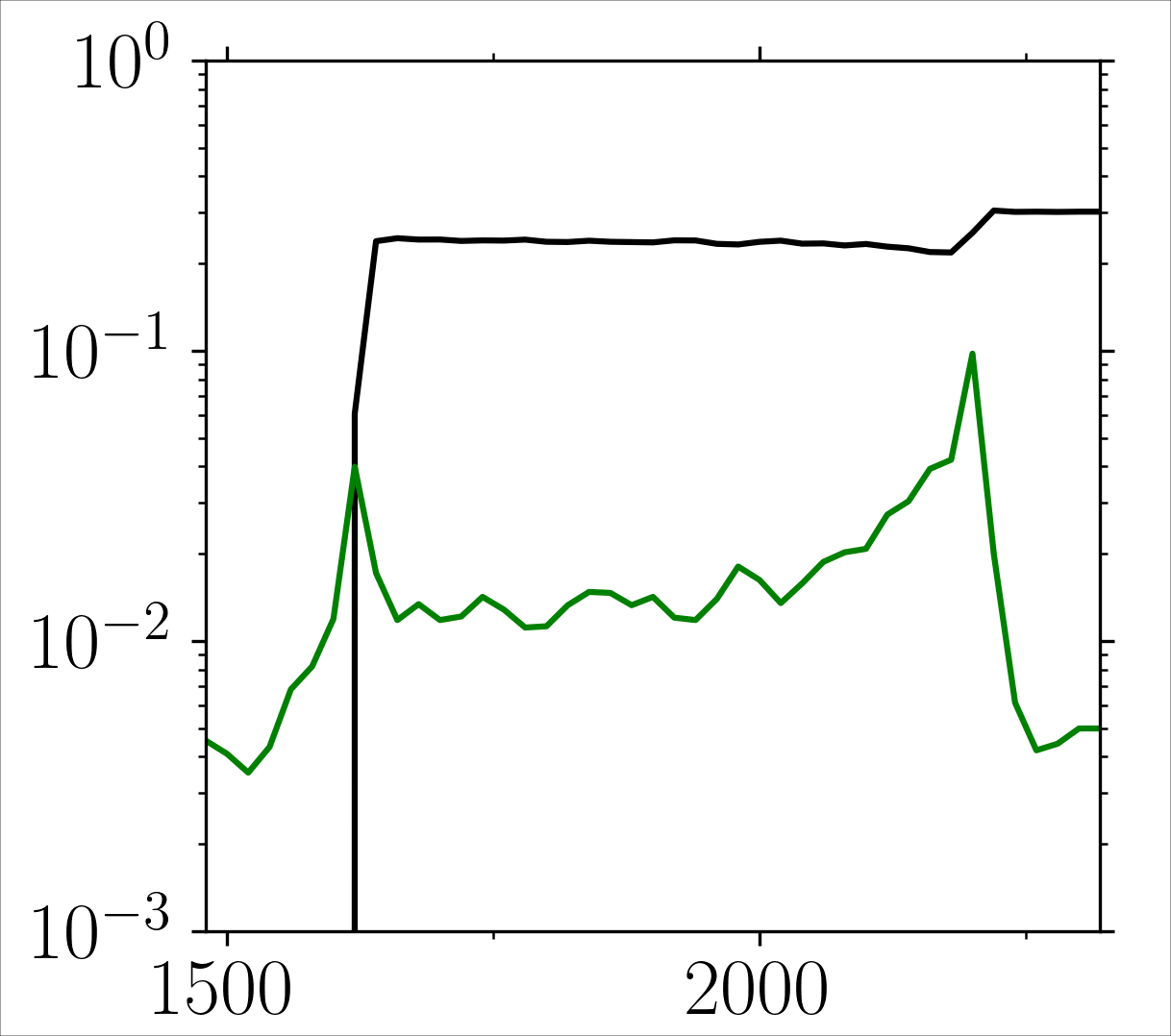}
        \LabelFig{18}{86}{$d)$}
        \Labelxy{50}{-8}{0}{$t$(ps)}
        \Labelxy{-4}{35}{90}{$\rho_\text{hcp},~\color{darkspringgreen}\rho_\text{bcc}$}
     \end{overpic}
    \vspace{6pt}

    \raggedright
    \begin{overpic}[width=0.23\textwidth]{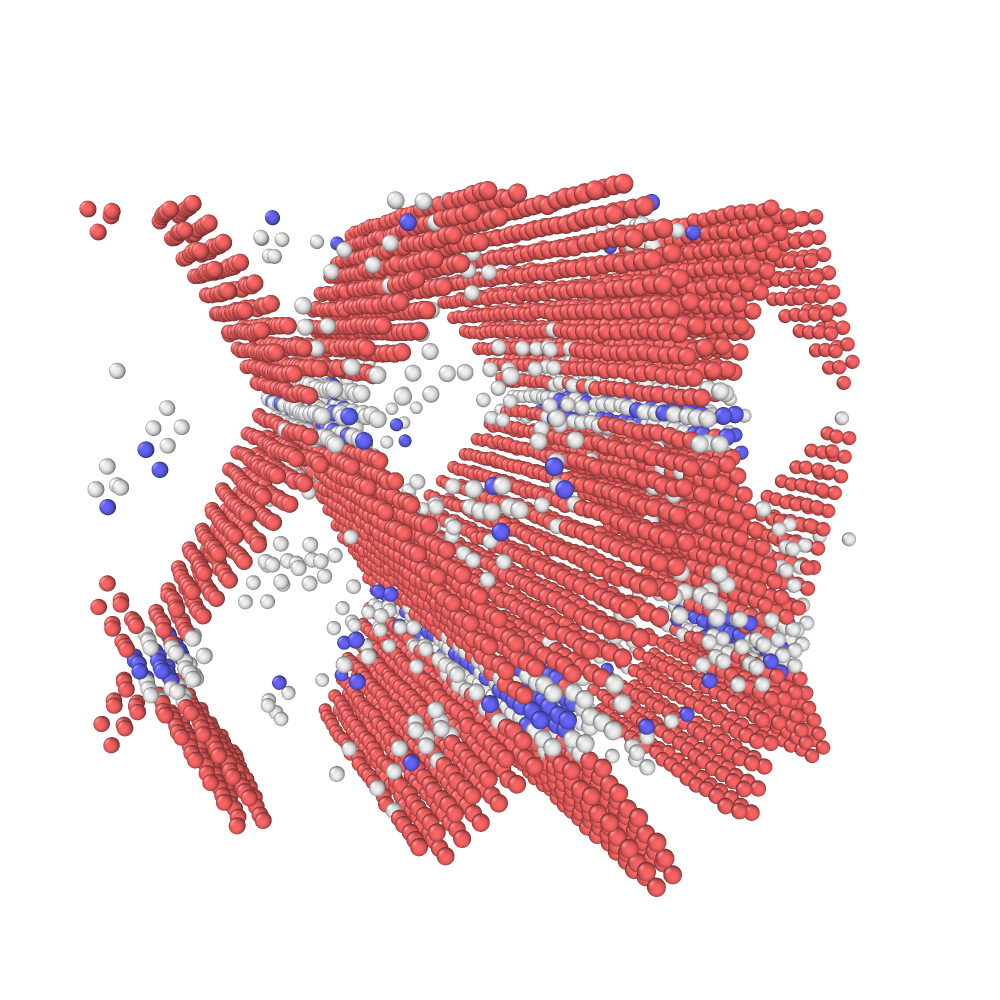}
        \put(6,4){\includegraphics[width=3.6cm,height=3.6cm]{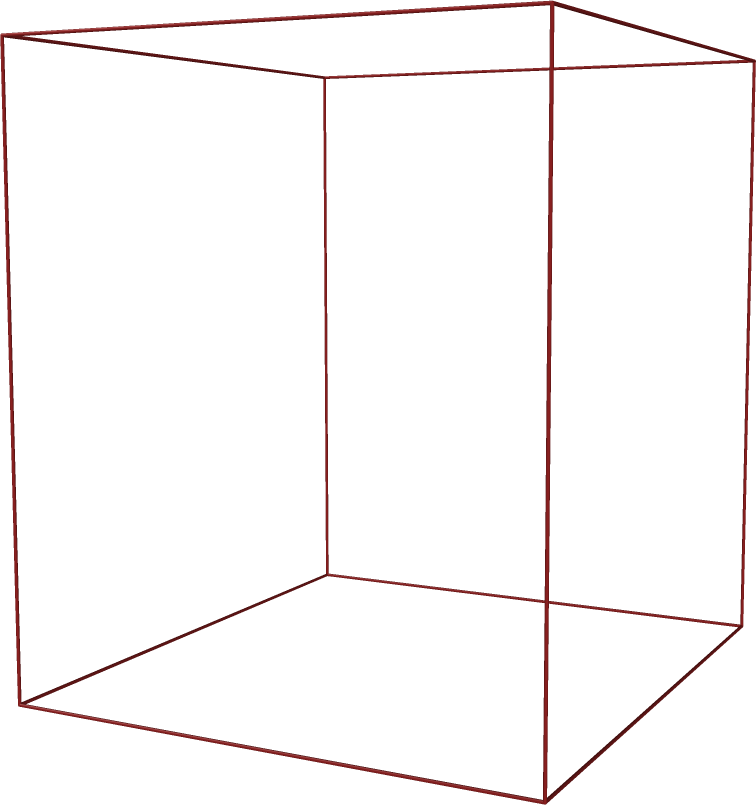}}
        \put(118,6){\includegraphics[width=3.6cm,height=3.6cm]{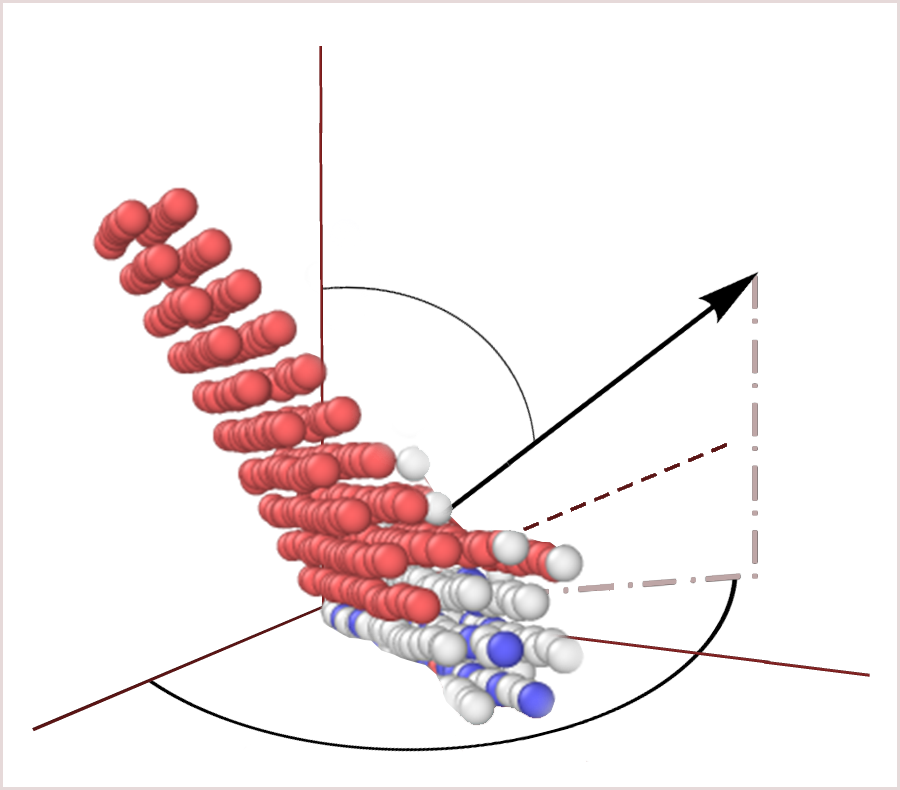}}
        \put(100,78){\color{red} hcp}
        \put(100,66){\color{blue} bcc}
        \put(100,54){\scriptsize \color{gray} unknown}
        \LabelFig{160}{63}{$\phi$}
        \LabelFig{180}{8}{$\theta$}
        \LabelFig{18}{93}{$e)$}
        \LabelFig{122}{93}{$f)$}
        \begin{tikzpicture}
            \coordinate (a) at (0,0); 
            \node[white] at (a) {\tiny.};               
            \arrow{1.0}{3.6}{0.4}{0}{+1}{$\hspace{10pt} \rate$}
            \arrow{2.2}{0.82}{0.4}{1}{-0.1}{\hspace{16pt}$y[010]$}
            \arrow{-1}{.34}{0.4}{-1}{-0.34}{\hspace{-38pt}$x[100]$}
		\draw[red,fill=red] (2.6,3.2) circle (1pt);
		\draw[blue,fill=blue] (2.6,2.7) circle (1pt);
		\draw[gray,fill=gray] (2.6,2.2) circle (1pt);
        \end{tikzpicture}
     \end{overpic}
    %
    %
    \caption{\textbf{a}) The evolution of the normal stress $\sigma_{zz}$ and spontaneous stress rate $\sigmadot$ with time $t$ at $T=300$ K and $\rate=10^8~\text{s}^{-1}$ corresponding to the Cantor alloy. \textbf{b}) A magnified view of a) with the $i$-th avalanche starting at $t_i$ and having the duration of $\mathcal{T}_i$. \textbf{c}) Fraction of atoms $\rho_\text{hcp}$ and $\rho_\text{bcc}$ with hcp and bcc arrangement versus time. \textbf{d}) A magnified view of c). \textbf{e}) Illustration of the uniaxial setup with the formation of the hcp clusters of atoms within $\{111\}$ slip planes during an avalanche. \textbf{f}) Crystallographic orientation of a hcp cluster of size $s_\text{hcp}$ described by the azimuthal angle $\theta$ and polar angle $\phi$.}
    \label{fig:loadCurve}
\end{figure}
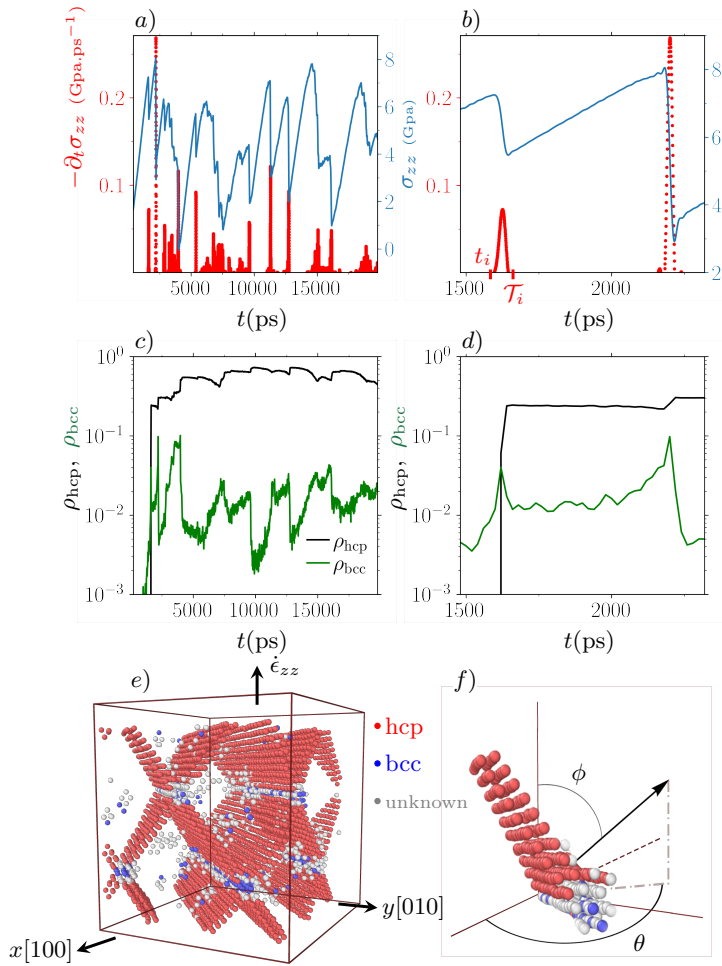

Overall, the aforementioned studies suggest certain \emph{indirect} (but insightful) links between seratted flow features and dislocation slip patterns in pure crystalline metals. 
Such correlations become even more challenging within the framework of high/medium-entropy alloys (HEAs/MEAs), bearing in mind inherent atomic-level complexities, due to underlying lattice distortions, which are known to be the dominant source of HEAs/MEAs' exceptional properties \cite{li2019mechanical,shang2021mechanical}. 
Unlike conventional alloys, these complex concentrated alloys possess a rugged energy landscape giving rise to intrinsic randomness in local Peierls stresses and, therefore, unusual pinning patterns and jerky glide dynamics of roughened dislocations \cite{Li2019,zhang2019effect}.
A relevant study by Hu et al. \cite{hu2018dislocation} demonstrated the accumulation of dislocation bands and pile-ups in a compressed HEA nanopillar, owing to the complex interplay with random obstacles, that significantly differs from the surface-induced annhilation mechanism observed in pure fcc metals.
Another complication arises from compositional/microstructural heterogeneities (local chemical ordering \cite{zhang2020short,wu2021short}, nano-precipitation \cite{ardell1985precipitation}, deformation-induced phase transition \cite{basu2020strengthening}) that interplay with dynamics of dislocations in often unpredictable and inextricable ways \cite{naghdi2022dislocation} in chemically complex alloys.
Prior applications of AE tests, as the gold standard in the field, mainly reported certain critical features of jerky plastic flow and associated strain bursts in HEAs \cite{chen2022acoustic,chen2022multiple,brechtl2023mesoscopic} but did not fully succeed in directly associating their temporal dynamics to bulk substructural features \cite{ahmed2021multiscale}. 
In-situ characterization of nanoscale deformation patterns in combination with AE experiments has been practically challenging due to complex microstructural origins of acoustic signals, highly-specialized instrumentation, and sample preparation difficulties.

Here in this study, our aim is to explore such microstructure-property correlations in the single-phase face-centered cubic (fcc) \glfour and \glsix as two exemplary HEA and MEA.
Our motivation for revisiting plasticity in deforming Cantor and \glsix alloys is two folds: \emph{\romannum{1}}) we aim to replicate laboratory-based investigations of dislocation avalanches using model simulation systems of \glfour and \glsix alloys \emph{\romannum{2}}) we seek for potential microstructural footprints in avalanche statistics to gain further understanding into underlying atomic-level deformation mechanisms.
To this end, we perform atomistic simulations of model Cantor and \glsix alloys under uniaxial tension and analyze serration features of the stress response together with the incurred slip patterns within the bulk sample.
As a comparative study, we further investigate avalanche properties in pure Ni which lacks the chemical/microstructural heterogeneity element (owing to lattice distortion) as in Cantor and \glsix alloys but still features nontrivial avalanche properties.
We, in particular, probe deformation rate effects on spatial-temporal evolution of slip events and their statistics at room temperature.
We find that dislocation avalanches in the slowly driven HEA and MEA exhibit a scale-free process characterized by asymptotic power-law regimes and critical scaling exponents that govern serrated plastic flow but show minimal variations with respect to the chemical composition.
Our findings indicate that the morphology of microstructural changes is strongly rate-dependent and exhibits meaningful correlations with avalanche size statistics at slow driving rates.

The paper's layout is as follows.
In Sec.~\ref{sec:methods}, we describe the numerical setup, sample preparation, loading protocols, and relevant simulation details including interatomic forces and shear test description. 
Section~\ref{sec:results} presents our simulation results relevant to investigations of dislocation avalanches followed by a  phase analysis of the microstructure as well as their potential correlations under different deformation rates.
In this context, Sec.~\ref{sec:avalanche} examines avalanche statistics (size and duration) to characterize their rate-dependence.
Microstructural signatures of dislocation avalanches will be discussed in Sec.~\ref{sec:microstructure} and \ref{sec:correlations}.
Section~\ref{sec:conclusions} presents relevant discussions and conclusions.

\section{\label{sec:methods}Methods \& Protocols}
We performed molecular dynamics simulations in LAMMPS \cite{LAMMPS} by implementing atomistic samples of size $N=10,000$ within a three-dimensional periodic cell.
We prepared cubic samples with dimension $L=40$~\r{A} along the $x[100]$, $y[010]$, and $z[001]$ directions.
The NPT ensembles were implemented via a Nose-Hoover thermostat and barostat with relaxation time scales $\tau_d^\text{therm}=10$ fs and $\tau_d^\text{bar}=100$ fs ($1~\text{fs}=10^{-15}$ s).
We also set the discretization time to $\Delta t= 1.0$ fs.
Samples were initially prepared via an energy minimization at $T=0$ K (at a fixed pressure) and subsequently thermalized at room temperature ($T=300$ K) {and constant pressure $P=0$ bar} for the duration of $100$ ps prior to loading. 
The interatomic forces were derived from the modified embedded-atom method potential developed recently by Choi et al. \cite{choi2018understanding}. 
Tensile tests were carried out by deforming the $z$ dimension of the simulation box at constant strain rates $\dot{\epsilon}_{zz} = 10^8-10^{10}~\text{s}^{-1}$ with $P_{xx}=P_{yy}=0$.
We also checked that the stress response and associated statistics are almost rate-independent below $\ratezero$  corresponding to a quasi-static limit. 

\section{\label{sec:results}Results}
We performed a series of tensile tests on model Cantor, \glsix, and Ni alloys at different deformation rates and room temperature. 
The evolution of the (normal) stress $\sigma_{zz}$ with the tensile strain $\epsilon_{zz}$ and the associated rate $\sigmadot$ are plotted in Fig.~\ref{fig:loadCurve}(a) at $T=300$ K and $\rate=10^8~\text{s}^{-1}$ corresponding to \glfour\!\!.
Upon yielding, the mechanical response is characterized by a stick-slip-type behavior with abrupt force drops preceded by longer stress build-up periods as in Fig.~\ref{fig:loadCurve}(b).
Similarly, the stress rate exhibits an intermittent dynamics with quiescent periods (i.e., $\sigmadot\simeq 0$) that are frequently interrupted by fairly short-lived bursts of events.
The latter are typically accompanied by slips across close-packed $\{111\}$ atomic planes in a fcc structure to form hcp layers within stacking fault regions as in Fig.~\ref{fig:loadCurve}(e) and (f).
Figure \ref{fig:loadCurve}(c) and (d) displays the fraction of atoms with hcp(bcc) structure $\rho_\text{hcp(bcc)}$ and its evolution with time $t$.
In what follows, we identify individual stress avalanches and probe their statistics (i.e., size and duration) showing non-trivial correlations with the structure of slip planes at low strain rates.

\subsection{\label{sec:avalanche}Avalanche Analysis: Size \& Duration}
We define the avalanche size as the magnitude of the stress drop $S=-\int_{t_i}^{t_i+\mathcal{T}_{i}} \sigmadot~dt$ corresponding to event $i$ initiated at $t_i$ with duration $\mathcal{T}_i$ as illustrated in Fig.~\ref{fig:loadCurve}(b).
During the avalanche period, the stress rate exceeds a threshold value set by the median rate, $-\sigmadot \ge \dot{\sigma}_\text{th}$ with $\dot{\sigma}_\text{th}=\text{median}(-\sigmadot)$.
We also checked the robustness of our results against variations in $\dot{\sigma}_\text{th}$ (data not shown).  
To remove the thermal noise from the stress signal, we use the optimal (Wiener) filtering (see the Supplementary Materials for further details) and ensure that our avalanche analysis is performed on sufficiently smooth timeseries. 
We systematically gather statistics of avalanches incurred at the strain interval $\epsilon_{zz}=0.2-1.0$ within the steady-state flow regime.
To improve the collected statistics, we consider fairly large statistical ensembles with order $10-100$ realizations per deformation rate. 
The avalanche analysis is performed on an extensive dataset, typically including around $10^3-10^4$ avalanches in each case, to ensure the robustness and accuracy of the estimated scaling exponents.

\begin{figure}[t]
    \centering
    \begin{overpic}[width=0.49\columnwidth]{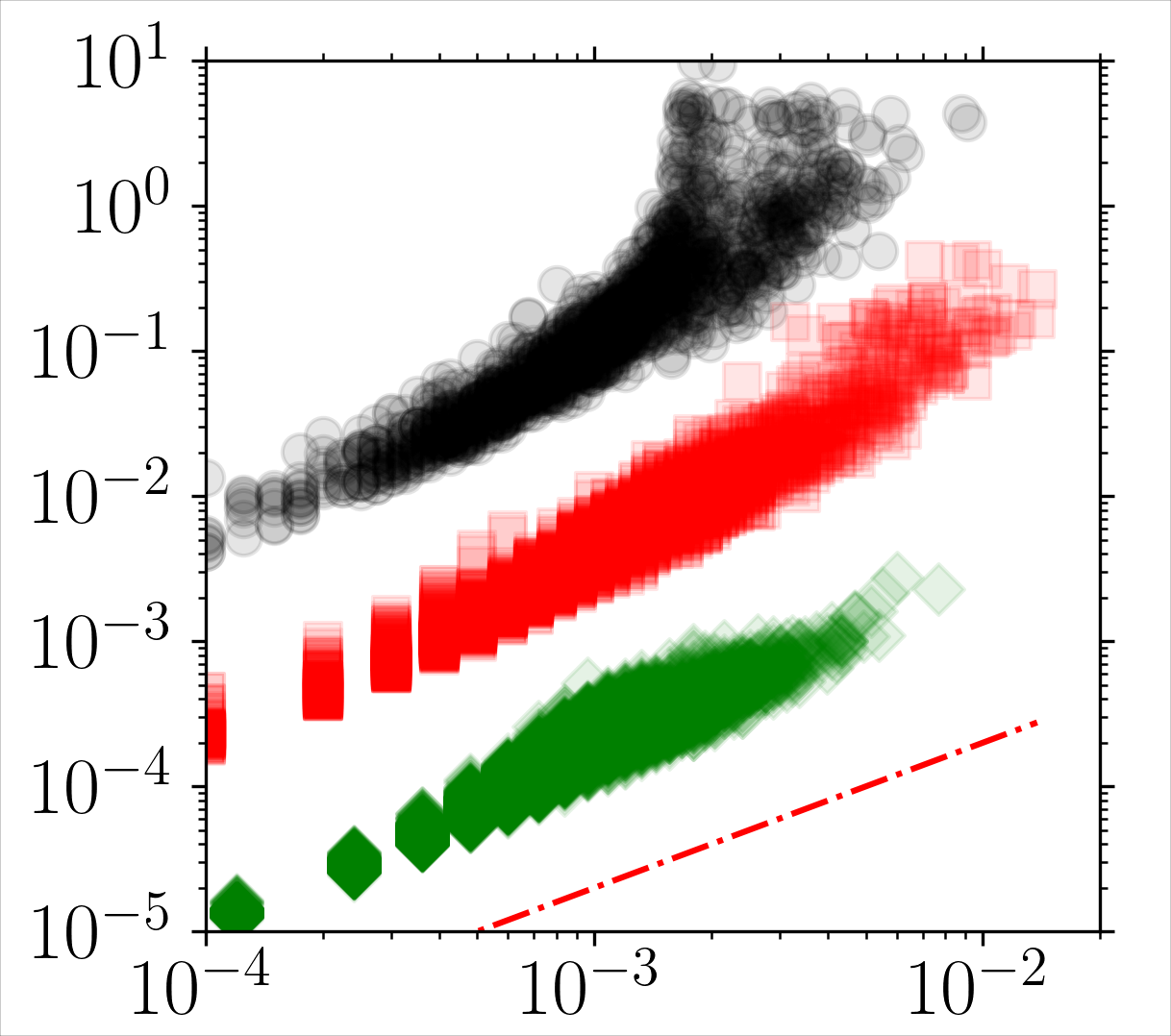}
        \LabelFig{16}{87}{$a)$~\scriptsize \glfour}
        \Labelxy{50}{-6}{0}{$\rate \mathcal{T}$}
        \Labelxy{-8}{45}{90}{{$S$~\scriptsize (Gpa)}}
        \begin{tikzpicture}
          \coordinate (a) at (0,0);
            \node[white] at (a) {\tiny.};               %
            \drawSlope{2.9}{.6}{0.35}{111}{red}{\hspace{-2pt}$\gamma$}
            \legLine{1.1}{3.1}{red}{\hspace{-10pt}$\scriptstyle S\propto \mathcal{T}^{\gamma}$}{0.5}
		\end{tikzpicture}
	\end{overpic}
    \begin{overpic}[width=0.49\columnwidth]{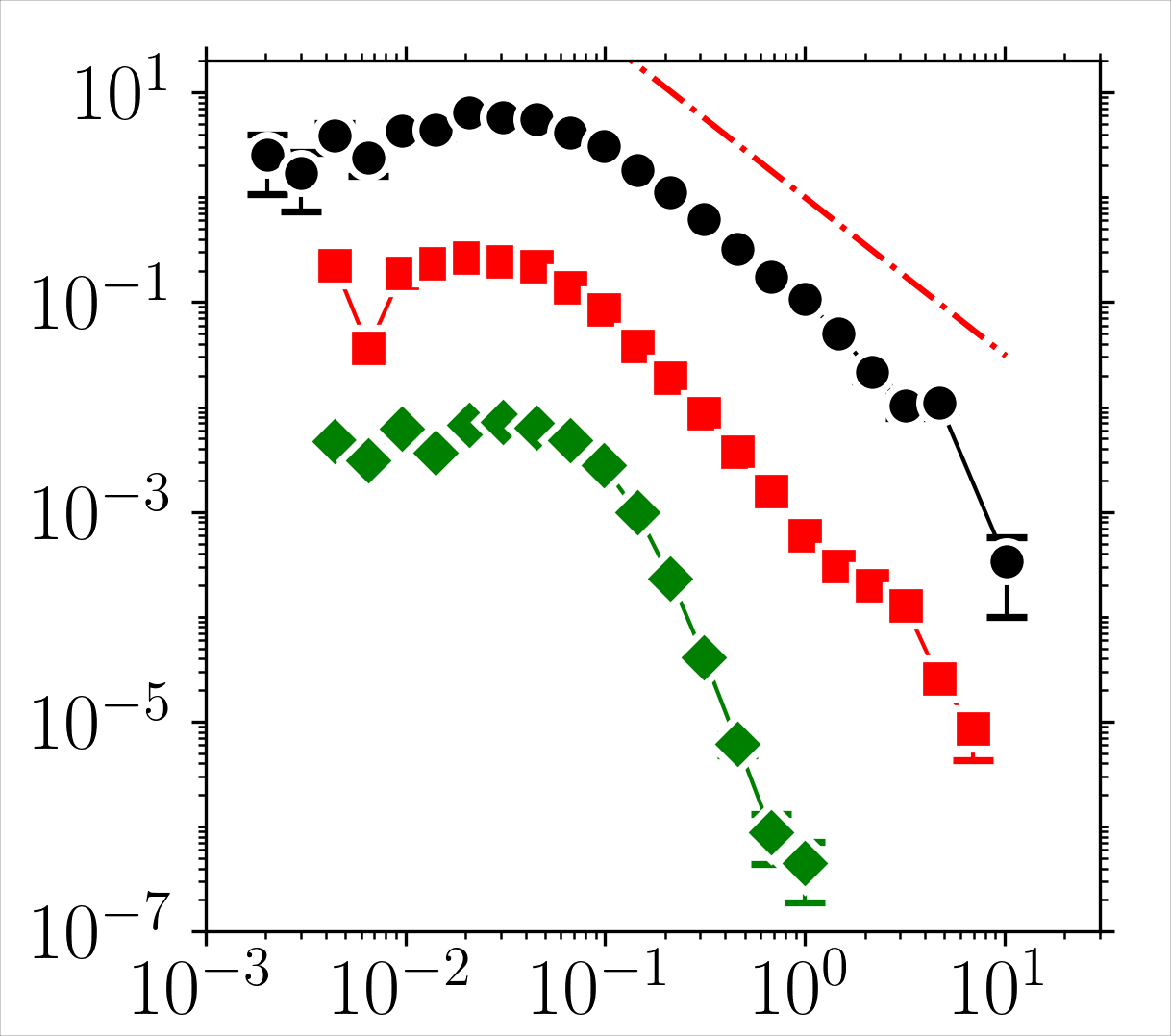}
        \put(96,53.5){\includegraphics[width=0.05\textwidth]{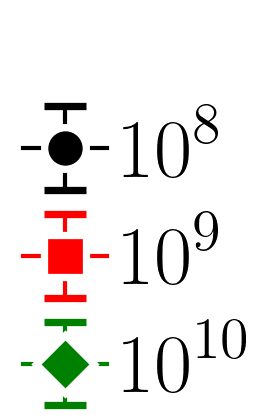}}
        \Labelxy{99}{82}{0}{$\rate ~\scriptstyle (\text{s}^{-1})$}
        \LabelFig{16}{87}{$b)$}
        \Labelxy{50}{-6}{0}{$S$~\scriptsize (Gpa)}
        \Labelxy{-6}{45}{90}{{$P(S)$}}
        \begin{tikzpicture}
          \coordinate (a) at (0,0);
            \node[white] at (a) {\tiny.};               %
            \drawSlope{2.9}{3.0}{0.35}{236}{red}{\hspace{-2pt}$\tau$}
            \legLine{1.0}{0.55}{red}{\hspace{8pt}$\scriptstyle P(S)\propto S^{-\tau}$}{0.5}
		\end{tikzpicture}
	\end{overpic}
    \vspace{+2pt}
    
    \begin{overpic}[width=0.49\columnwidth]{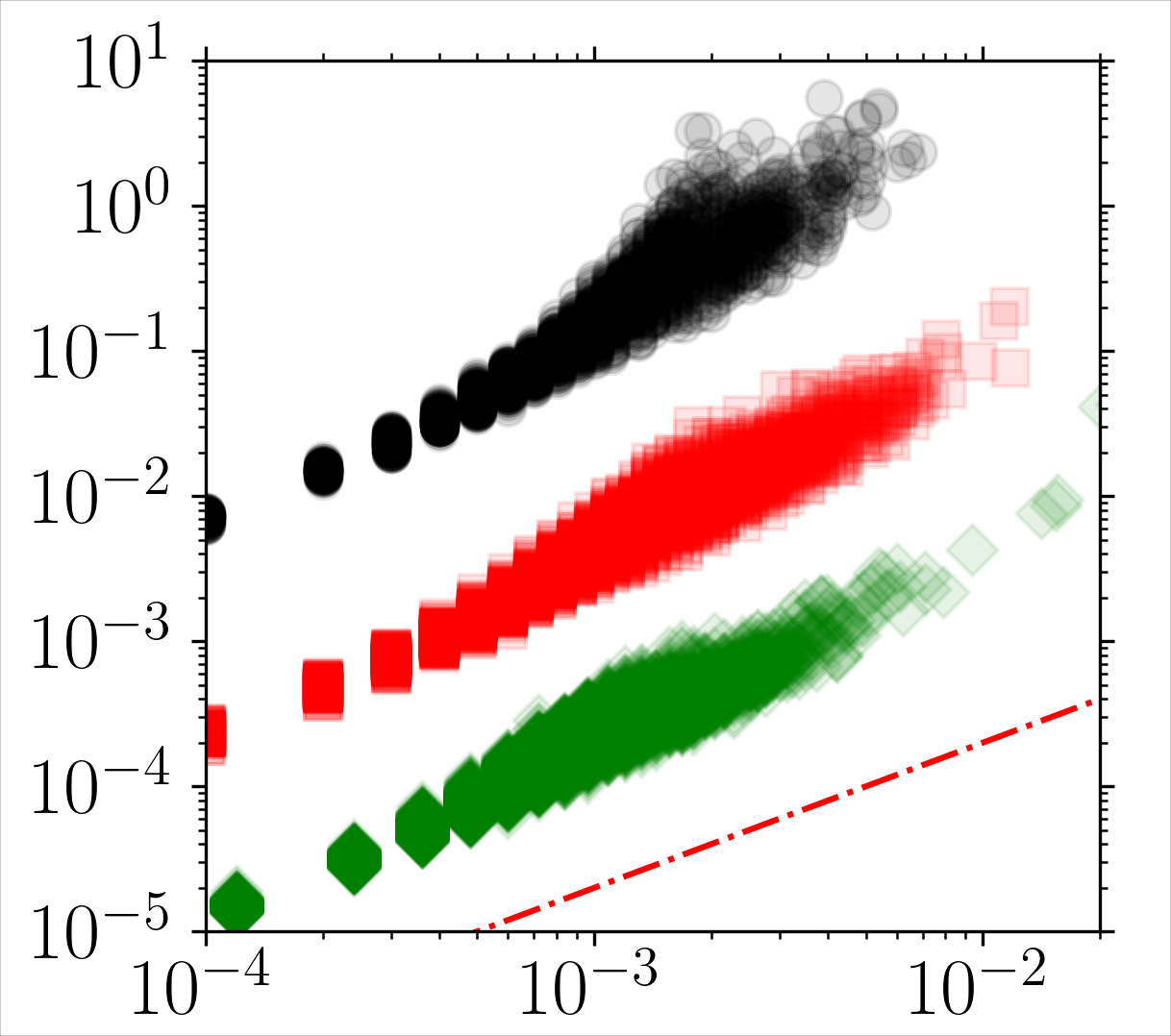}
        \LabelFig{16}{87}{$c)$ \scriptsize NiCoCr}
        \Labelxy{50}{-6}{0}{$\rate \mathcal{T}$}
        \Labelxy{-8}{45}{90}{{$S$~\scriptsize (Gpa)}}
        \begin{tikzpicture}
          \coordinate (a) at (0,0);
            \node[white] at (a) {\tiny.};               %
            %
		\end{tikzpicture}
	\end{overpic}
    \begin{overpic}[width=0.49\columnwidth]{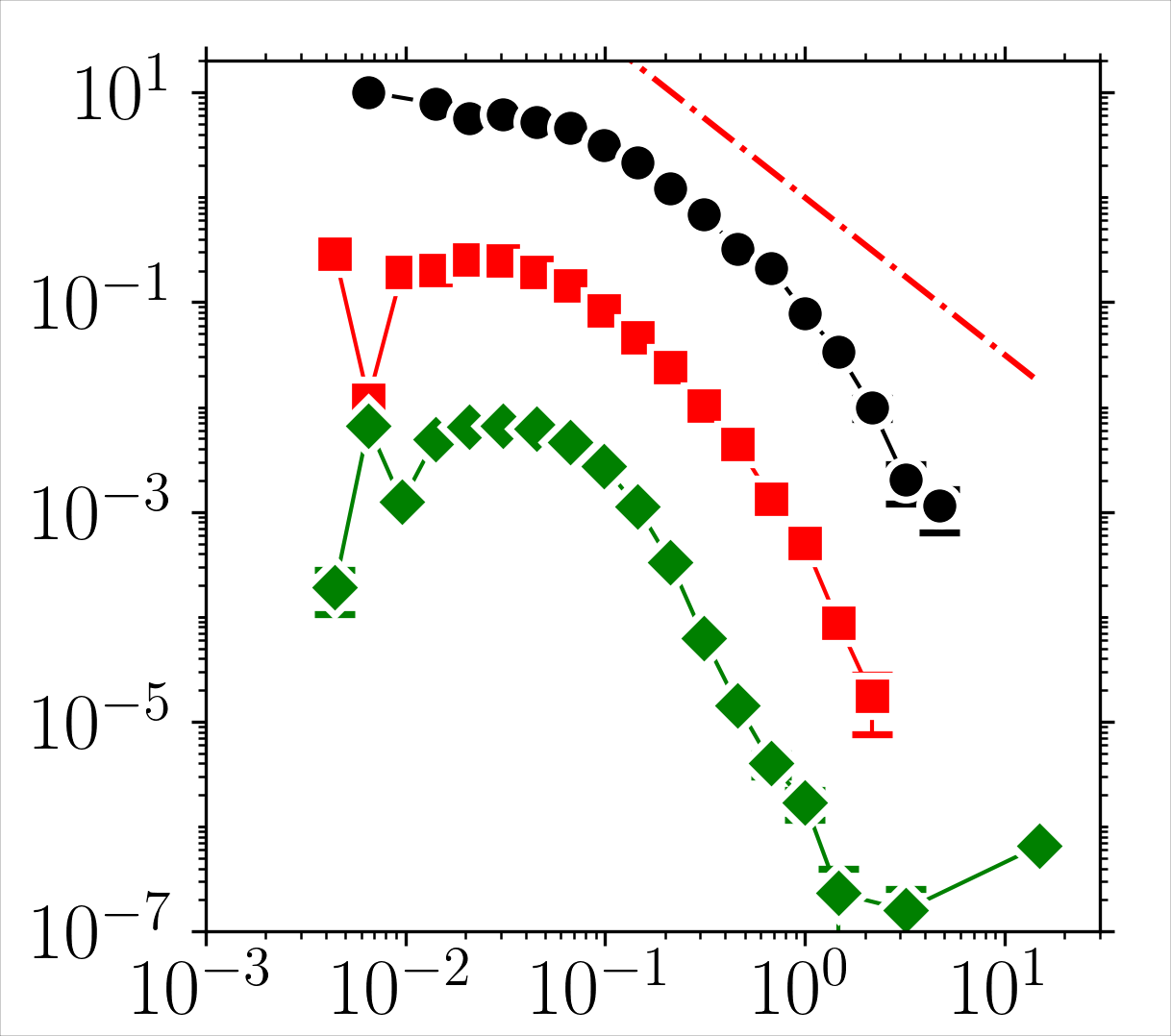}
        \LabelFig{16}{87}{$d)$}
        \Labelxy{50}{-6}{0}{$S$~\scriptsize (Gpa)}
        \Labelxy{-6}{45}{90}{{$P(S)$}}
        \begin{tikzpicture}
          \coordinate (a) at (0,0);
            \node[white] at (a) {\tiny.};               %
            %
		\end{tikzpicture}
	\end{overpic}
    \vspace{+2pt}
    
    \begin{overpic}[width=0.49\columnwidth]{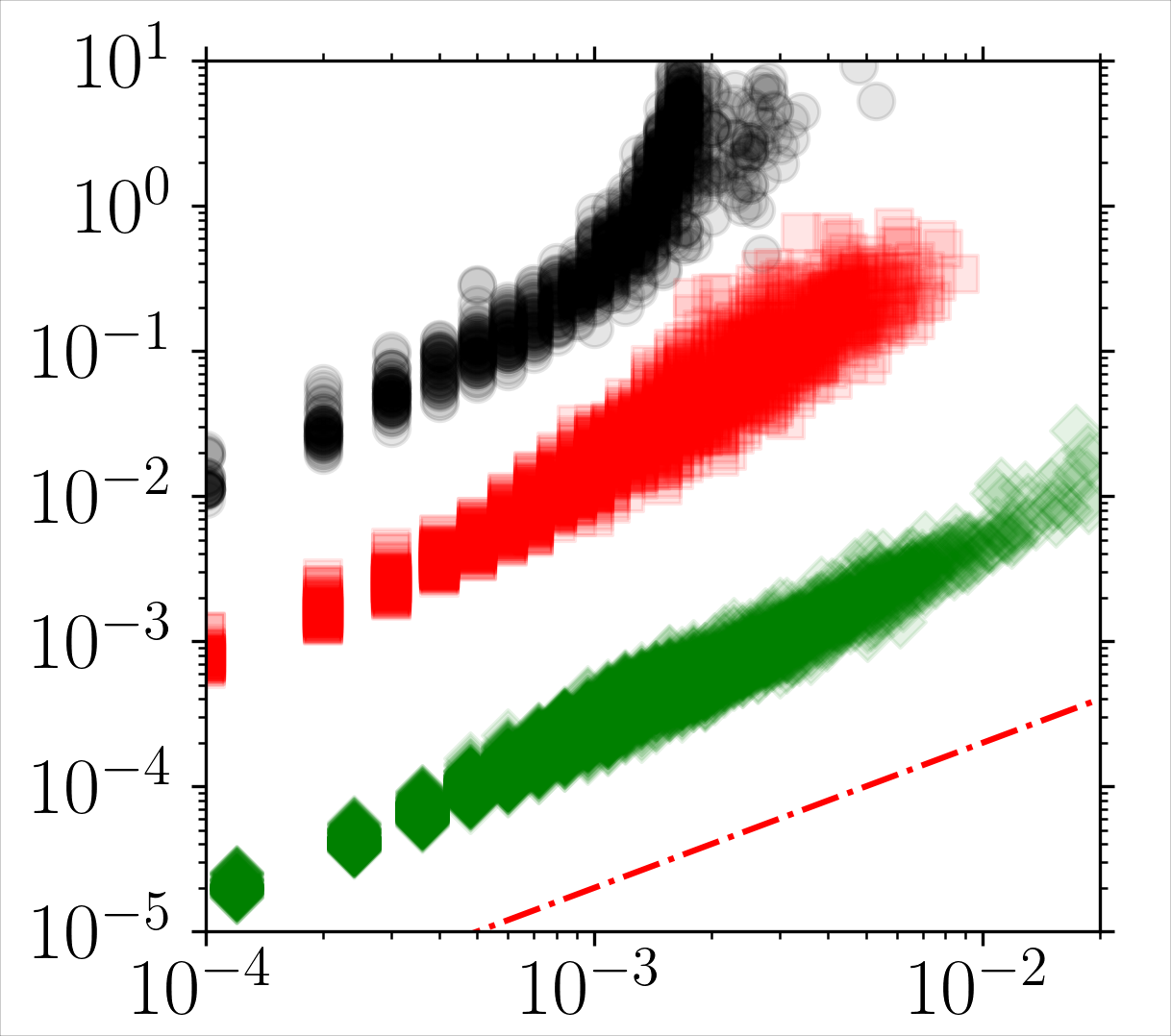}
        \LabelFig{16}{87}{$e)$ \scriptsize Ni}
        \Labelxy{50}{-6}{0}{$\rate \mathcal{T}$}
        \Labelxy{-8}{45}{90}{{$S$~\scriptsize (Gpa)}}
        \begin{tikzpicture}
          \coordinate (a) at (0,0);
            \node[white] at (a) {\tiny.};               %
            %
		\end{tikzpicture}
	\end{overpic}
    \begin{overpic}[width=0.49\columnwidth]{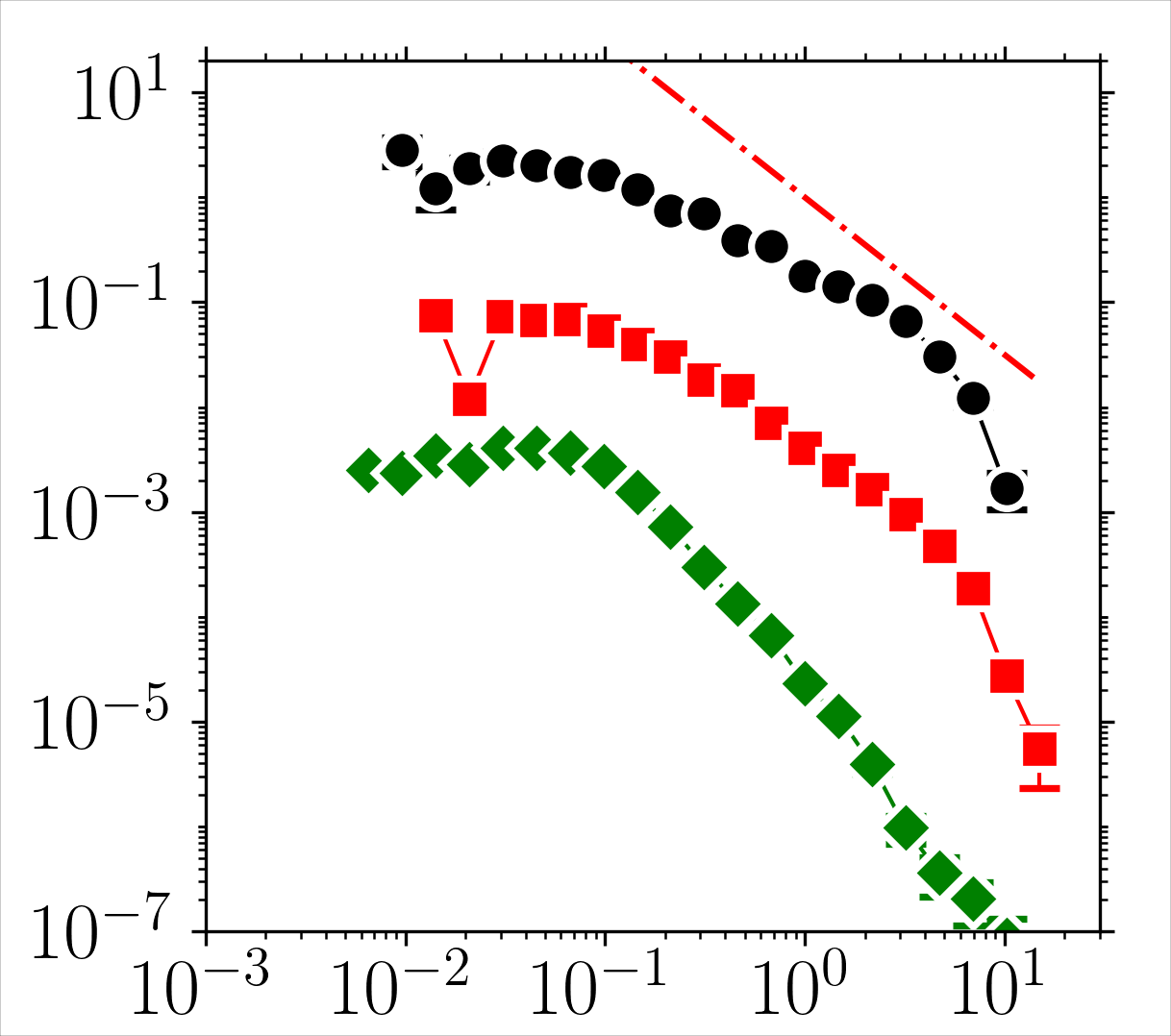}
        \LabelFig{16}{87}{$f)$}
        \Labelxy{50}{-6}{0}{$S$~\scriptsize (Gpa)}
        \Labelxy{-6}{45}{90}{{$P(S)$}}
        \begin{tikzpicture}
          \coordinate (a) at (0,0);
            \node[white] at (a) {\tiny.};               %
            %
		\end{tikzpicture}
	\end{overpic}
    \caption{Avalanche statistics at $T=300$ K and multiple strain rates $\rate$ corresponding to \glfour\!\!, \glsix\!\!, and Ni. \textbf{a, c, d}) Scatter plot of avalanche size $S$ and scaled duration $\rate\mathcal{T}$ \textbf{b, d, f}) avalanche size distributions $P(S)$. The dashdotted lines denote power laws \textbf{a, c, e}) $S\propto \mathcal{T}^{\gamma}$ with  $\gamma=1.0$ \textbf{b, d, f}) $P(S)\propto S^{-\tau}$ with $\tau=3/2$. The error bars indicate standard errors. The data are shifted vertically for the sake of clarity.}
    \label{fig:statistics}
\end{figure}

Figure~\ref{fig:statistics}(a-f) displays the scatter plot of the avalanche size $S$ and event duration $\mathcal{T}$, scaled by $\rate^{-1}$, as well as size distributions $P(S)$ at $T=300$ K and various rates $\rate$ corresponding to Cantor alloy, \glsix\!\!\!, and pure Ni.
The scatter plots in Fig.~\ref{fig:statistics}(a), (c), and (e) demonstrate that, statistically speaking, larger avalanches tend to have longer duration with a scaling behavior that may be described on average as $\langle S \rangle \propto \mathcal{T}^{\gamma}$ with $\gamma\simeq1.0$.
The observed scaling regime appears to be fairly limited at the slowest rate $\ratezero$ with the duration that tends to saturate at large avalanche sizes, possibly due to size effects.
In Fig.~\ref{fig:statistics}(b), the size distribution associated with the slowest rate in Cantor alloy decays as a power-law $P(S)\propto S^{-\tau}$ (above a certain cut-off size $S_c \simeq 10^{-1}$ Gpa) which spans at least two decades in $S$ and seems to be well-predicted by the mean-field estimate $\tau=3/2$ \cite{fisher1998collective}.
As $\rate$ is increased toward $10^{10}~\text{s}^{-1}$, the size distributions tend to exhibit a steeper fall-off  with a nearly exponential-like drop at the fastest deformation rate.
We observe very similar trends for avalanche statistics in \glsix as in Fig.~\ref{fig:statistics}(c) and (d) in terms of the rate dependence, except for a comparatively limited power-law scaling regime associated with $P(S)$ at $\ratezero$.  
The scaling between the size and duration in the case of pure Ni closely resembles that of the two alloys as in Fig.~\ref{fig:statistics}(e).
However, the avalanche size exponent in Fig.~\ref{fig:statistics}(f) is notably shallower than the mean-field prediction $\tau<3/2$ at the slowest rate but tends to become more mean-field like at the intermediate strain rate.
We note that the observed power-law behavior in Ni appears to be quite sensitive to the filtering process and slight variations in the relevant parameters lead to a better agreement with theoretical predictions (refer to Fig.~S3(c)). 

\subsection{\label{sec:microstructure}Microstructural Analysis: Crystal phase, Cluster Statistics, and Slip Planes}
As a structural metric associated with dislocation avalanches, we identified atomic structure types via the common neighbor analysis implemented in OVITO \cite{stukowski2009visualization}, seeking for atoms in hexagonal close-packed (hcp) and body-centered cubic (bcc) arrangements.
The hcp atoms are associated with stacking faults which are bounded by partial dislocations in a face-centered cubic (fcc) structure.

\begin{figure}[t]
    \centering
    \begin{overpic}[width=0.49\columnwidth]{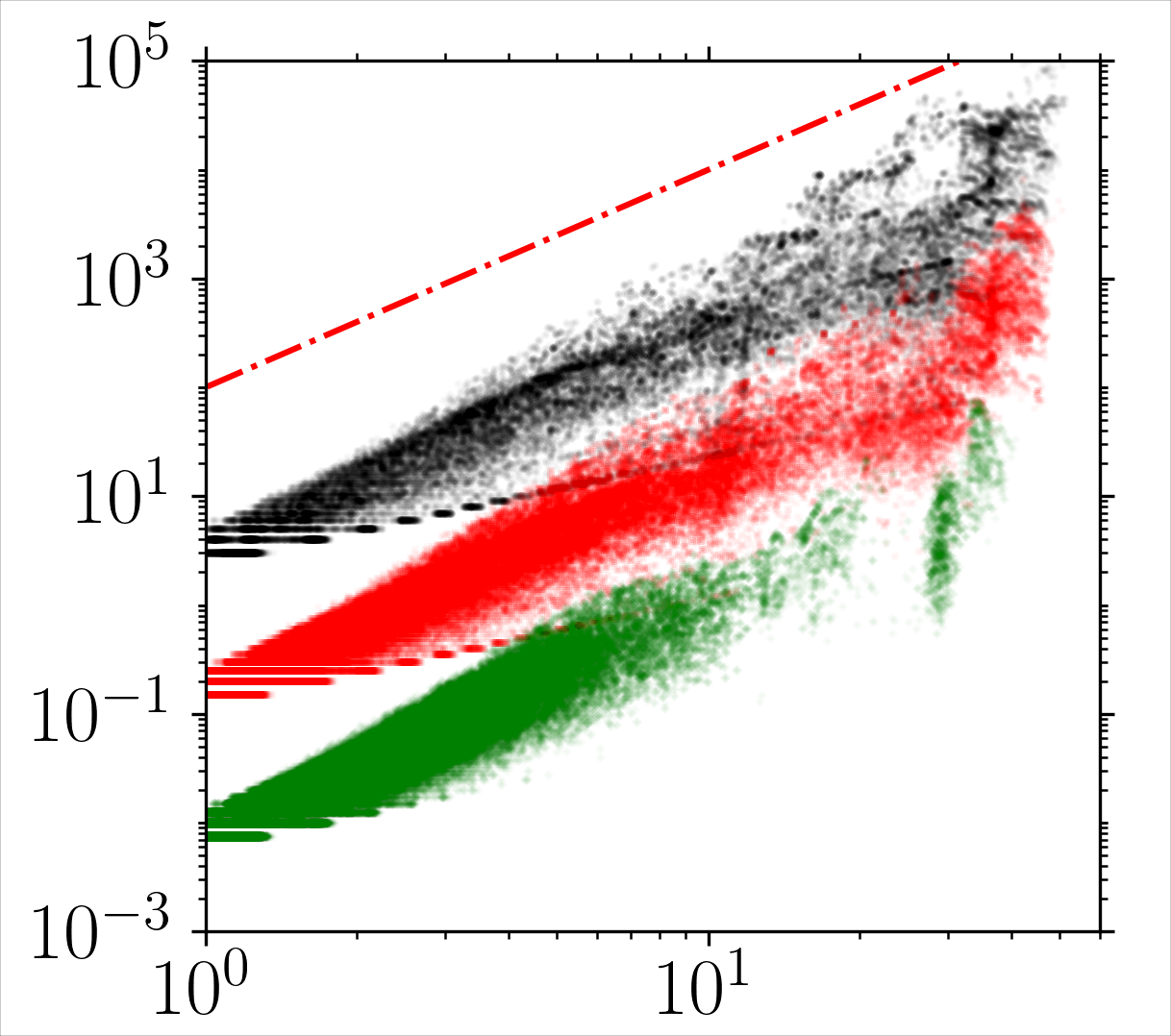}
        \LabelFig{16}{87}{$a)$~\scriptsize \glfour}
        \Labelxy{50}{-6}{0}{$r_g$~\scriptsize (\r{A})}
        \Labelxy{-4}{41}{90}{$s_\text{hcp}$}
        \begin{tikzpicture}
          \coordinate (a) at (0,0);
            \node[white] at (a) {\tiny.};               %
            \drawSlope{1.6}{2.8}{0.35}{-65}{red}{\hspace{1pt}$d_f$}
		\end{tikzpicture}
    \end{overpic}
    \begin{overpic}[width=0.49\columnwidth]{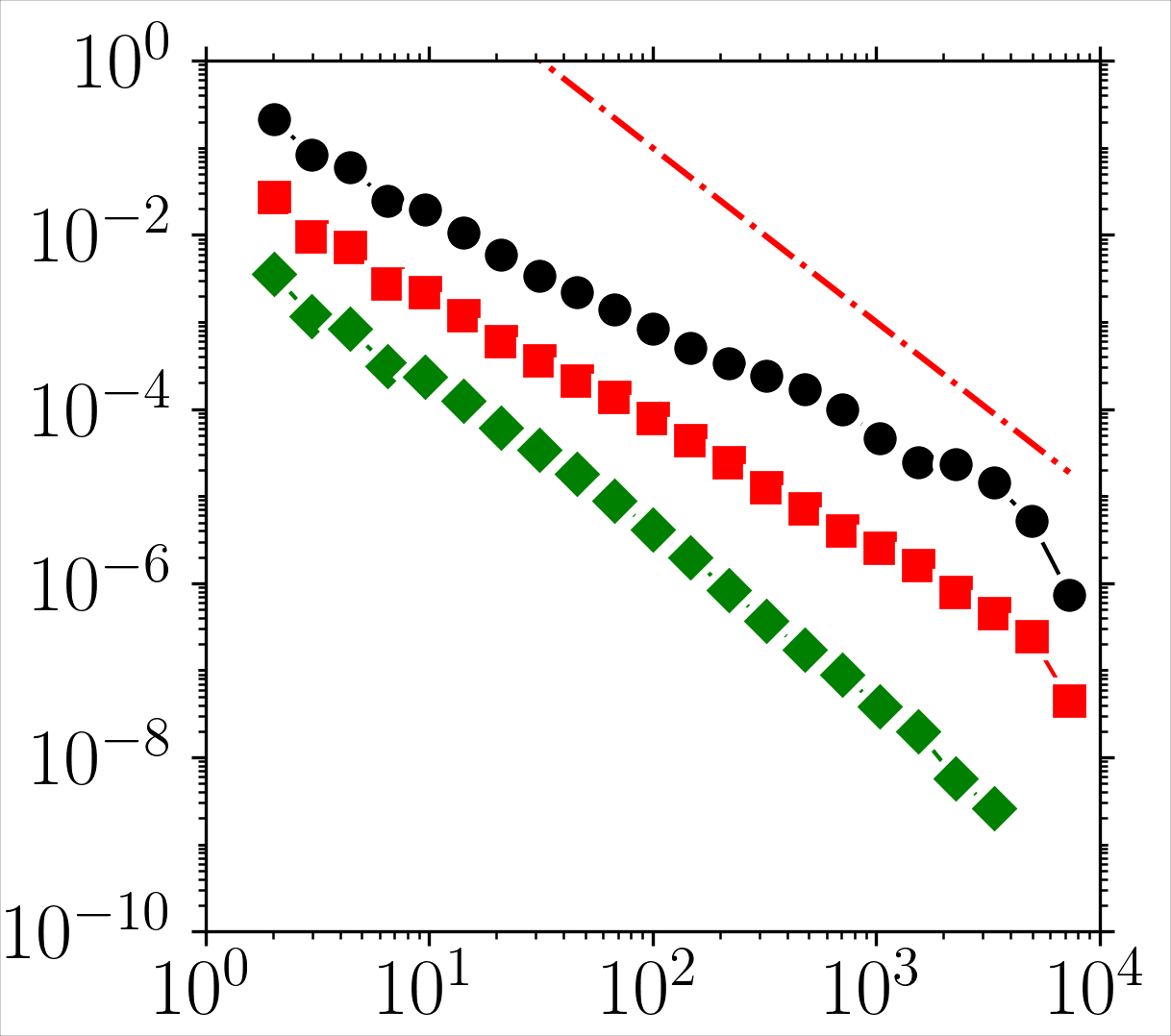}
        \LabelFig{16}{87}{$b)$} 
        \Labelxy{50}{-6}{0}{$s_\text{hcp}$}
        \Labelxy{-6}{36}{90}{$P(s_\text{hcp})$}
        \put(96,53.5){\includegraphics[width=0.05\textwidth]{Figs/legend_E1-4.png}}
        \Labelxy{99}{82}{0}{$\rate ~\scriptstyle (\text{s}^{-1})$}
        \begin{tikzpicture}
          \coordinate (a) at (0,0);
            \node[white] at (a) {\tiny.};               %
            \drawSlope{2.6}{3}{0.35}{234}{red}{\hspace{-2pt}$\tau_c$}
		\end{tikzpicture}
    \end{overpic}
     \vspace{2pt}

    \begin{overpic}[width=0.49\columnwidth]{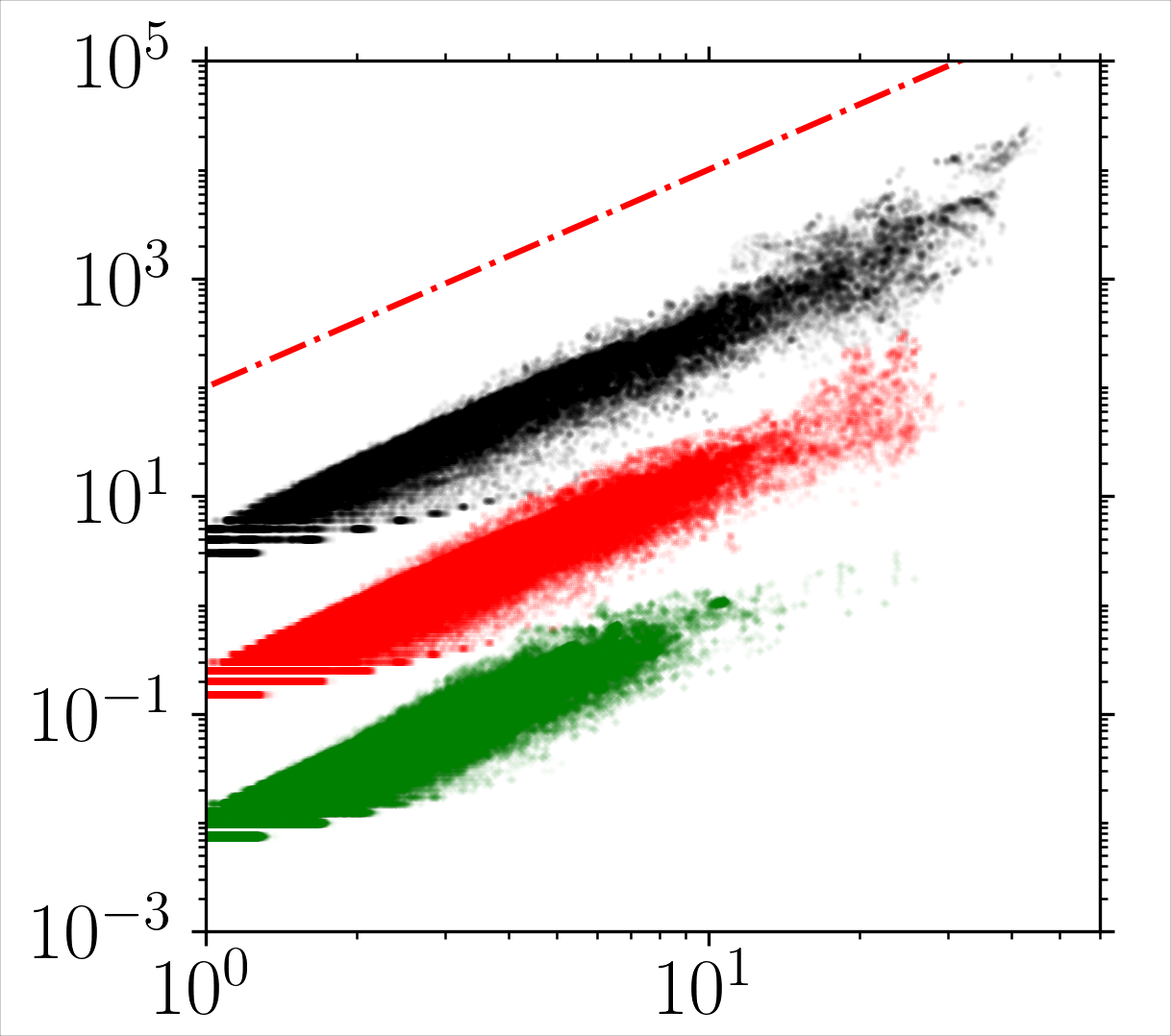}
        \LabelFig{16}{87}{$c)$~\scriptsize \glsix}
        \Labelxy{50}{-6}{0}{$r_g$~\scriptsize (\r{A})}
        \Labelxy{-4}{41}{90}{$s_\text{hcp}$}
    \end{overpic}
    \begin{overpic}[width=0.49\columnwidth]{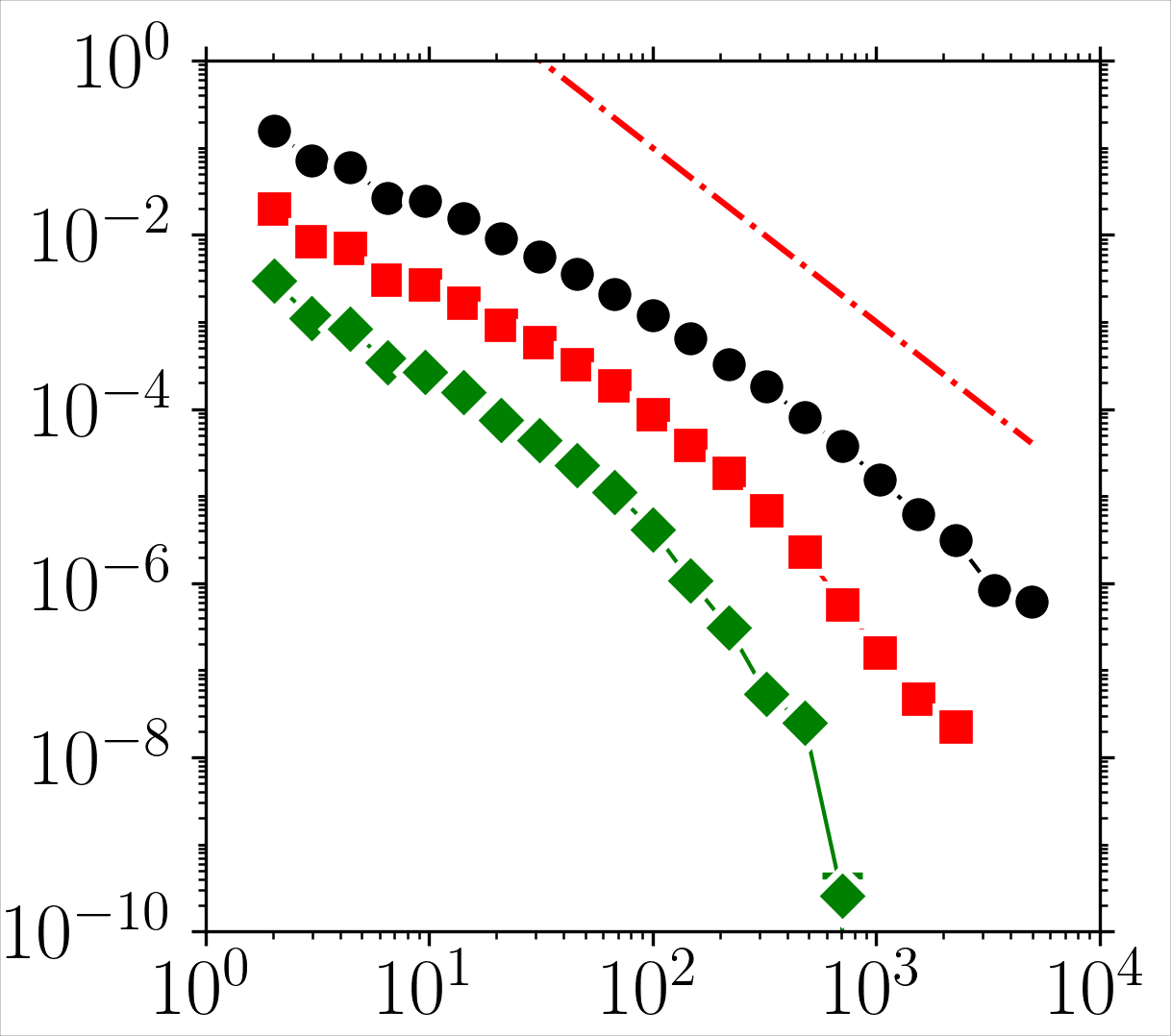}
        \LabelFig{16}{87}{$d)$} 
        \Labelxy{50}{-6}{0}{$s_\text{hcp}$}
        \Labelxy{-6}{36}{90}{$P(s_\text{hcp})$}
    \end{overpic}
     \vspace{2pt}

    \begin{overpic}[width=0.49\columnwidth]{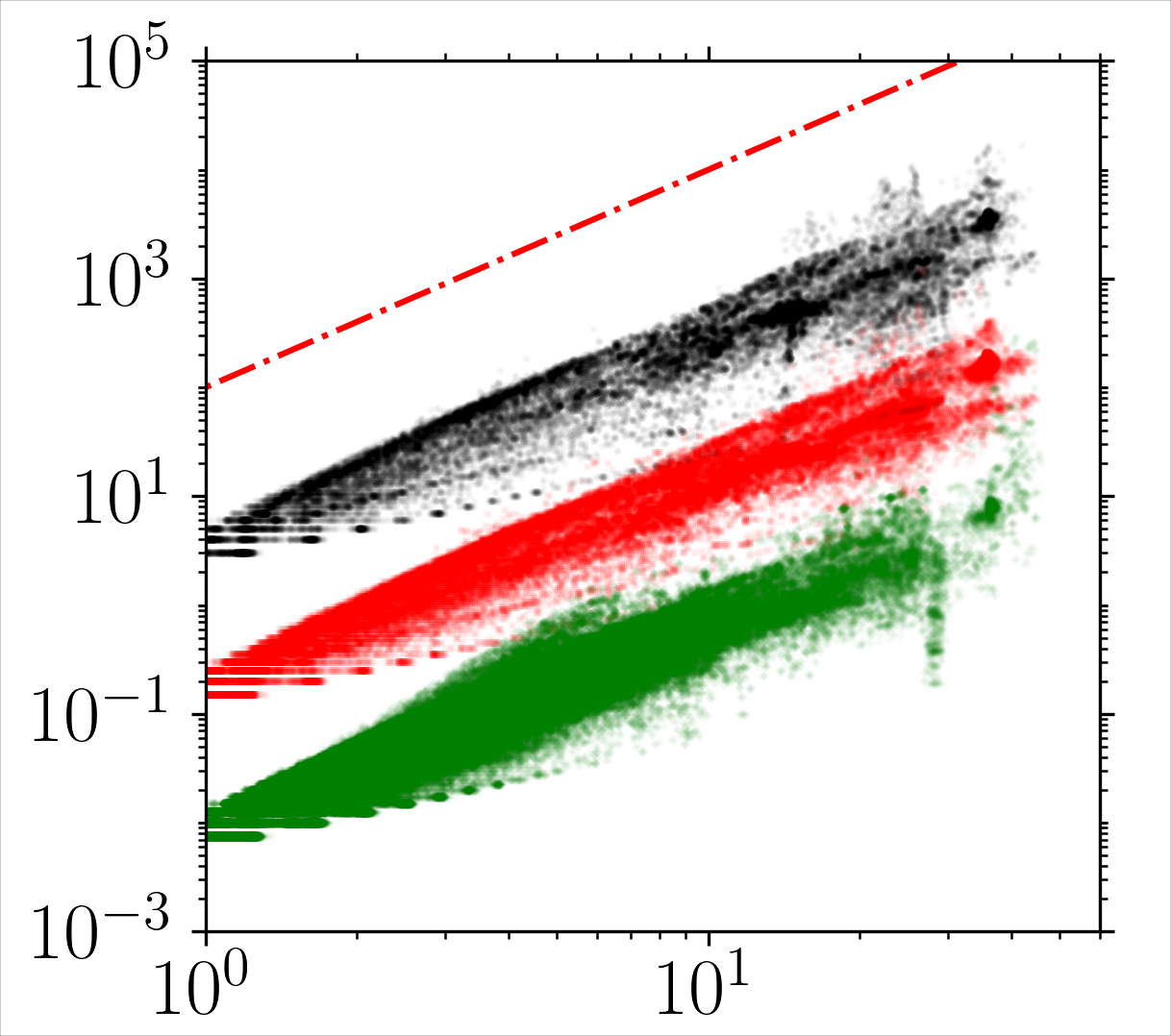}
        \LabelFig{16}{87}{$e)$~\scriptsize Ni}
        \Labelxy{50}{-6}{0}{$r_g$~\scriptsize (\r{A})}
        \Labelxy{-4}{41}{90}{$s_\text{hcp}$}
    \end{overpic}
    \begin{overpic}[width=0.49\columnwidth]{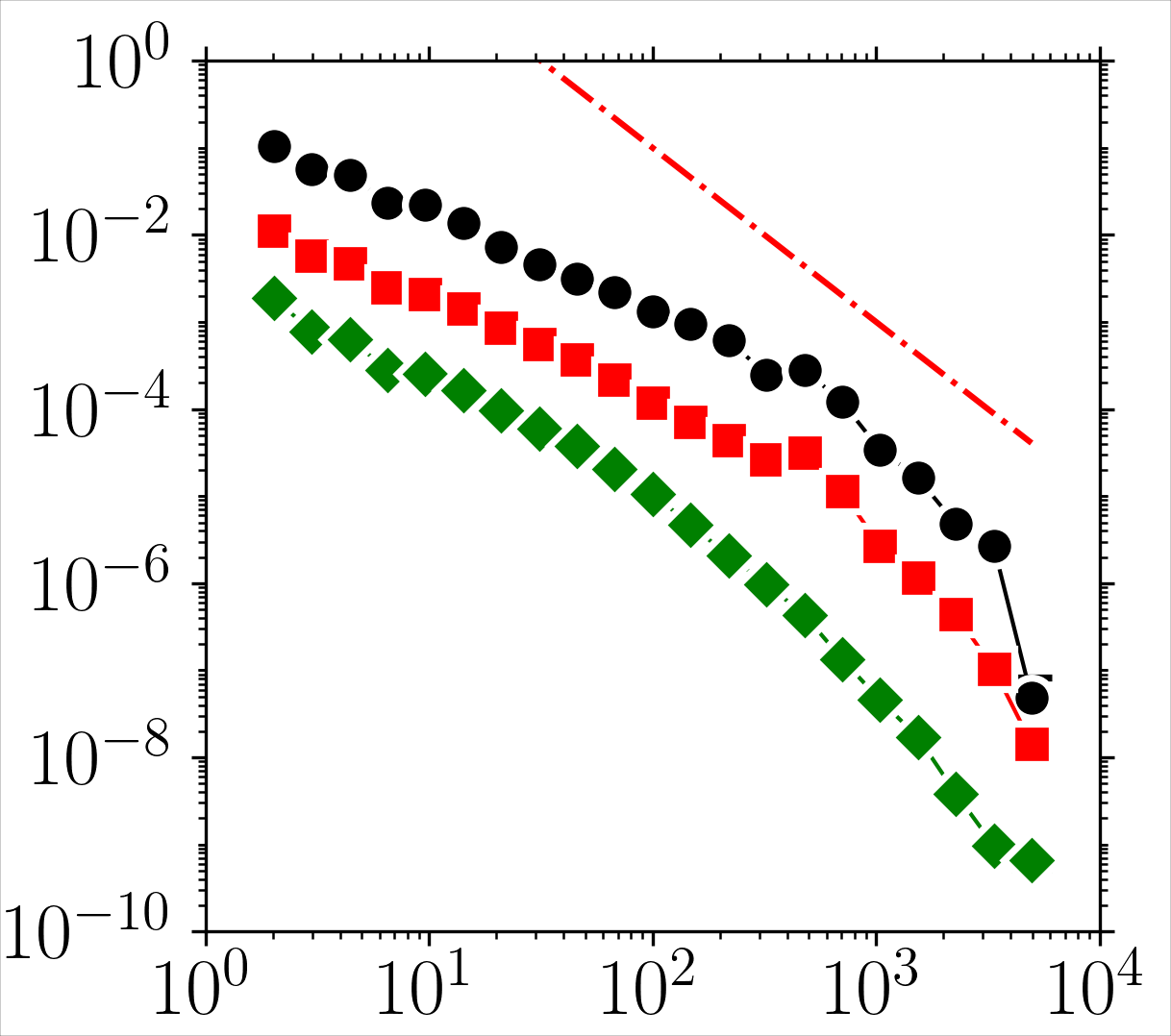}
        \LabelFig{16}{87}{$f)$} 
        \Labelxy{50}{-6}{0}{$s_\text{hcp}$}
        \Labelxy{-6}{36}{90}{$P(s_\text{hcp})$}
    \end{overpic}
    \caption{Cluster size statistics at different strain rates $\rate$ associated with \glfour, \glsix, and Ni. \textbf{a, c, e}) Scatter plot of cluster size $s_\text{hcp}$ and associated radius of gyration $r_g$ \textbf{b, d, f}) Cluster size distribution $P(s_\text{hcp})$. The dashdotted lines denote power laws \textbf{a, c, e}) $s_\text{hcp}\propto r_g^{d_f}$ with $d_f=2$ \textbf{b, d, f}) $P(s_\text{hcp})\propto s_\text{hcp}^{-\tau_c}$ with $\tau_c=2$. The error bars indicate standard errors. The data are shifted vertically for the sake of clarity.}
    \label{fig:cluster_size}
\end{figure}

\begin{figure*}[t]
    \centering
    \begin{overpic}[width=0.49\columnwidth]{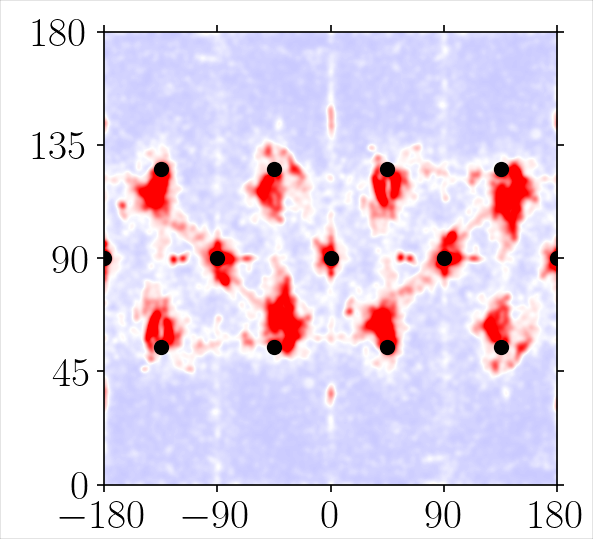}
        \LabelFig{16}{89}{$a)$~\scriptsize \glfour\!\!, $\epsilon_{zz}=\scriptstyle 10^8~ \text{s}^{-1}$}
        \Labelxy{50}{-6}{0}{$\theta$\scriptsize ~(deg)}
        \Labelxy{-4}{41}{90}{$\phi$\scriptsize ~(deg)}
        \put(80,26){\tiny $(\bar{1}11)$}
        \put(60,26){\tiny $(111)$}
        \put(40,26){\tiny $(1\bar{1}1)$}
        \put(20,26){\tiny $(\bar{1}\bar{1}1)$}

        \put(80,66){\tiny $(\bar{1}1\bar{1})$}
        \put(60,66){\tiny $(11\bar{1})$}
        \put(40,66){\tiny $(1\bar{1}\bar{1})$}
        \put(20,66){\tiny $(\bar{1}\bar{1}\bar{1})$}
        
        \put(30,40){\tiny $(0\bar{1}0)$}
        \put(50,40){\tiny $(100)$}
        \put(70,40){\tiny $(010)$}
    \end{overpic}
    \begin{overpic}[width=0.49\columnwidth]{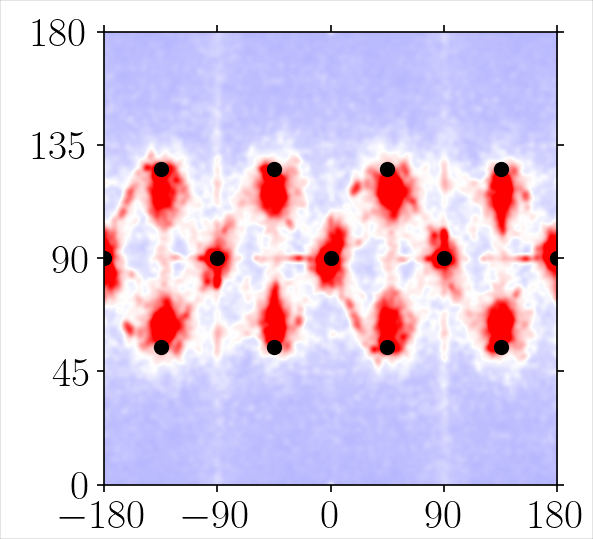}
        \LabelFig{16}{89}{$b)~\scriptstyle 10^{9}~ \text{s}^{-1}$}
        \Labelxy{50}{-6}{0}{$\theta$\scriptsize ~(deg)}
    \end{overpic}
    \begin{overpic}[width=0.49\columnwidth]{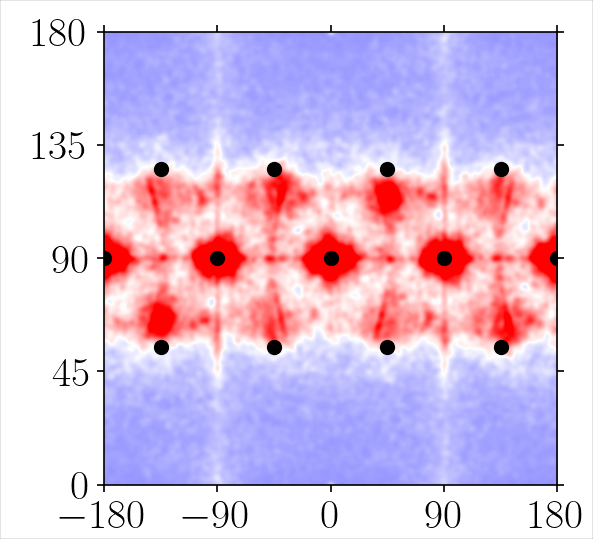}
        \LabelFig{16}{89}{$c)~\scriptstyle 10^{10}~ \text{s}^{-1}$}
        \Labelxy{50}{-6}{0}{$\theta$\scriptsize ~(deg)}
    \end{overpic}
    %
    \vspace{+16pt}
    \hspace{-2pt}

    \begin{overpic}[width=0.49\columnwidth]{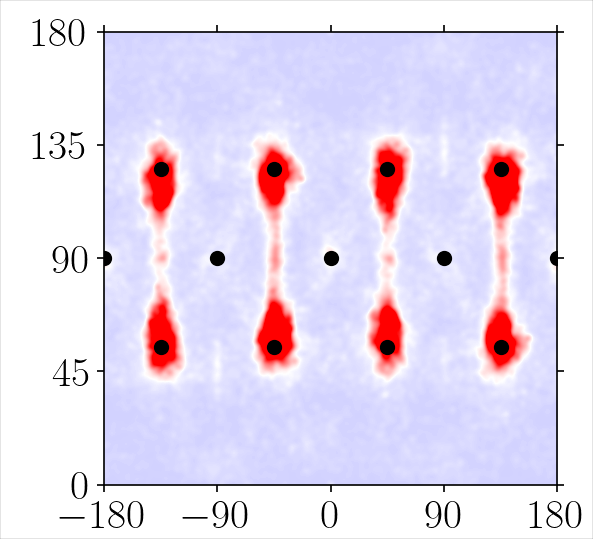}
        \LabelFig{16}{89}{$d)$~\scriptsize \glsix, $\epsilon_{zz}=\scriptstyle 10^8~ \text{s}^{-1}$}
        \Labelxy{50}{-6}{0}{$\theta$\scriptsize ~(deg)}
        \Labelxy{-4}{41}{90}{$\phi$\scriptsize ~(deg)}
    \end{overpic}
    \begin{overpic}[width=0.49\columnwidth]{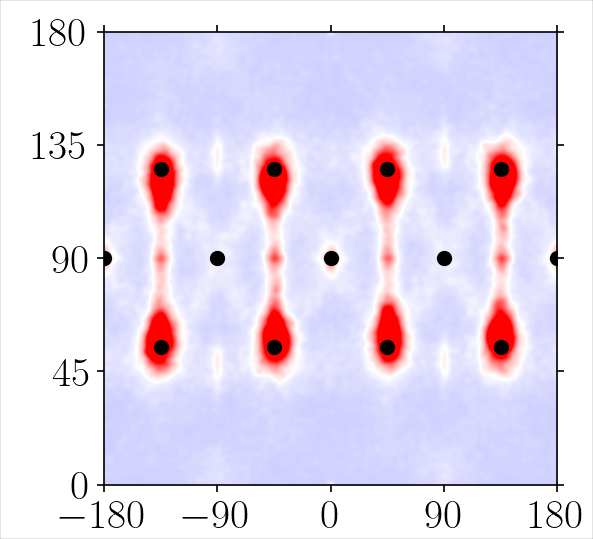}
        \LabelFig{16}{89}{$e)~\scriptstyle 10^{9}~ \text{s}^{-1}$}
        \Labelxy{50}{-6}{0}{$\theta$\scriptsize ~(deg)}
    \end{overpic}
    \begin{overpic}[width=0.49\columnwidth]{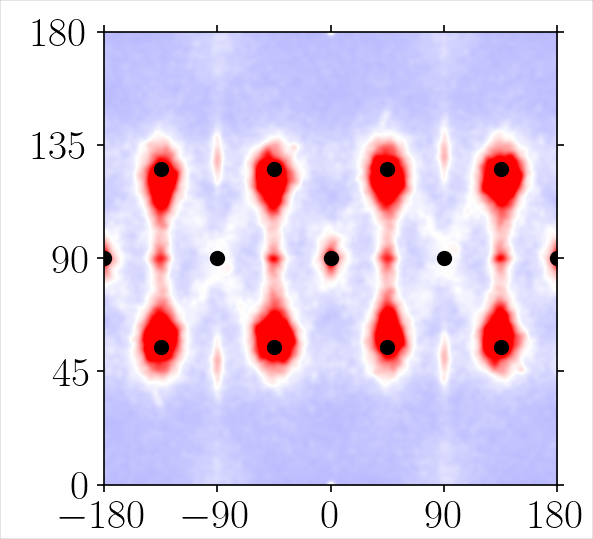}
        \LabelFig{16}{89}{$f)~\scriptstyle 10^{10}~ \text{s}^{-1}$}
        \Labelxy{50}{-6}{0}{$\theta$\scriptsize ~(deg)}
    \end{overpic}
    \vspace{+16pt}
    \hspace{-2pt}

    \begin{overpic}[width=0.49\columnwidth]{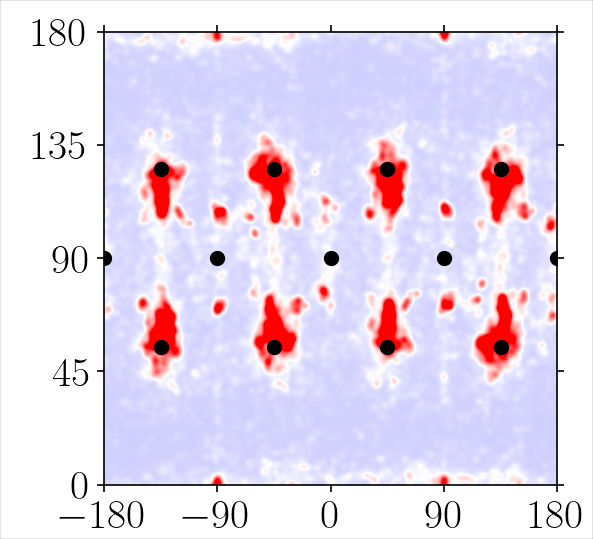}
        \LabelFig{16}{89}{$g)$~\scriptsize Ni, $\epsilon_{zz}=\scriptstyle 10^8~ \text{s}^{-1}$}
        \Labelxy{50}{-6}{0}{$\theta$\scriptsize ~(deg)}
        \Labelxy{-4}{41}{90}{$\phi$\scriptsize ~(deg)}
    \end{overpic}
    \begin{overpic}[width=0.49\columnwidth]{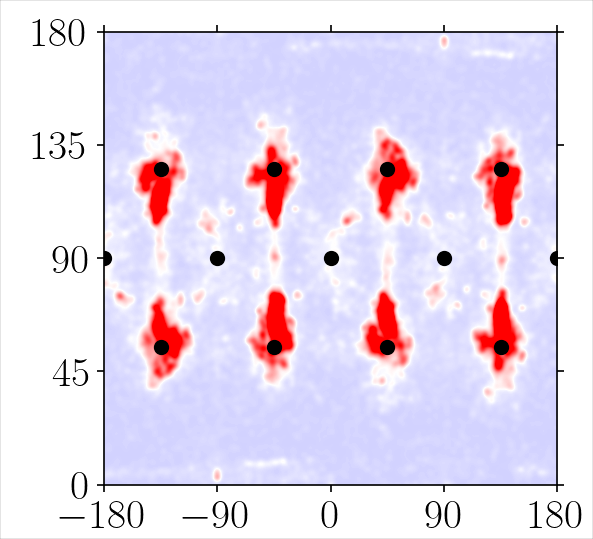}
        \LabelFig{16}{89}{$h)~\scriptstyle 10^{9}~ \text{s}^{-1}$}
        \Labelxy{50}{-6}{0}{$\theta$\scriptsize ~(deg)}
    \end{overpic}
    \begin{overpic}[width=0.49\columnwidth]{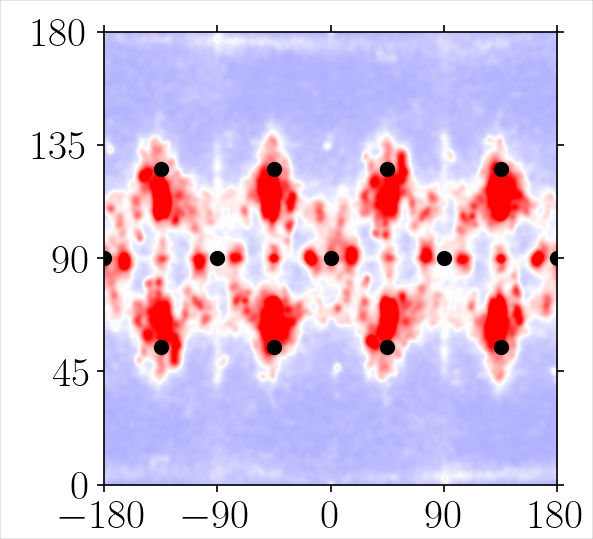}
        \LabelFig{16}{89}{$i)~\scriptstyle 10^{10}~ \text{s}^{-1}$}
        \Labelxy{50}{-6}{0}{$\theta$\scriptsize ~(deg)}
    \end{overpic}
    \caption{Orientation density maps $(\theta, \phi)$ associated with hcp glide planes at various deformation rates $\rate$ corresponding to \textbf{a-c}) Cantor alloy, \textbf{d-f}) \glsix, and \textbf{g-i}) pure Ni. The (black) symbols denote different crystallographic planes as in Fig.~\ref{fig:loadCurve}. The red and blue colors denote high and low densities, respectively.}
    \label{fig:orientation}
\end{figure*}

We investigated statistics of hcp clusters (including size and orientation) and sought for their correlations with stress avalanches.
Here a cluster is defined as a set of adjacent atoms with the same structural type ---hcp in this study.
As a basic statistical property, $P(s_\text{hcp})$ denotes the probability distribution function associated with the number of clusters containing $s_\text{hcp}$ atoms.
The radius of gyration associated with a cluster of size $s_\text{hcp}$ may be also defined as $r_g=\sum_{i=1}^{s_\text{hcp}}|\vec{r}_i-\vec{r}_0|^2/s_\text{hcp}$ with the center of mass $\vec{r}_0=\sum_{i=1}^{s_\text{hcp}}\vec{r_i}/s_\text{hcp}$.
Figure~\ref{fig:cluster_size}(a), (c), and (e) illustrates that $s_\text{hcp}\propto r_g^{d_f}$ with fractal dimension $d_f\simeq 2.0$.  
This almost agrees with Fig.~\ref{fig:loadCurve}(e) and (f) in the sense that, on average, hcp-type clusters tend to form fairly planar structures.
At the slowest rate $\ratezero$, the proposed scaling seems to be quite consistent with observations for \glsix and Ni in Fig.~\ref{fig:cluster_size}(c) and (e)  whereas Cantor alloy in Fig.~\ref{fig:cluster_size}(a) exhibits a slightly larger scatter in measurements likely due to microstructural heterogeneities.

Figure~\ref{fig:cluster_size}(b), (d), and (f) plots $P(s_\text{hcp})$ at different strain rates.
We note that the cluster size distributions develop fairly long tails with decreasing $\rate$, due to system-spanning slip planes, with a decay that can be best described by a power-law $P(s_\text{hcp})\propto s_\text{hcp}^{-\tau_c}$ over at least two decades in $s_\text{hcp}$. Here the distributions show a meaningful rate-dependence with the trend already observed for avalanche size distributions in Fig.~\ref{fig:statistics}(b) and (d).
Nevertheless, we see a slower-than-predicted decay of $P(s_\text{hcp})$ at the slowest rate ---$\tau_c < \tau_\text{pred}=2$ as inferred from percolation theory \cite{stauffer2018introduction}--- for the three metals which crossovers to the predicted (mean-field) behavior at faster deformation rates.

Having analyzed the size distributions of hcp clusters, we now turn to their crystallographic orientation relationship.
The latter is described based on the azimuthal angle $\theta$ and polar angle $\phi$ measured from the lattice coordinate frame as in Fig.~\ref{fig:loadCurve}(f).
The density plots presented in Fig.~\ref{fig:orientation} correspond to orientation maps $(\theta, \phi)$ at three different strain rates.
Within these maps, the black (solid) circles denote the $\{111\}$ and $\{100\}$ orientations corresponding to the undeformed crystals as in Fig.~\ref{fig:loadCurve}(e). 

Our data in Fig.~\ref{fig:orientation}(a) confirm that, statistically speaking, hcp glide planes tend to align with four different sets of $\{111\}$ closed-packed planes in Cantor alloy.
The density maps also suggest a fair amount of activation in the vicinity of the $\{100\}$ family which are mostly due to the loading-induced \emph{reorientation} of the slip planes.
We note that uniaxial tension is performed normal to the $(001)$ plane with $\phi=0^\circ, 180^\circ$. 
There exists a certain amount of data scatter possibly attributed to clusters being of very small size $s_\text{hcp} \ll 10$ and/or numerical artifacts because of non-planar topology of hcp clusters.
An increase of the deformation rate in Fig.~\ref{fig:orientation}(b) and (c) tends to reorient hcp planes to a larger extent as illustrated by the broader and/or denser distributions around $\{100\}$. 
As for \glsix in Fig.~\ref{fig:orientation}(d), a relatively insignificant reorientation of slip planes appears to be relevant at the slowest rate but intensifies with increasing rates in Fig.~\ref{fig:orientation}(e) and (f).
Another observation is the preferential reorientation around $\{110\}$ family planes (i.e. $\phi=90^\circ$ and $\theta=\pm 45^\circ, \pm 135^\circ$).
In the case of pure Ni in Fig.~\ref{fig:orientation}(g), (h), and (i), the metal appears to indicate fairly consistent features with Cantor alloy but with a slightly weaker reorientation of slip planes.

\subsection{\label{sec:correlations}Correlation Analysis: Dislocation Avalanches \& Microstructure}
\begin{figure}[t]
    \centering
    \begin{overpic}[width=0.49\columnwidth]{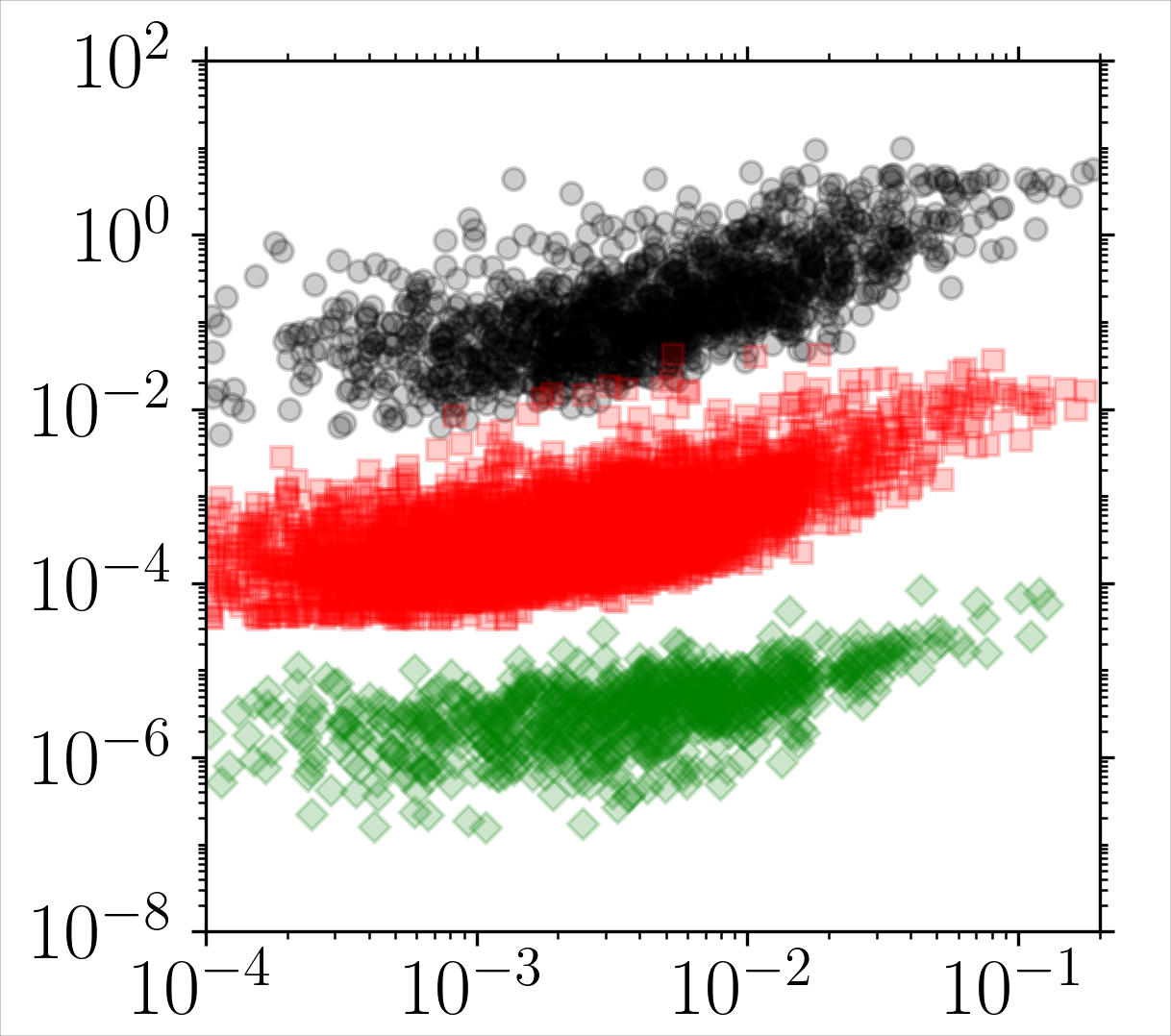}
        \LabelFig{16}{89}{$a)$~\scriptsize \glfour}
        \Labelxy{50}{-6}{0}{$\Delta\rho_\text{hcp}$}
        \Labelxy{-5}{41}{90}{$S$\scriptsize ~(Gpa)}
        \Labelxy{20}{12}{0}{$\scriptstyle c_{XY}=0.67~{\color{red}0.62}~{\color{darkspringgreen}0.57}$}        
    \end{overpic}
    \begin{overpic}[width=0.49\columnwidth]{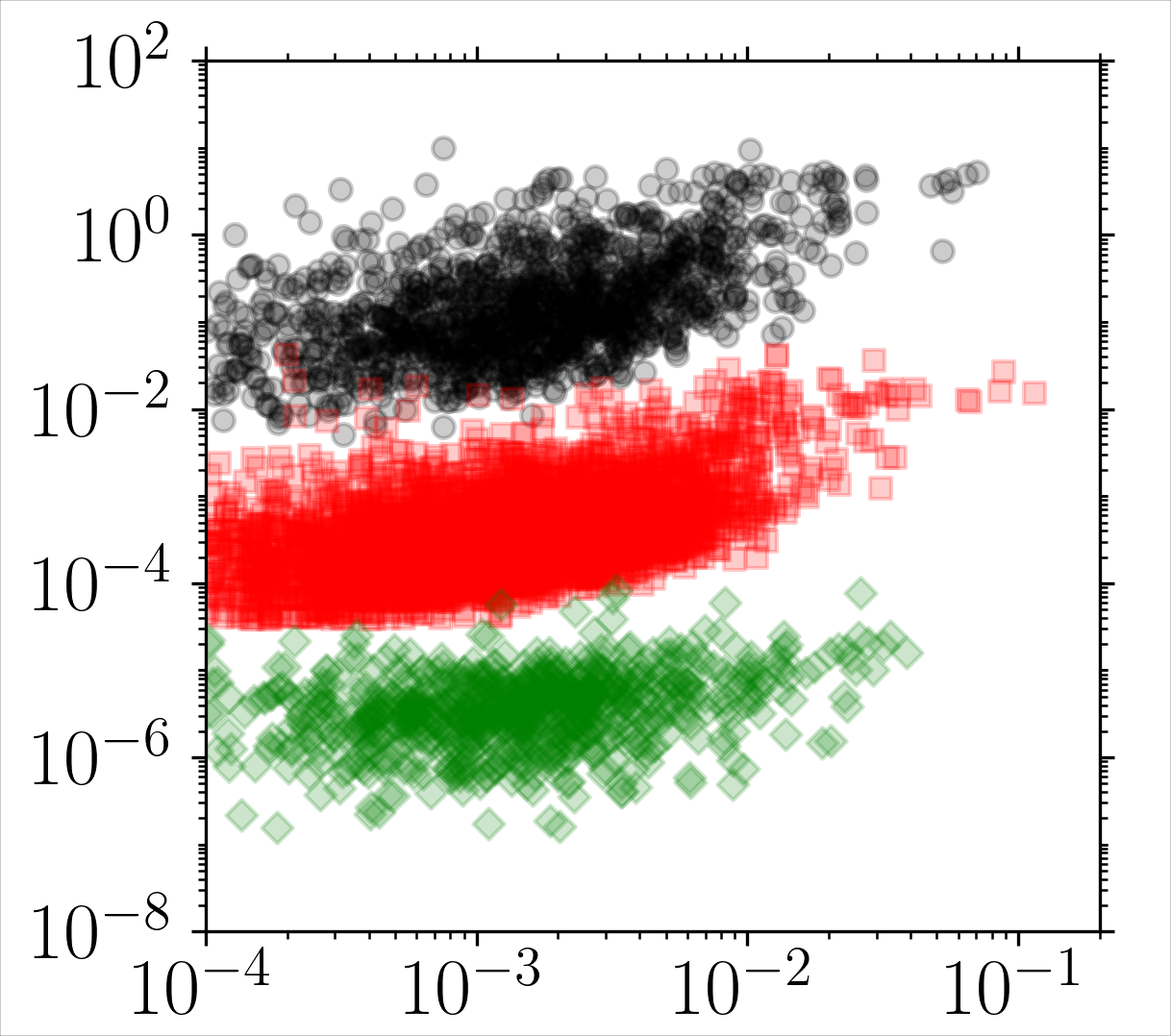}
        \LabelFig{16}{87}{$b)$}
        \Labelxy{50}{-6}{0}{$\Delta\rho_\text{bcc}$}
        \Labelxy{-5}{41}{90}{$S$\scriptsize ~(Gpa)}
        \Labelxy{20}{12}{0}{$\scriptstyle c_{XY}=0.52~{\color{red}0.51}~{\color{darkspringgreen}0.29}$}        
        \put(96,53.5){\includegraphics[width=0.05\textwidth]{Figs/legend_E1-4.png}} 
        \Labelxy{99}{82}{0}{$\rate ~\scriptstyle (\text{s}^{-1})$}
    \end{overpic}    
    \vspace{+2pt}

    \begin{overpic}[width=0.49\columnwidth]{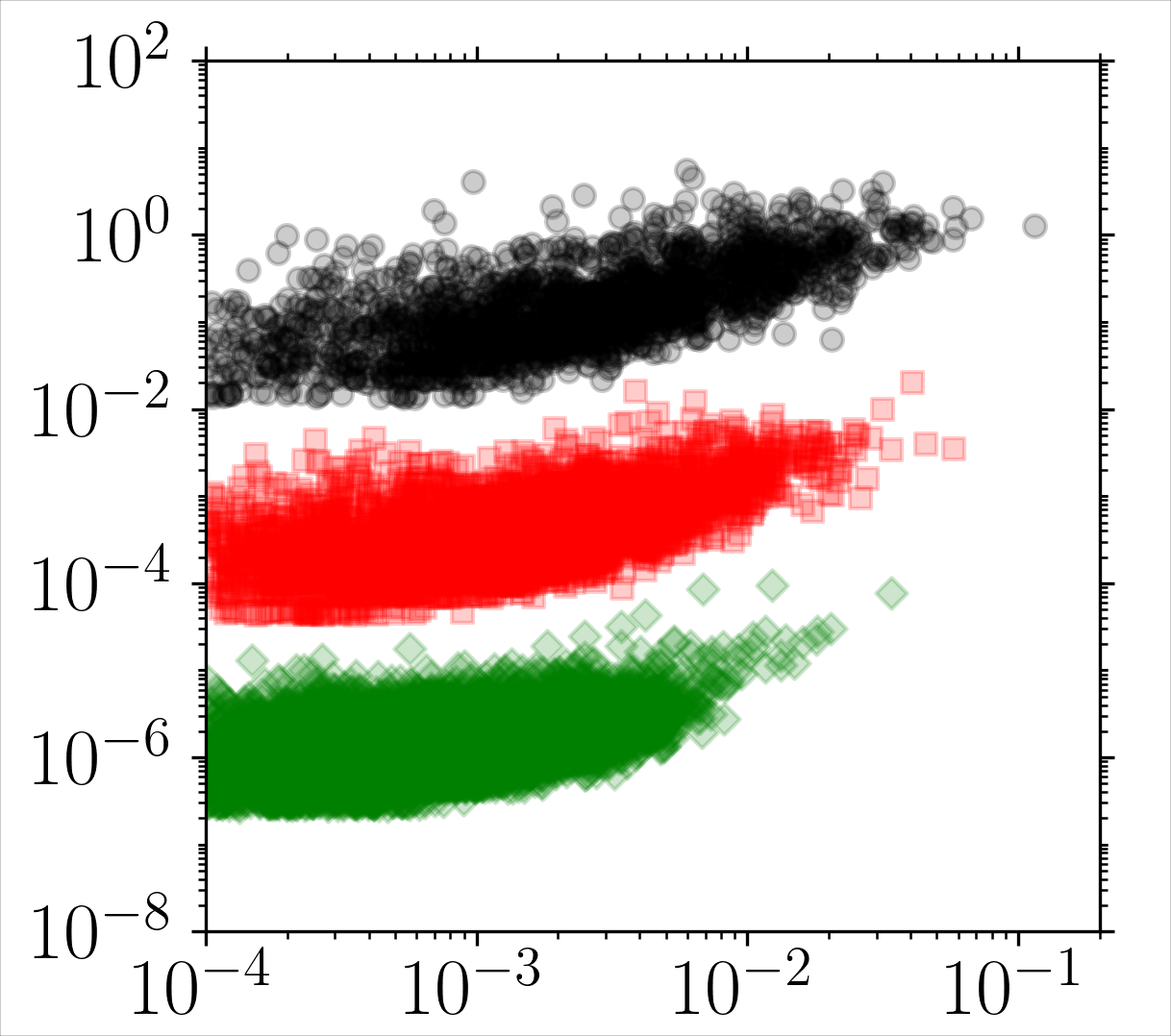}
        \LabelFig{16}{89}{$c)$~\scriptsize \glsix}
        \Labelxy{50}{-6}{0}{$\Delta\rho_\text{hcp}$}
        \Labelxy{-5}{41}{90}{$S$\scriptsize ~(Gpa)}
        \Labelxy{20}{12}{0}{$\scriptstyle c_{XY}=0.67~{\color{red}0.58}~{\color{darkspringgreen}0.43}$}        
    \end{overpic}
    \begin{overpic}[width=0.49\columnwidth]{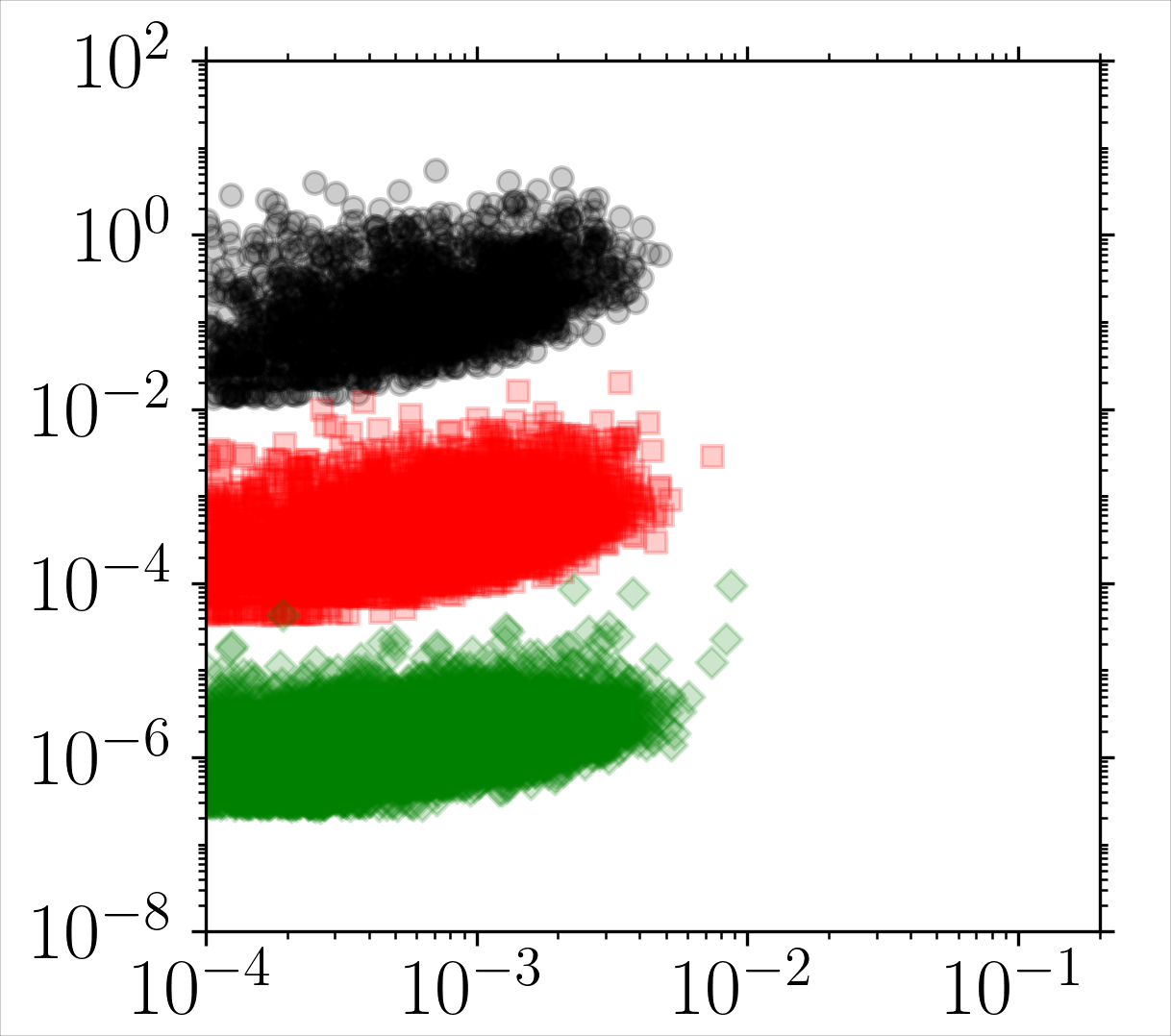}
        \LabelFig{16}{87}{$d)$}
        \Labelxy{50}{-6}{0}{$\Delta\rho_\text{bcc}$}
        \Labelxy{-5}{41}{90}{$S$\scriptsize ~(Gpa)}
        \Labelxy{20}{12}{0}{$\scriptstyle c_{XY}=0.43~{\color{red}0.40}~{\color{darkspringgreen}0.37}$}        
    \end{overpic}    
    \vspace{+2pt}

    \begin{overpic}[width=0.49\columnwidth]{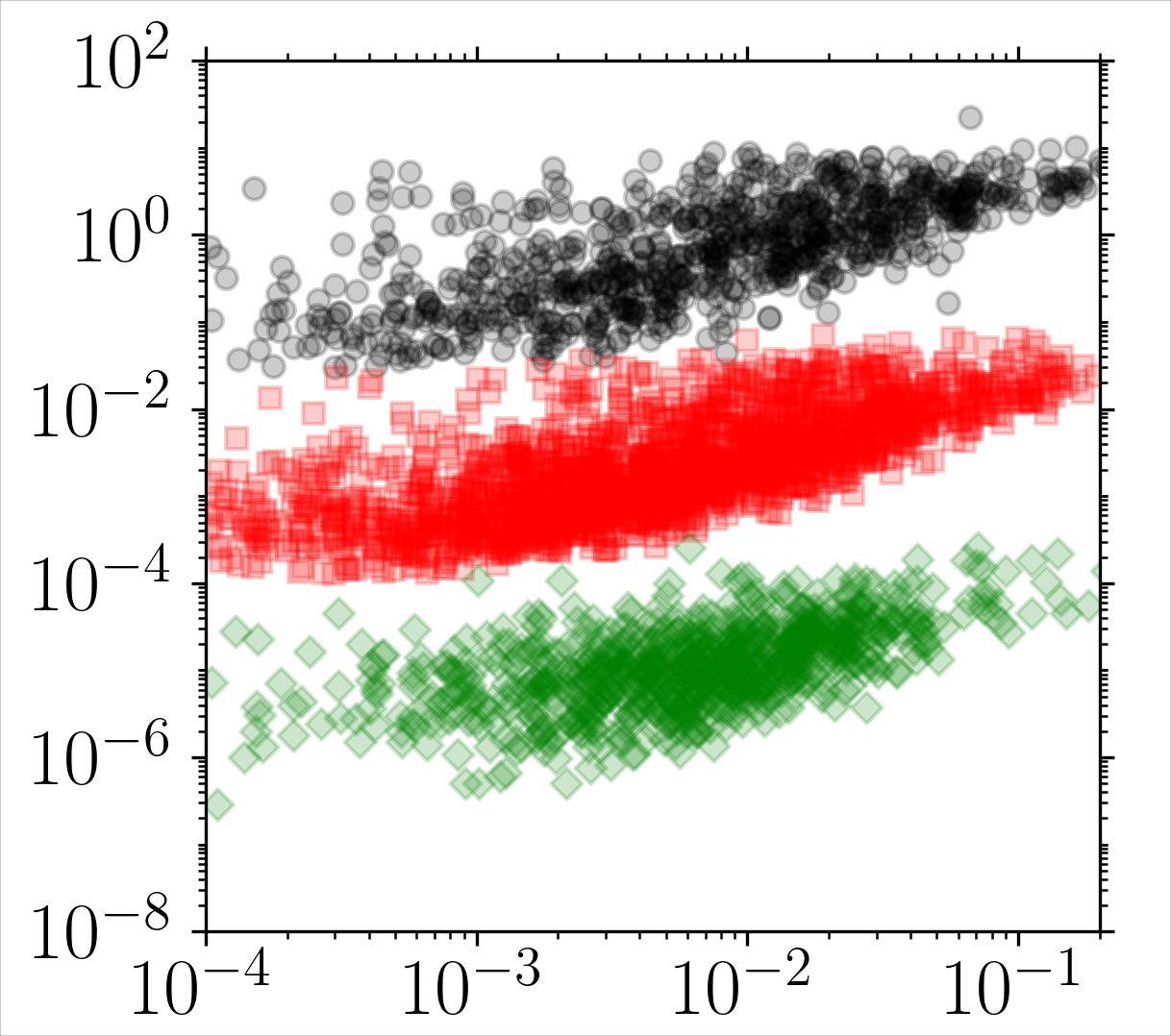}
        \LabelFig{16}{89}{$e)$~\scriptsize Ni}
        \Labelxy{50}{-6}{0}{$\Delta\rho_\text{hcp}$}
        \Labelxy{-5}{41}{90}{$S$\scriptsize ~(Gpa)}
        \Labelxy{20}{12}{0}{$\scriptstyle c_{XY}=0.68~{\color{red}0.71}~{\color{darkspringgreen}0.58}$}        
    \end{overpic}
    \begin{overpic}[width=0.49\columnwidth]{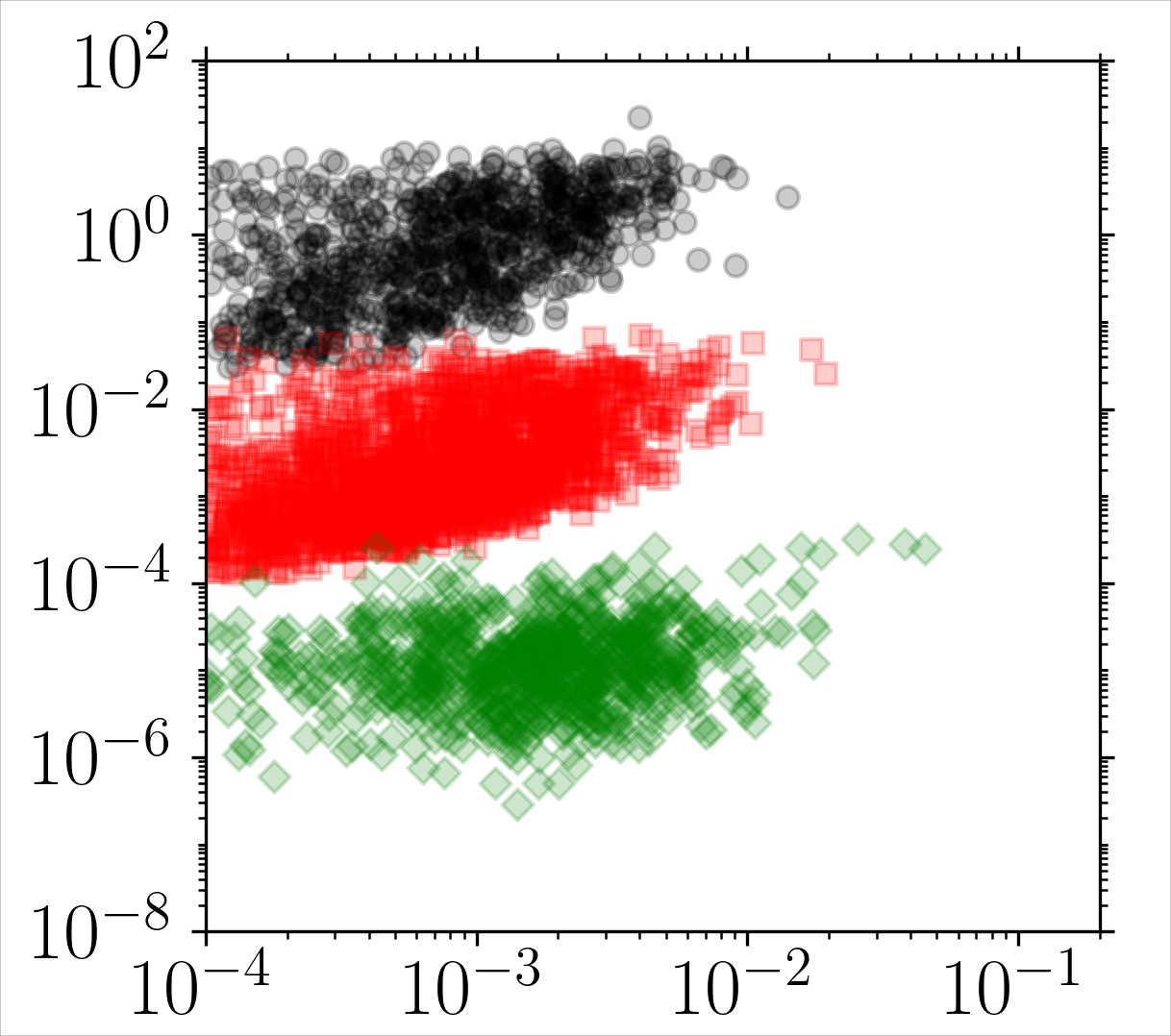}
        \LabelFig{16}{87}{$f)$}
        \Labelxy{50}{-6}{0}{$\Delta\rho_\text{bcc}$}
        \Labelxy{-5}{41}{90}{$S$\scriptsize ~(Gpa)}
        \Labelxy{20}{12}{0}{$\scriptstyle c_{XY}=0.50~{\color{red}0.52}~{\color{darkspringgreen}0.17}$}        
    \end{overpic}    
    \caption{Correlations between avalanche size and associated change in crystal structures corresponding to \glfour\!\!, \glsix\!\!, and pure Ni. Scatter plot of avalanche size $S$ and the creation/annihilation ratio of \textbf{a, c, e}) hcp structure $\Delta\rho_\text{hcp}$ \textbf{b, d, f}) bcc structure $\Delta\rho_\text{bcc}$ at various rates $\rate$. Here $c_{XY}$ denotes the corresponding correlation coefficient between $\Delta\rho_\text{hcp}$ and $\Delta\rho_\text{bcc}$ with $S$.} 
    \label{fig:correlations}
\end{figure}

We carried out a correlation analysis between avalanche sizes $S$ and associated changes in the fraction of hcp atoms $\rho_\text{fcc}$ incurred over the duration of individual avalanches (refer to Fig.~\ref{fig:loadCurve}(c) and (d)).
The latter is defined as $\Delta\rho_\text{hcp}=\int_{t_i}^{t_i+\mathcal{T}_{i}} |\partial_t\rho_\text{hcp}|~dt$ corresponding to the $i_\text{th}$ avalanche at $t_i$ with duration $\mathcal{T}_i$. 
We also obtained the (linear) correlation coefficient $c_{XY}=\langle \hat{X} \hat{Y} \rangle$ between the two observables $X=\text{log}_{10}S$ and $Y=\text{log}_{10}\rho_\text{hcp}$. 
Here $\hat{X}$ indicates the deviation from the mean $\langle X\rangle$, normalized by the standard deviation $\text{std}(X)$ associated with each variable.
The above analysis was repeated for the bcc arrangement with the results shown as the scatter plots of Fig.~\ref{fig:correlations} at multiple $\rate$.

Our data in Fig.~\ref{fig:correlations}(a) and (b) exhibit a large scatter in Cantor alloy but the observed trend indicates meaningful variations between the two sets of observables at the slowest driving rate $\ratezero$.  
Here, (positive) correlations between $S$ and $\Delta\rho_\text{hcp}$ (or $\Delta\rho_\text{bcc}$) imply that avalanches of large size typically correspond to considerable microstructural changes, most likely associated with fcc-to-hcp and fcc-to-bcc phase transformations.
$\Delta\rho_\text{hcp}$ features a relatively stronger association with $S$, as demonstrated by larger correlation coefficients $c_{XY}$, suggesting that plastic avalanches are presumably rooted in partial slips and formation of layers of hcp stacking in fcc \glfour\!\!. 
With increasing $\rate$ toward $\ratefour$, such correlations become less pronounced, in particular, between avalanche sizes $S$ and $\Delta\rho_\text{bcc}$.  
Our correlation analysis associated with \glsix and Ni as in Fig.~\ref{fig:correlations}(c-f) reveal similar correlation patterns compared with Cantor alloy.

\section{\label{sec:conclusions}Conclusions \& Discussions }
We initiated this study with the aim of \emph{\romannum{1}}) replicating empirical observations on serrated stress response in plastically deforming Cantor and \glsix alloys \emph{\romannum{2}}) inferring structural signatures of dislocation avalanches and associated rate effects.
As for \emph{\romannum{1}}), our atomistic simulations have closely reproduced experimental data on the scale-free nature of dislocation avalanches in HEAs.
This scale-invariance has been evidenced by robust scaling features (under slow ``quasi-static" drive) described by asymptotic power-law distributions and associated critical exponents that, in general, match empirical evaluations as well as mean-field predictions.
More specifically, the avalanche size exponents we measure are fairly compatible with mean-field estimates ($\tau=\frac{3}{2}$ \cite{fisher1998collective,friedman2012statistics,antonaglia2014temperature}) and within the experimentally reported range $1.3-2.0$ \cite{chen2022acoustic,antonaglia2014temperature,rizzardi2021microstructural} in chemically complex alloys.
Notably, \glsix exhibits a fairly restricted scaling regime associated with avalanche size distributions that is possibly attributed to strong heterogeneities and large atomic misfits \cite{esfandiarpour2022edge}, driving this alloy away from criticality.
One should also take the above estimates with grain of salt as various metrics have been utilized as avalanche size in the literature including (but not limited to) slip magnitude, stress drop, and emitted (acoustic) energies. 
The observed rate effects on statistics of dislocation avalanches are in line with experimentation on HEAs/MEAs that indicate dynamical cross-overs between different serration types at a certain range of deformation rates (see \cite{brechtl2020review} and references therein).
Beyond a certain rate threshold, the three deforming metals undergo a dynamical transition into a subcritical state characterized by non-critical exponential-like statistics of avalanches.

To discern the role of local lattice distortions associated with the Cantor and \glsix alloys, we have additionally probed statistics of avalanches in pure Ni showing fairly consistent features with the former metals in terms of the overall rate dependency.
Nevertheless, we have found a relatively shallow decay of avalanche size distributions within the quasi-static regime exhibiting a non-mean-field behavior $\tau<\frac{3}{2}$ which was also reported in two-dimensional dislocation dynamics simulations \cite{ispanovity2014avalanches}.
This might be indicative of the relative abundance of big avalanches over small ones in the pure metal possibly due to the absence of chemical/structural disorder.
In certain alloys, this heterogeneity element may act as effective obstacles against propagating avalanches and, therefore, strongly influence their statistical properties, often resulting in the dominance of smaller-size avalanches.
We conjecture that the disorder strength and associated length might be relevant parameters that govern the critical exponent $\tau$ and its deviation from the mean-field prediction.

In line with \emph{\romannum{2}}), our microstructural analyses have demonstrated nontrivial scaling features associated with the dynamics and topology of slip planes that could be better understood in the context of percolation transition. 
In this framework, we find robust power-law distributed cluster sizes with scaling exponent $\tau_c\simeq 1.0$ at slow driving rates that is shallower than the mean-field prediction $\tau_c^\text{mf}=2$ \cite{stauffer2018introduction} in the studied metals but cross-overs to the latter exponent at intermediate rates beyond which the cluster size distribution enters a subcritical regime, analogous to avalanche size distributions.
We note that Ni features a non-mean-field scaling for both avalanche size $S$ and cluster size $s_\text{hcp}$ distributions in the rate-independent regime. 
As for Cantor alloy (and to some degree \glsix alloy), the former distribution indicates an asymptotic mean-field behavior. 
This is rather counter-intuitive as one would naively expect the two observables to intercorrelate strongly.
Our speculation is that, due to underlying disorder in the HEA (and/or MEA), plastic avalanches may primarily occur as a result of the accumulation of broadly-distributed yet \emph{randomly}-triggered individual slips. 
By contrast, the observed trends associated with pure Ni might be indicative of spatial and/or temporal correlations between the latter.
Our correlation analysis may establish a direct mapping between avalanches of certain size and incurred slip patterns at slow strain rates.
Such a close correspondence between the two variables is not maintained at finite deformation rates which could be taken as another indication that high strain rates and/or stresses tend to drive these systems away from criticality.

Our findings on the strong coupling between temporal and spatial evolution of dislocation avalanches may contribute to ongoing efforts within the material science and physics communities that aim to infer underlying morphology and microstructural changes by solely probing the mechanical signals. 
Recently, the state-of-the-art machine learning models have emerged as robust computational tools to classify/reconstruct the bulk micro-structure based on feature extraction of surface measurements (e.g. frequency content, magnitudes, signal duration, and energy scales).
Combined with in-situ imaging techniques, such surrogate models and their predictions can provide a valuable insight into the  microstructural origins of plasticity in an timely, efficient, and accurate manner.

\begin{acknowledgments}
This research was funded by the European Union Horizon 2020 research and innovation program under grant agreement no. 857470 and from the European Regional Development Fund via Foundation for Polish Science International Research Agenda PLUS program grant no. MAB PLUS/2018/8.
\end{acknowledgments}

\clearpage
\bibliography{references}

\end{document}


\title{Serrated plastic flow in slowly-deforming complex concentrated alloys: universal signatures of dislocation avalanches}

\author{Kamran Karimi$^1$}
\author{Amin Esfandiarpour$^1$}%
\author{Stefanos Papanikolaou$^1$}
\affiliation{%
 $^1$ NOMATEN Centre of Excellence, National Center for Nuclear Research, ul. A. Sołtana 7, 05-400 Swierk/Otwock, Poland\\
}%

\maketitle

\section*{Supplementary Materials}

\begin{figure*}
    \begin{overpic}[width=0.24\textwidth]{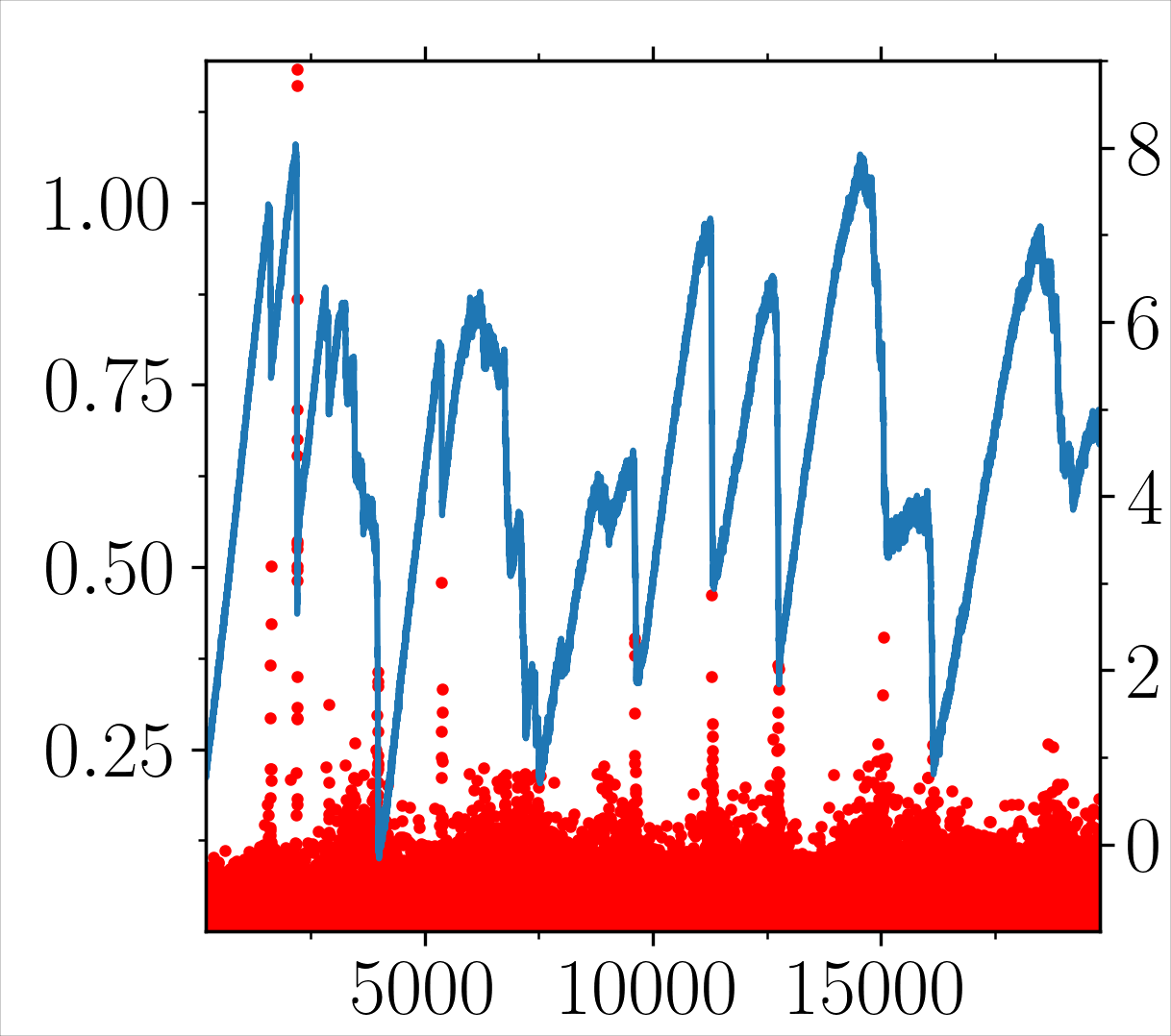}
        \LabelFig{16}{87}{$a)$ raw signal}
        \Labelxy{-6}{35}{90}{{$-\sigmadot \scriptstyle ~(\text{Gpa.ps}^{-1})$}}
        \Labelxy{50}{-8}{0}{$t$(ps)}
    \end{overpic}
    \begin{overpic}[width=0.24\textwidth]{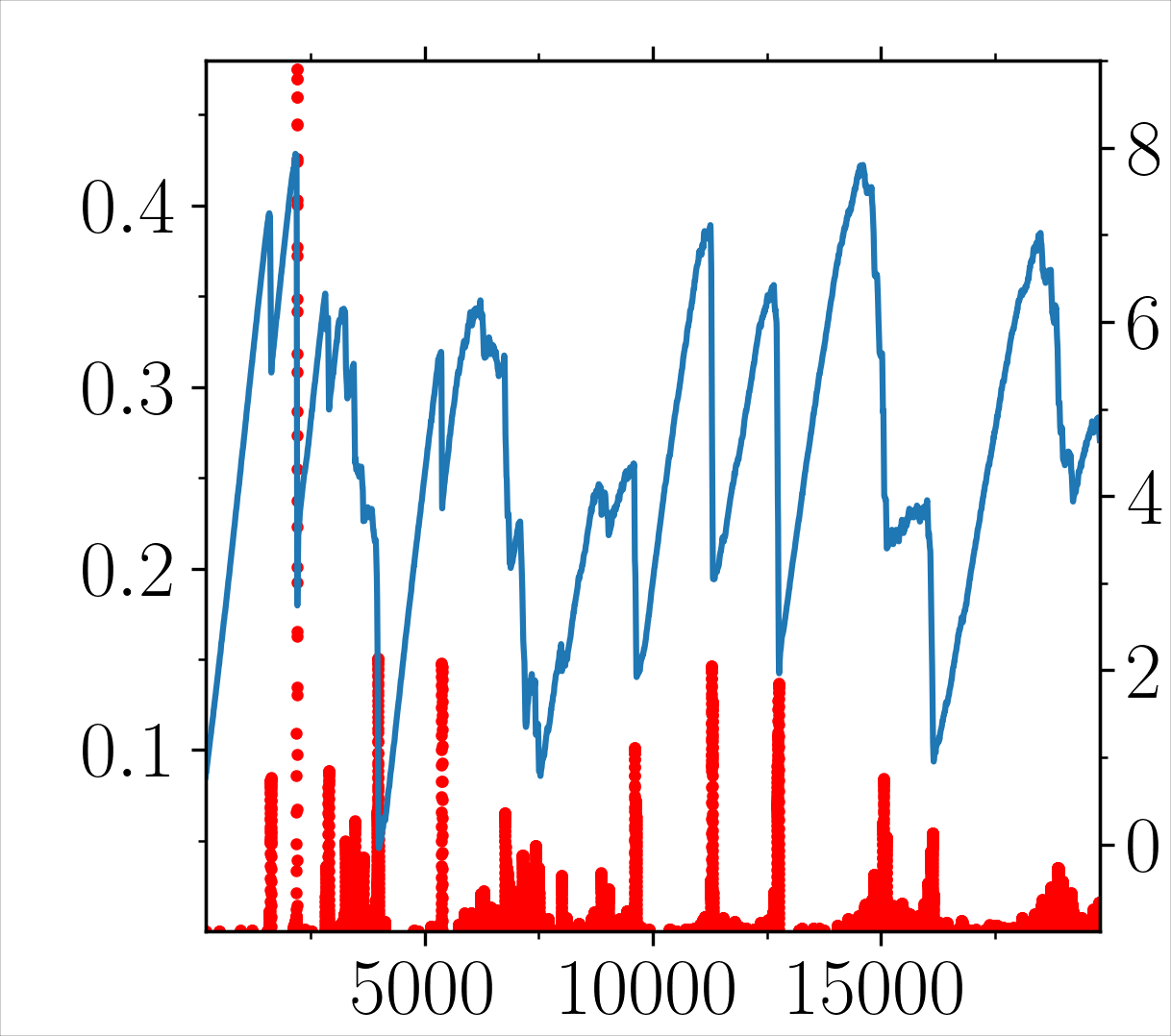}
        \LabelFig{16}{87}{$b)~\omega_c=1600$}
        \Labelxy{50}{-8}{0}{$t$(ps)}
    \end{overpic}
    \begin{overpic}[width=0.24\textwidth]{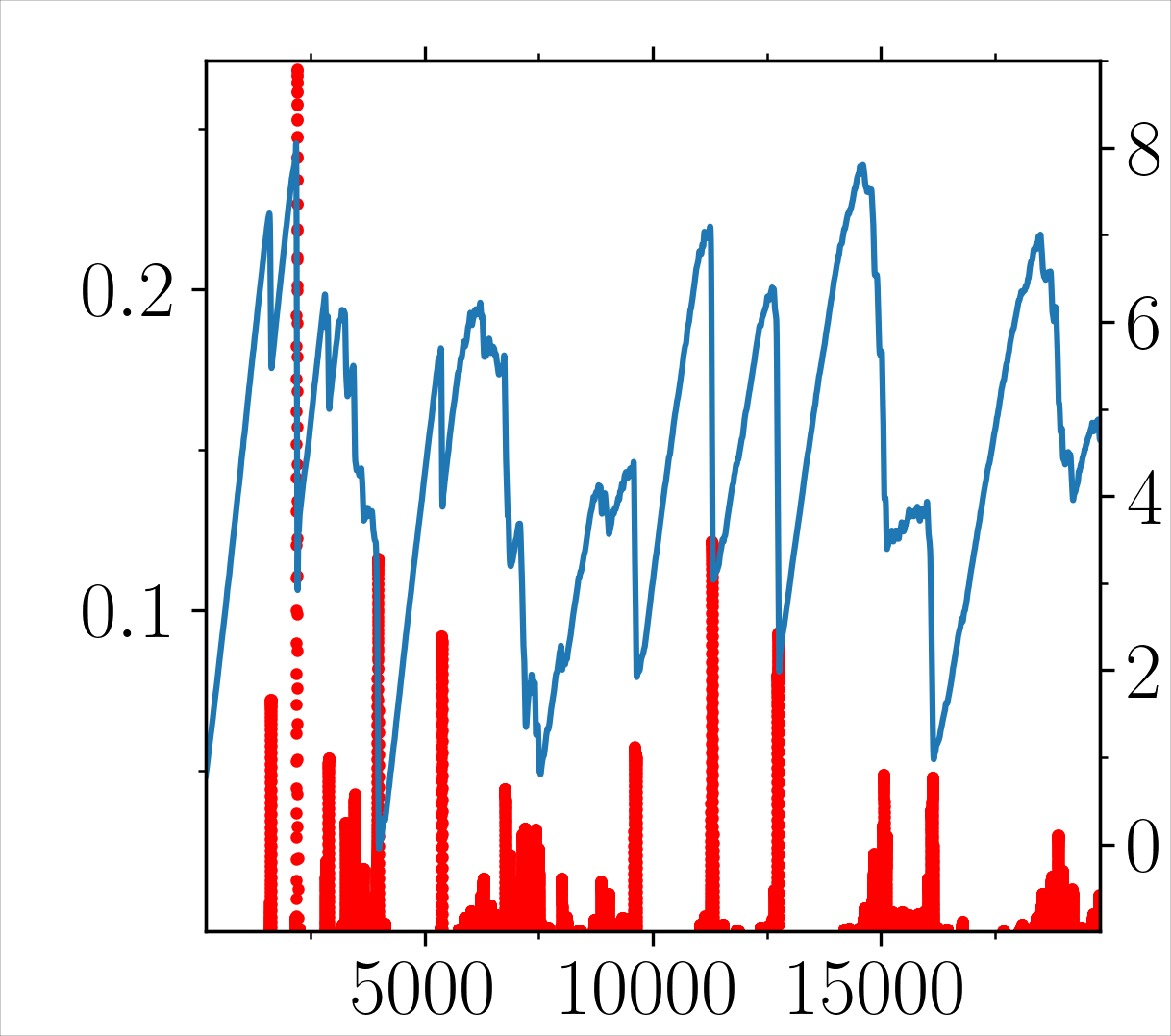}
        \LabelFig{16}{87}{$c)~\omega_c=800$}
        \Labelxy{50}{-8}{0}{$t$(ps)}
    \end{overpic}
    \begin{overpic}[width=0.24\textwidth]{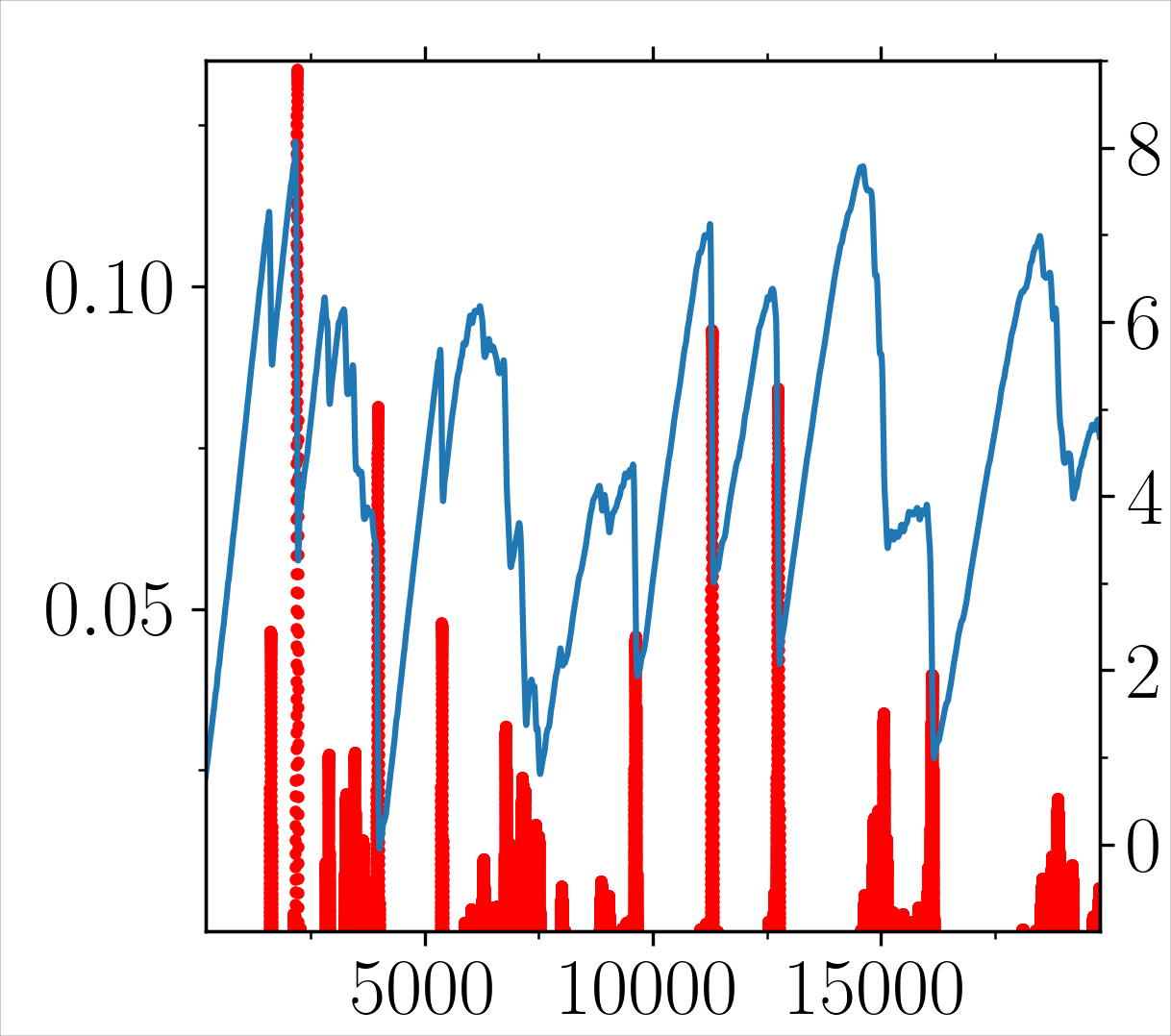}
        \LabelFig{16}{87}{$d)~\omega_c=400$}
        \Labelxy{104}{35}{90}{{$\sigma_{zz}$ \tiny(Gpa)}}
        \Labelxy{50}{-8}{0}{$t$(ps)}
    \end{overpic}
    \vspace{18pt}

    \begin{overpic}[width=0.24\textwidth]{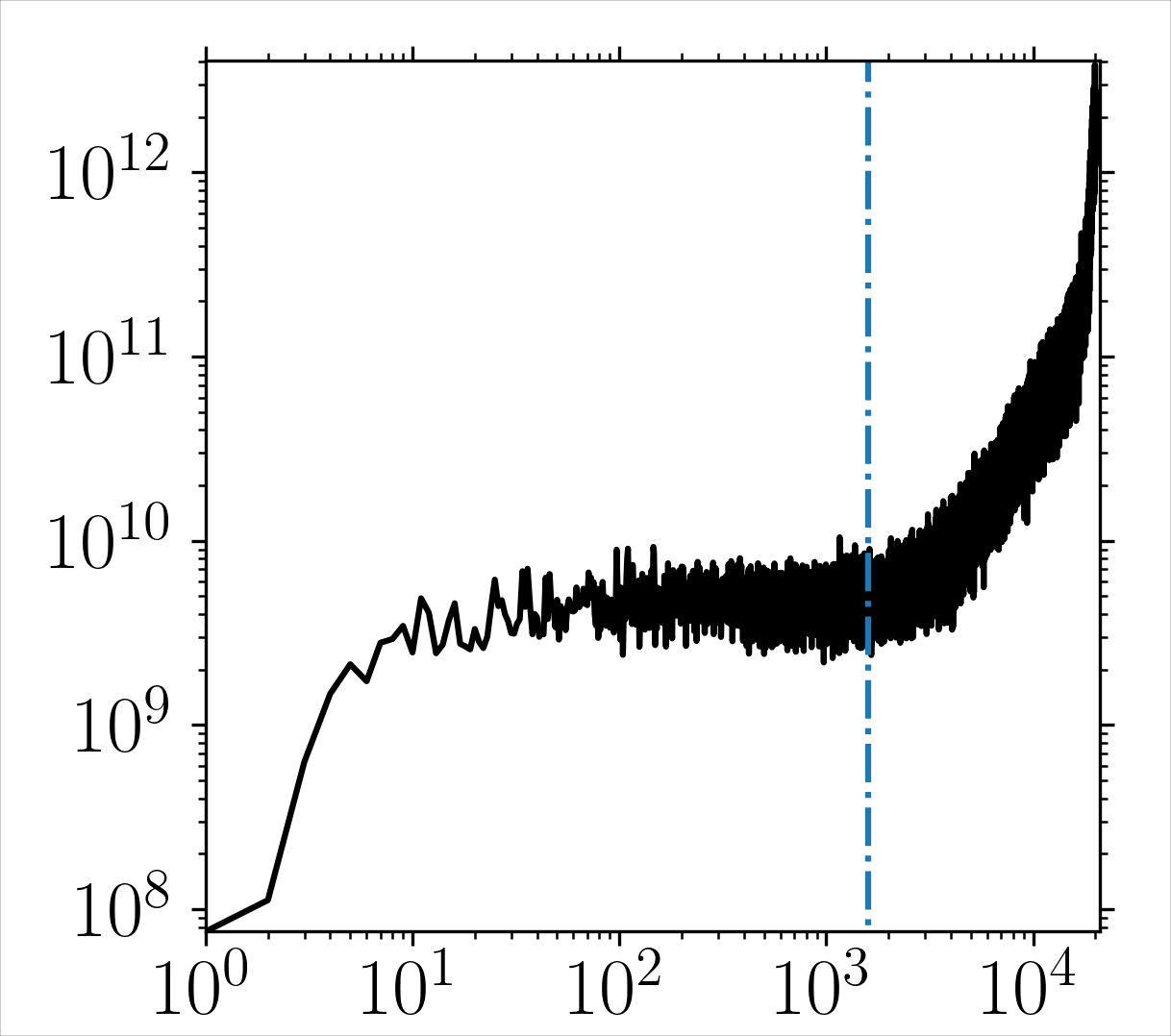}
        \LabelFig{16}{87}{$e)$}
        \Labelxy{50}{-5}{0}{$\omega$}
        \Labelxy{-6}{44}{90}{$\omega^{2.3}|\Sigma(\omega)|^2$}
        \put(76,12){\color{darkblue}$\omega_c$}
    \end{overpic}
    \begin{overpic}[width=0.24\textwidth]{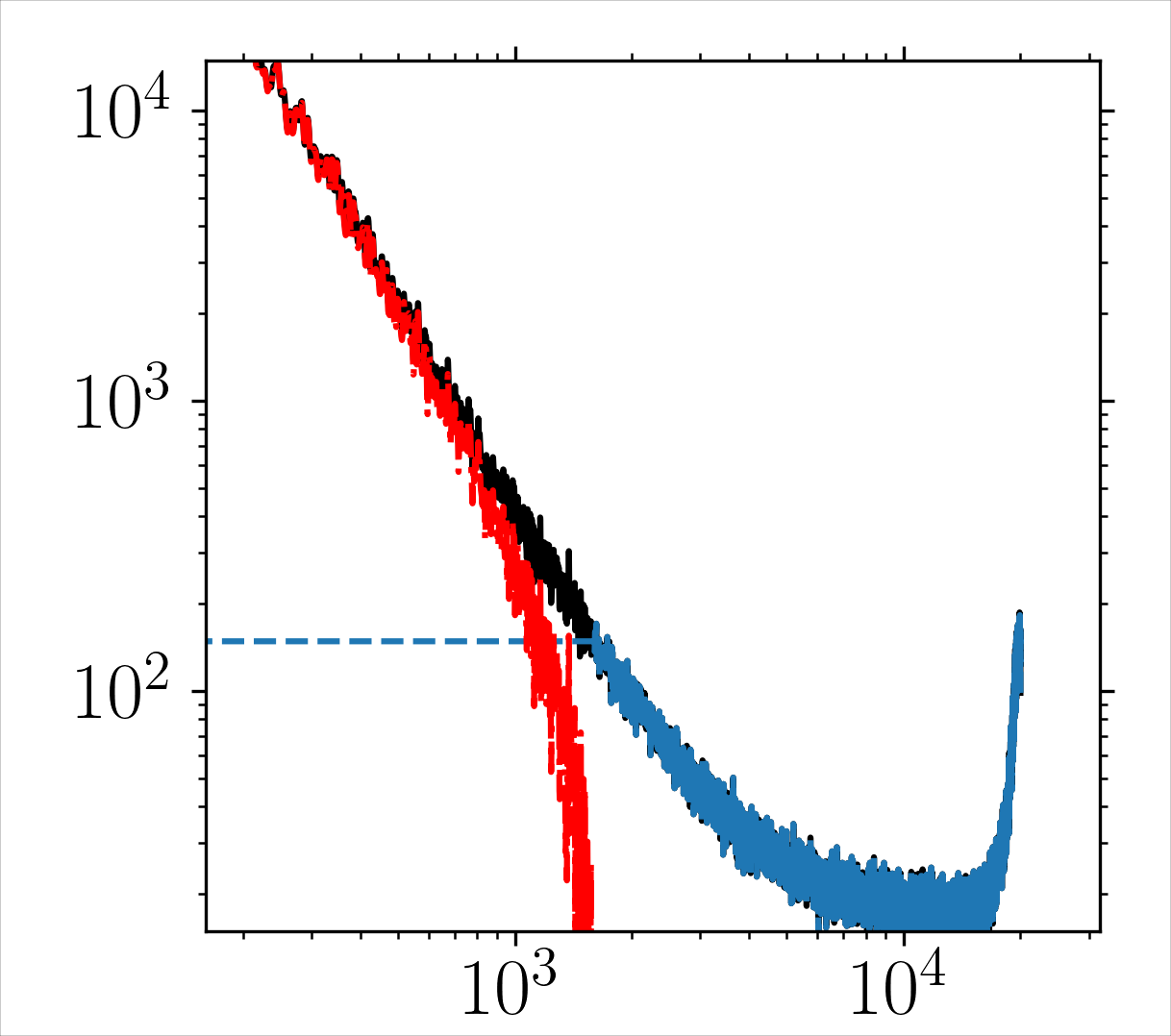}
        \LabelFig{16}{87}{$f)$}
        \Labelxy{50}{-5}{0}{$\omega$}
        \Labelxy{-6}{44}{90}{$|\Sigma(\omega)|^2$}
        \put(58,52){$\scriptstyle |\Sigma(\omega)|^2$}
        \put(20,38){\color{darkblue}$\scriptstyle |\mathcal{N}(\omega)|^2$}
        \put(18,19){\color{red}$\scriptstyle |\Sigma_\text{ath}(\omega)|^2$}
        \begin{tikzpicture}
          \coordinate (a) at (0,0);
            \node[white] at (a) {\tiny.};               %
             \carrow{2.4}{2.4}{0.5}{-1}{-1.0}{}{black}
            \carrow{1.1}{0.5}{0.7}{1.0}{0.0}{}{red}
	\end{tikzpicture}
    \end{overpic}
    \caption{Smeared stress signals associated with Cantor alloy at $\rate=10^8$ s$^{-1}$ using optimal (Wiener) filtering with multiple cut-off frequencies corresponding to \textbf{a}) the raw signal \textbf{b}) $\omega_c=1600$ \textbf{c}) $\omega_c=800$ \textbf{d}) $\omega_c=400$. \textbf{e}) Power spectrum $|\Sigma(\omega)|^2$ scaled by $\omega^{-2.3}$ associated with the raw signal. \textbf{f}) Wiener filtering with power spectra corresponding to the noise $|\mathcal{N}(\omega)|^2$ and athermal stress $|\Sigma_\text{ath}(\omega)|^2$ defined based on the extrapolation of the former into the signal region below cut-off frequency $\omega_c$.}
    \label{fig:kernels}
\end{figure*}

\begin{figure*}
    \begin{overpic}[width=0.24\textwidth]{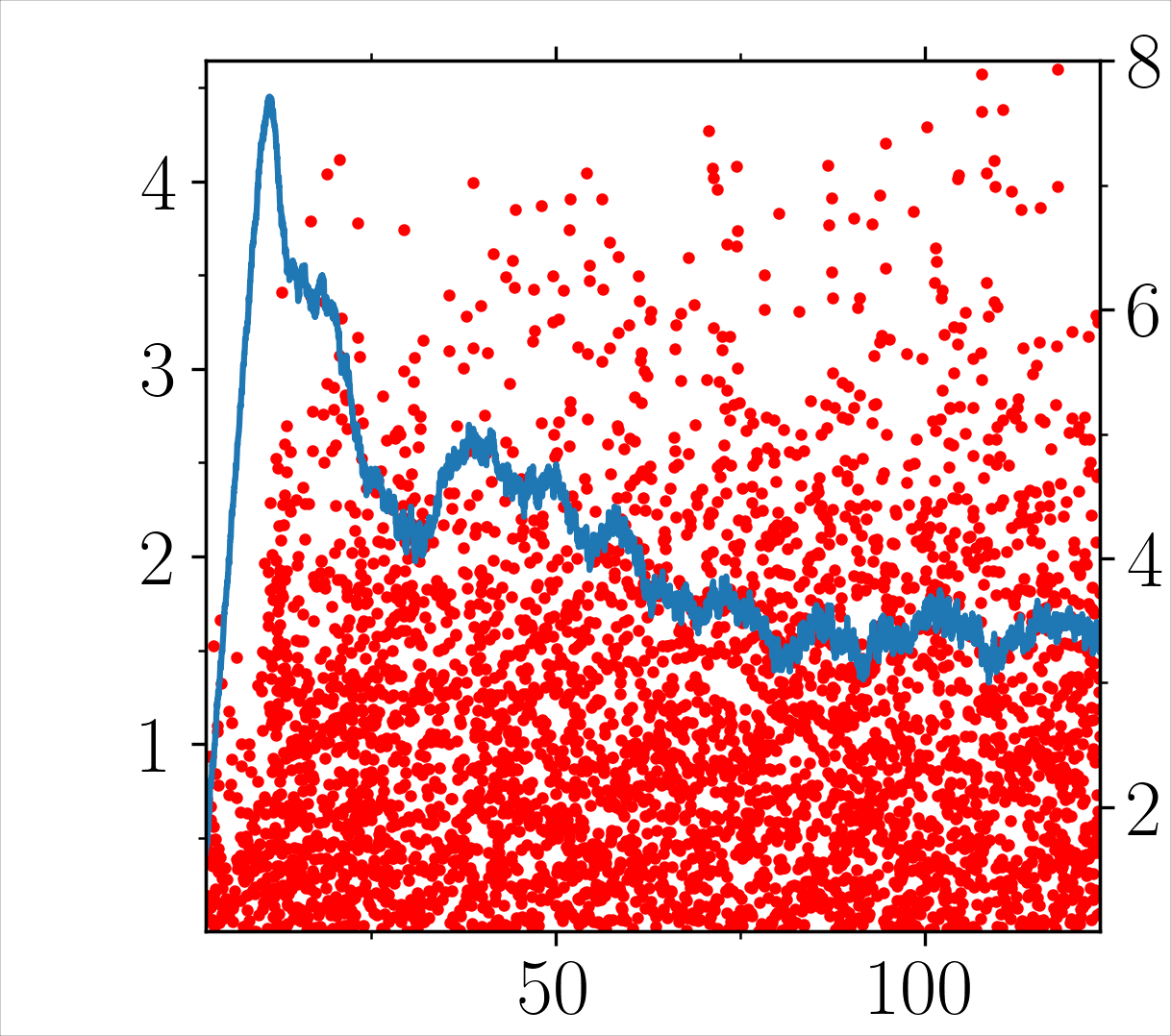}
        \LabelFig{16}{87}{$a)$ raw signal}
        \Labelxy{-6}{35}{90}{{$-\sigmadot \scriptstyle ~(\text{Gpa.ps}^{-1})$}}
        \Labelxy{50}{-8}{0}{$t$(ps)}
    \end{overpic}
    \begin{overpic}[width=0.24\textwidth]{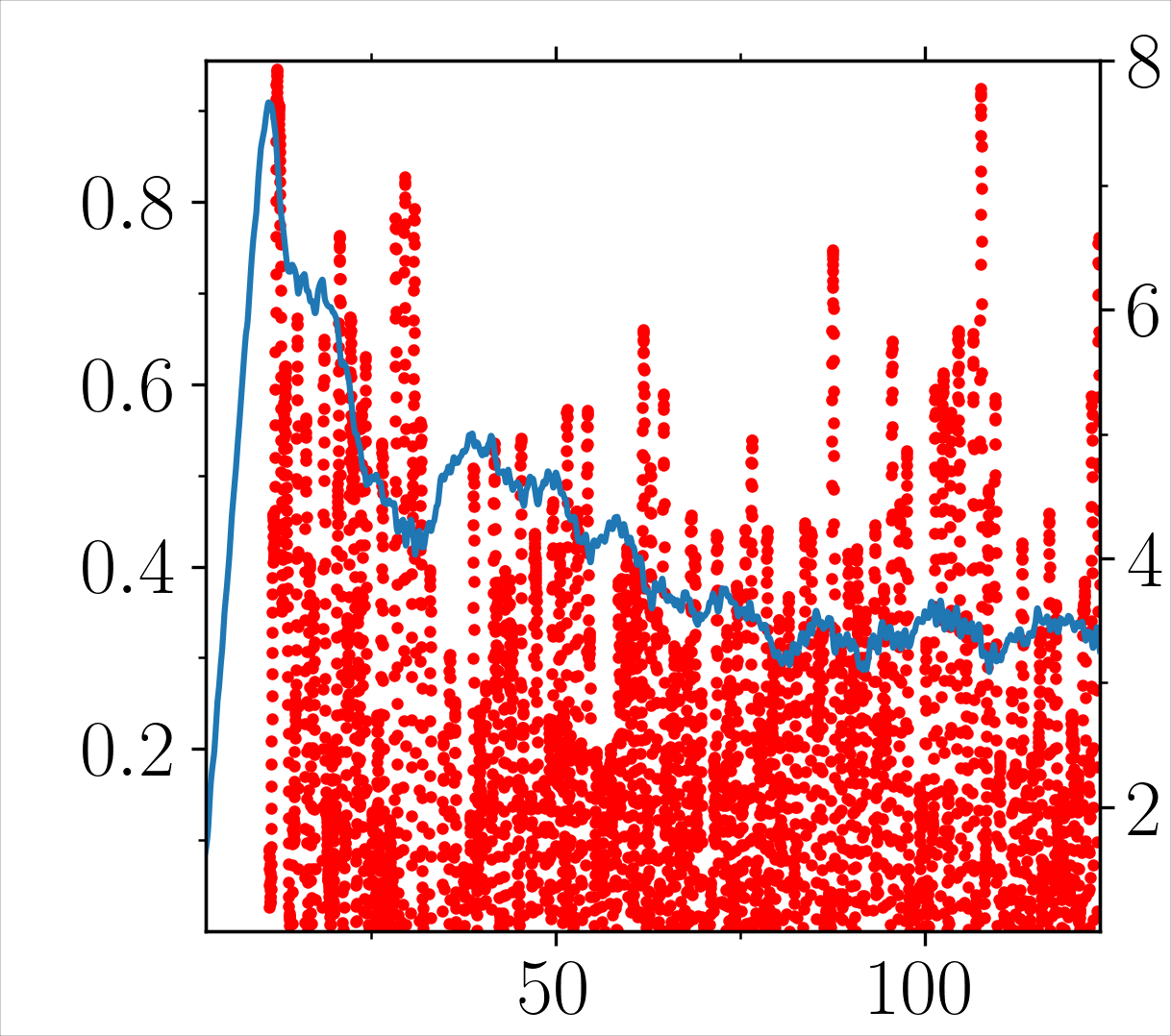}
        \LabelFig{16}{87}{$b)~\omega_c=400$}
        \Labelxy{50}{-8}{0}{$t$(ps)}
    \end{overpic}
    \begin{overpic}[width=0.24\textwidth]{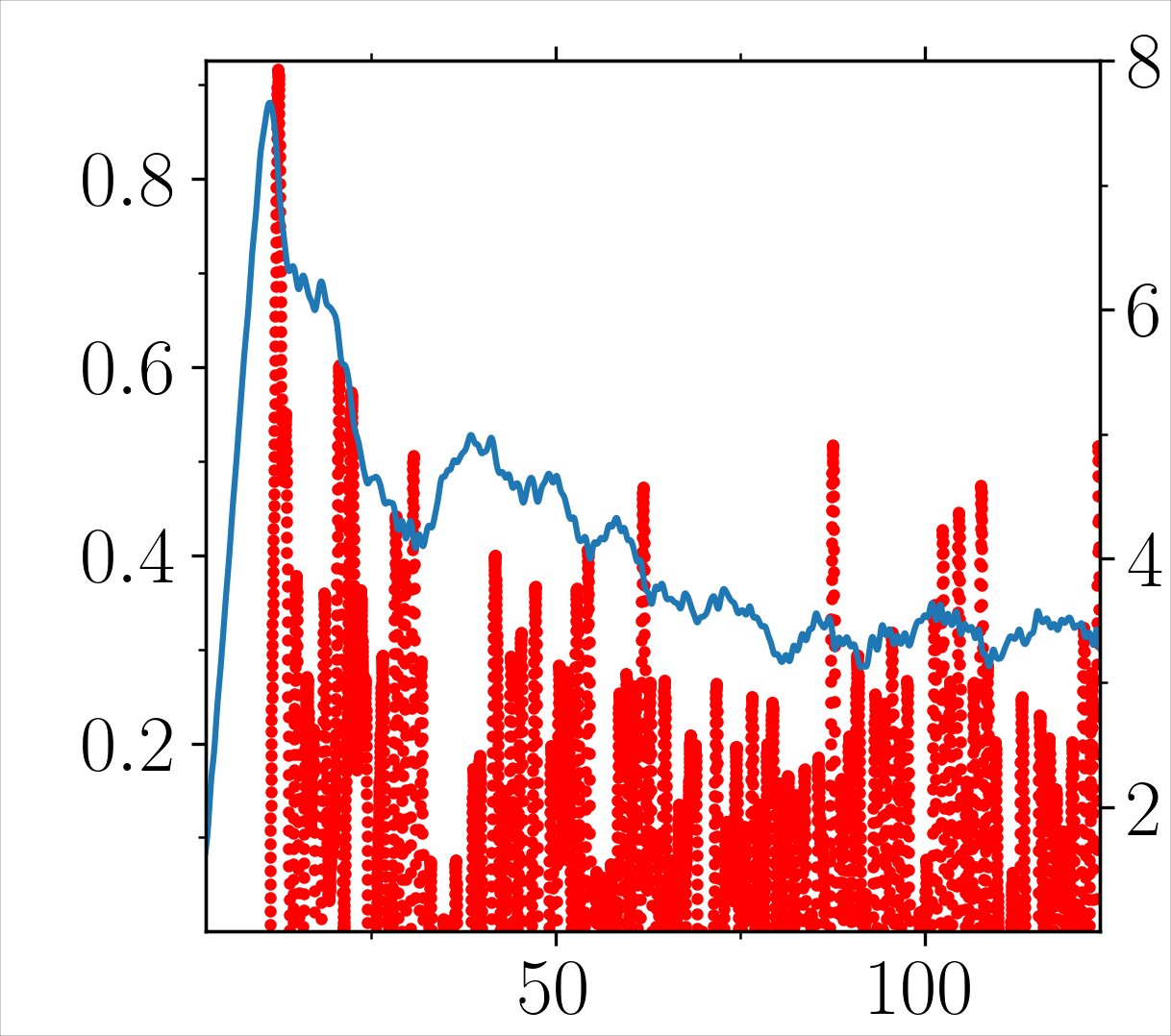}
        \LabelFig{16}{87}{$c)~\omega_c=200$}
        \Labelxy{50}{-8}{0}{$t$(ps)}
    \end{overpic}
    \begin{overpic}[width=0.24\textwidth]{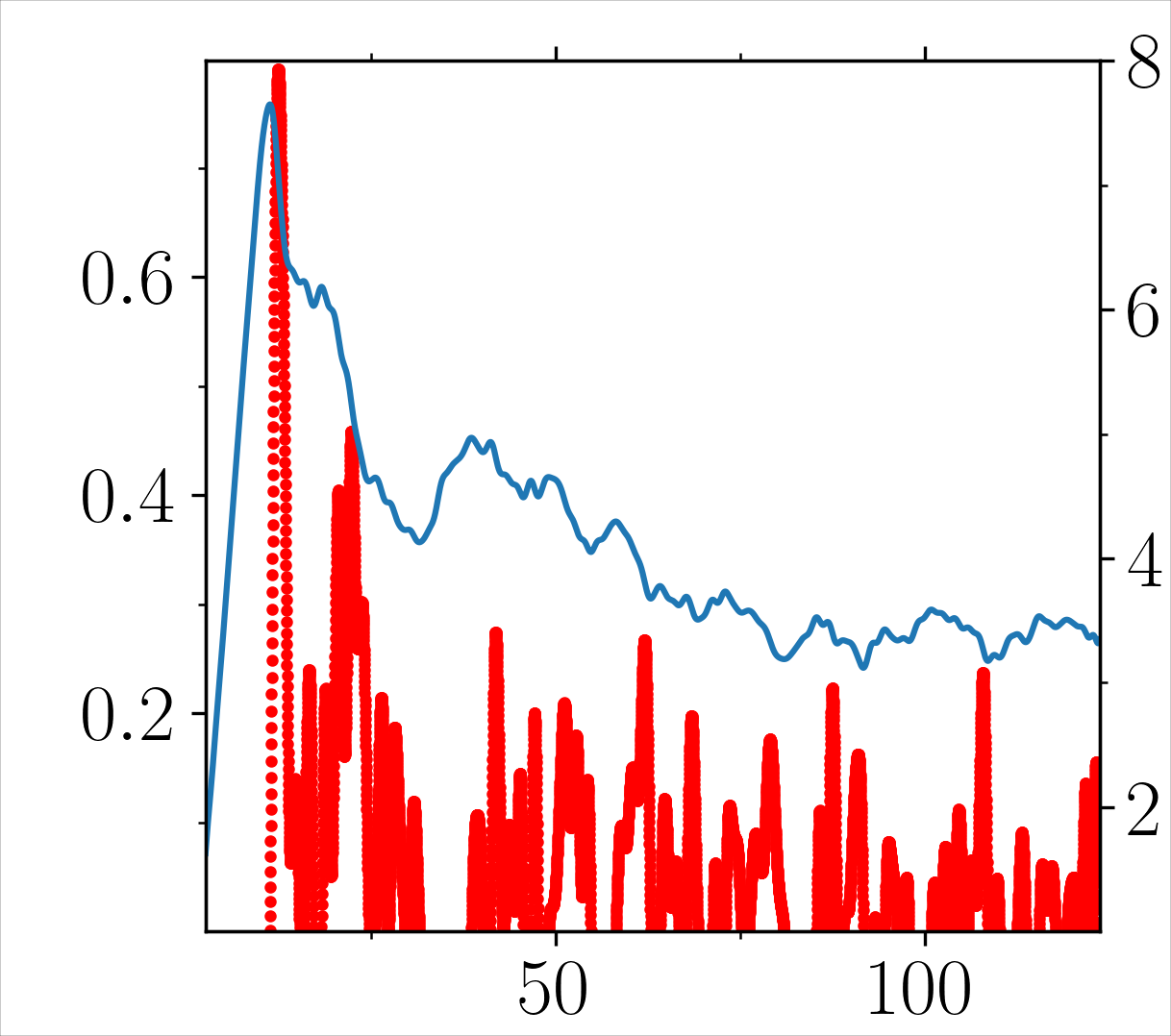}
        \LabelFig{16}{87}{$d)~\omega_c=100$}
        \Labelxy{104}{35}{90}{{$\sigma_{zz}$ \tiny(Gpa)}}
        \Labelxy{50}{-8}{0}{$t$(ps)}
    \end{overpic}
    \vspace{18pt}

    \begin{overpic}[width=0.24\textwidth]{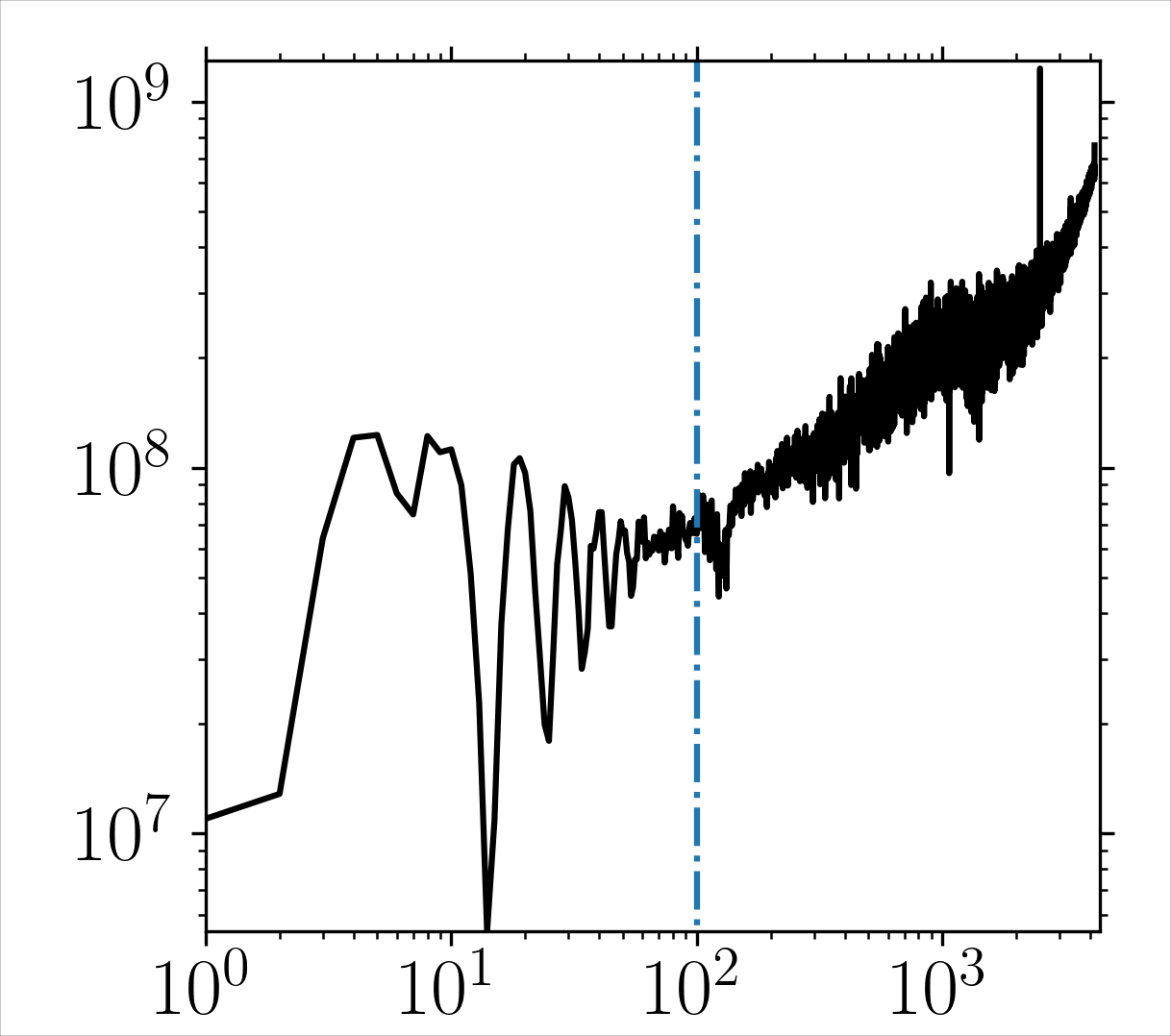}
        \LabelFig{16}{87}{$e)$}
        \Labelxy{50}{-5}{0}{$\omega$}
        \Labelxy{-6}{44}{90}{$\omega^{2.3}|\Sigma(\omega)|^2$}
        \put(76,12){\color{darkblue}$\omega_c$}
    \end{overpic}
    \begin{overpic}[width=0.24\textwidth]{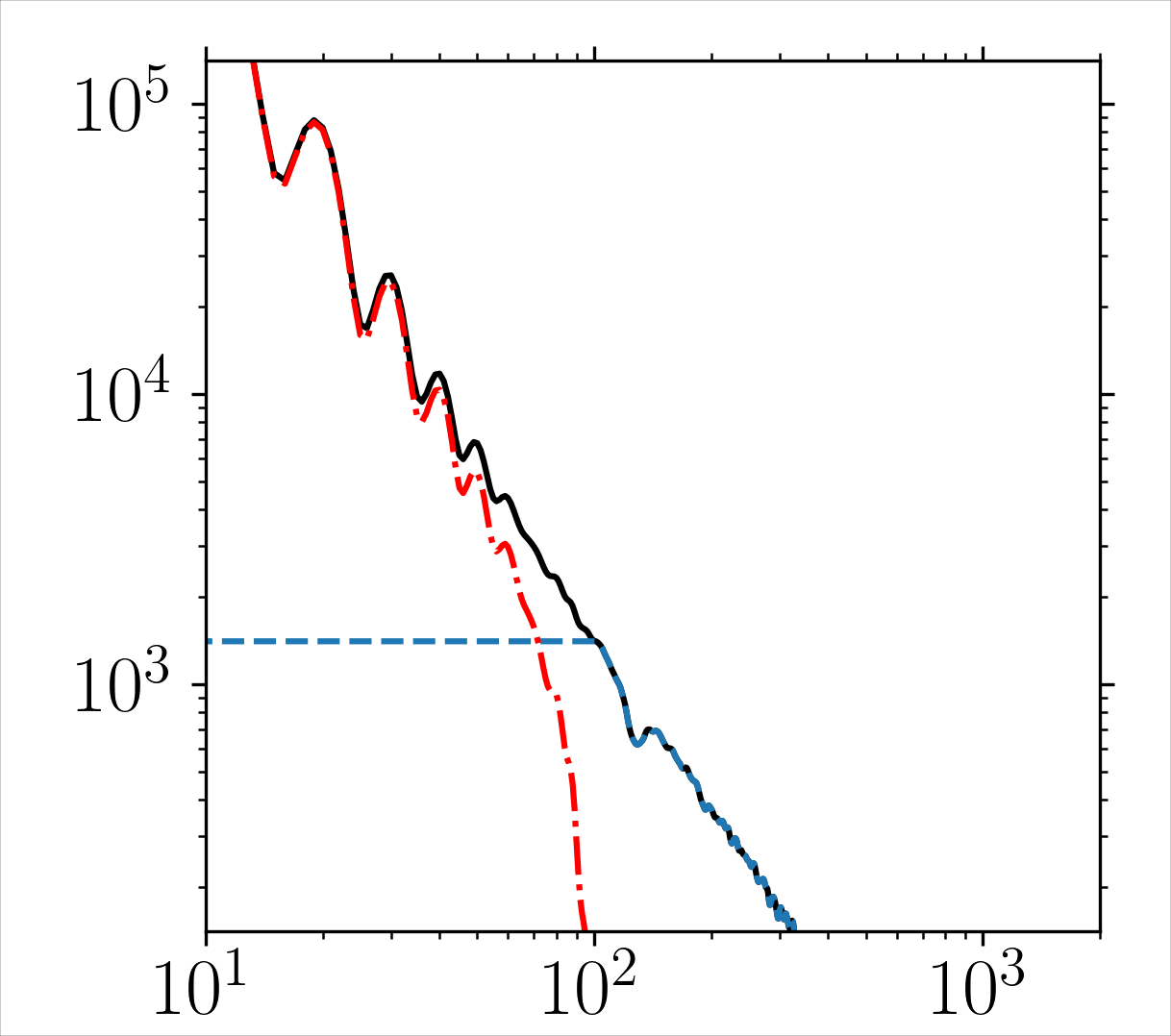}
        \LabelFig{16}{87}{$f)$}
        \Labelxy{50}{-5}{0}{$\omega$}
        \Labelxy{-6}{44}{90}{$|\Sigma(\omega)|^2$}
        \put(58,52){$\scriptstyle |\Sigma(\omega)|^2$}
        \put(20,38){\color{darkblue}$\scriptstyle |\mathcal{N}(\omega)|^2$}
        \put(18,19){\color{red}$\scriptstyle |\Sigma_\text{ath}(\omega)|^2$}
        \begin{tikzpicture}
          \coordinate (a) at (0,0);
            \node[white] at (a) {\tiny.};               %
             \carrow{2.4}{2.4}{0.5}{-1}{-1.0}{}{black}
            \carrow{1.1}{0.5}{0.7}{1.0}{0.0}{}{red}
	\end{tikzpicture}
    \end{overpic}
    \caption{Smeared stress signals associated with Cantor alloy at $\ratefour$ using optimal (Wiener) filtering with multiple cut-off frequencies corresponding to \textbf{a}) the raw signal \textbf{b}) $\omega_c=400$ \textbf{c}) $\omega_c=200$ \textbf{d}) $\omega_c=100$. \textbf{e}) Power spectrum $|\Sigma(\omega)|^2$ scaled by $\omega^{-2.3}$ associated with the raw signal. \textbf{f}) Wiener filtering with power spectra corresponding to the noise $|\mathcal{N}(\omega)|^2$ and athermal stress $|\Sigma_\text{ath}(\omega)|^2$ defined based on the extrapolation of the former into the signal region below cut-off frequency $\omega_c$.}
    \label{fig:kernelsRateThree}
\end{figure*}

To perform the avalanche analysis, we consider a set of different stress timeseries $\sigma_{zz}(t)$ including $10^4$ sample points each. 
To filter the thermal noise from the stress signal, we employ optimal (Wiener) filtering with the Fourier transforms.
In this context, the measured stress signal may be expressed as 
\begin{equation}
\sigma_{zz}(t)=\sigma_\text{ath}(t)+n(t),
\end{equation}
with the athemral component $\sigma_\text{ath}(t)$ and additive noise term $n(t)$.
We will estimate the \emph{true} signal $\sigma_\text{ath}(t)$ or $\Sigma_\text{ath}(\omega)$ by 
\begin{equation}
\Sigma_\text{ath}(\omega)=\Phi(\omega)\Sigma(\omega),
\end{equation}
where $\Sigma(\omega)$ is the Fourier transform of $\sigma_{zz}(t)$.
Following the procedure described in \cite{press1992numerical}, the optimal filter $\phi(t)$ or $\Phi(\omega)$ is given by
\begin{equation}
\Phi(\omega)=\frac{|\Sigma(\omega)|^2-|\mathcal{N}(\omega)|^2}{|\Sigma(\omega)|^2}.
\end{equation}
To estimate the noise spectrum $\mathcal{N}(\omega)|^2$, we follow a proper extrapolation by drawing a flat line across the region with the dominant power below a cut-off frequency $\omega_c$.
The difference between the measured power $|\Sigma(\omega)|^2$ and the former curve is $|\Sigma_\text{ath}(\omega)|^2$ as illustrated in Fig.~\ref{fig:kernels}(f) corresponding to deforming Cantor alloy at $\ratezero$.
Notably, the signal power scales as a power-law $|\Sigma(\omega)|^2 \propto \omega ^ {-\beta}$ with $\beta \simeq 2.3$ which differs from the scaling exponent associated with the Brownian noise ($\beta=2$). 
The power-law regime spans almost two decades in frequency, as shown by the flat region in Fig.~\ref{fig:kernels}(e), up to a characteristic value (i.e. $\omega_c$) beyond which the data shows a clear uptick.   
We applied the Wiener filter by varying the cut-off frequency $\omega_c$ on the input signals with the results shown in Fig.~\ref{fig:kernels}(a-d).
Effects of filtering on the rate of stress change $-\sigmadot$, shown by the red dots, are quite pronounced in terms of effectively removing the background noise due to thermal fluctuations.
We repeated the above analysis corresponding to Cantor alloy at $\ratefour$ as illustrated in Fig.~\ref{fig:kernelsRateThree}.
The overall softening behavior is significant, as opposed to the steady plastic flow associated with the slow rates.
We remark a relatively smooth evolution of stress avalanches in this case which seem to lack the observed intermittency associated with the slowly-deforming metal. 

\begin{figure*}
    \begin{overpic}[width=0.24\textwidth]{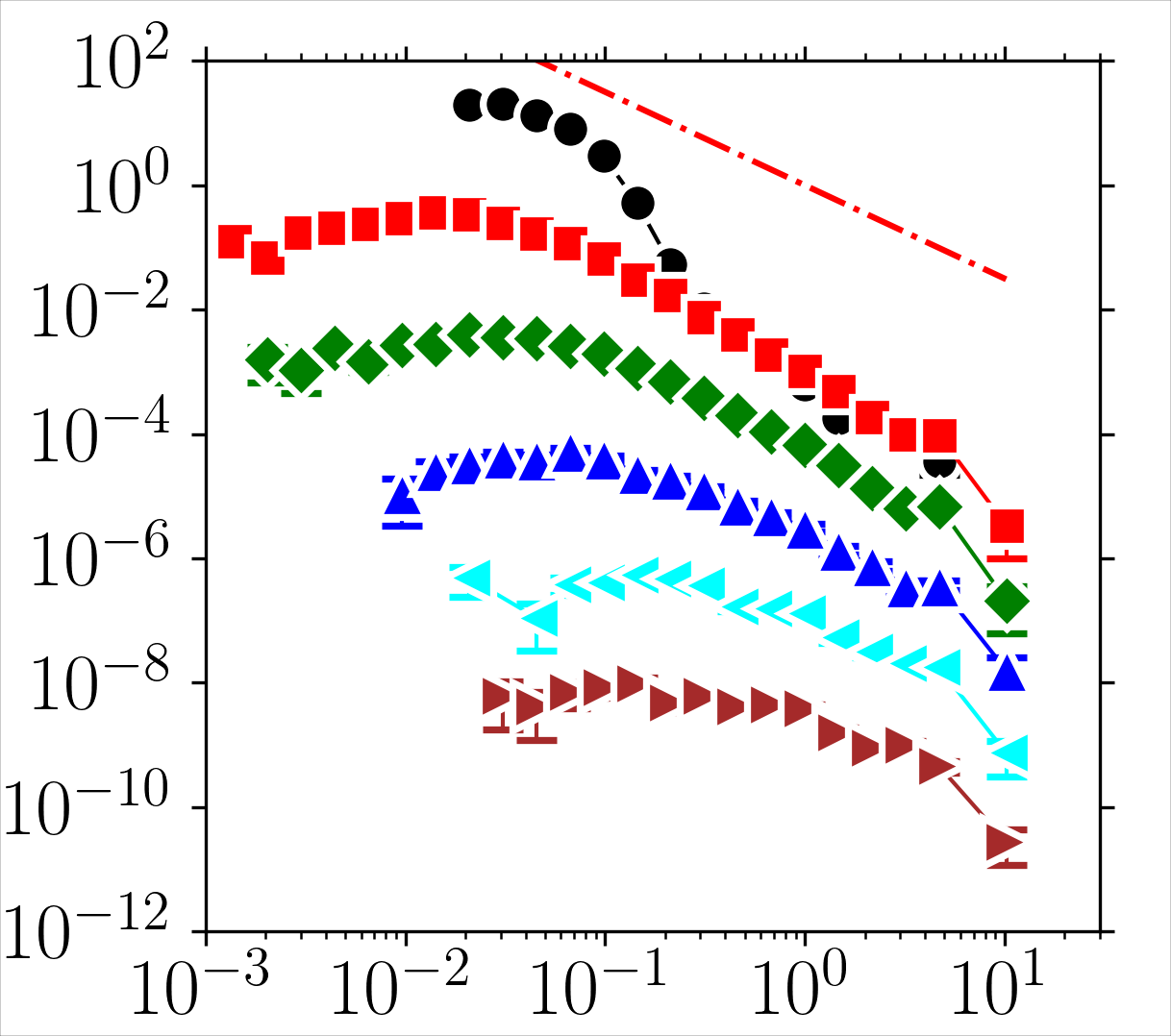}
        \LabelFig{16}{87}{$a)$ \scriptsize\glfour}
        \Labelxy{50}{-6}{0}{$S$~\scriptsize (Gpa)}
        \Labelxy{-6}{44}{90}{$P(S)$}
    \end{overpic}
    \begin{overpic}[width=0.24\textwidth]{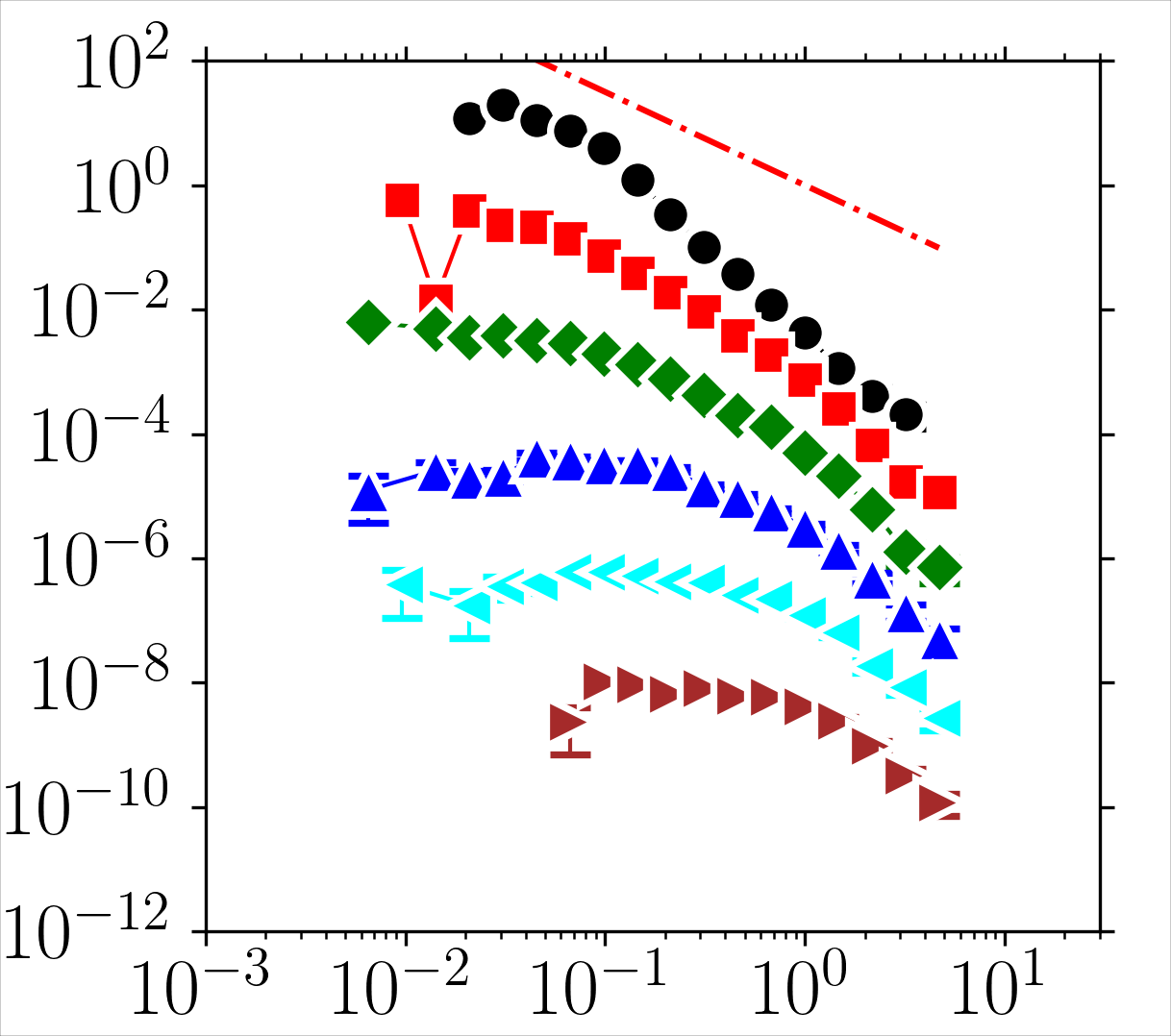}
        \LabelFig{16}{87}{$b)$ \scriptsize\glsix}
        \Labelxy{50}{-6}{0}{$S$~\scriptsize (Gpa)}
        \Labelxy{-6}{44}{90}{$P(S)$}
    \end{overpic}
    \begin{overpic}[width=0.24\textwidth]{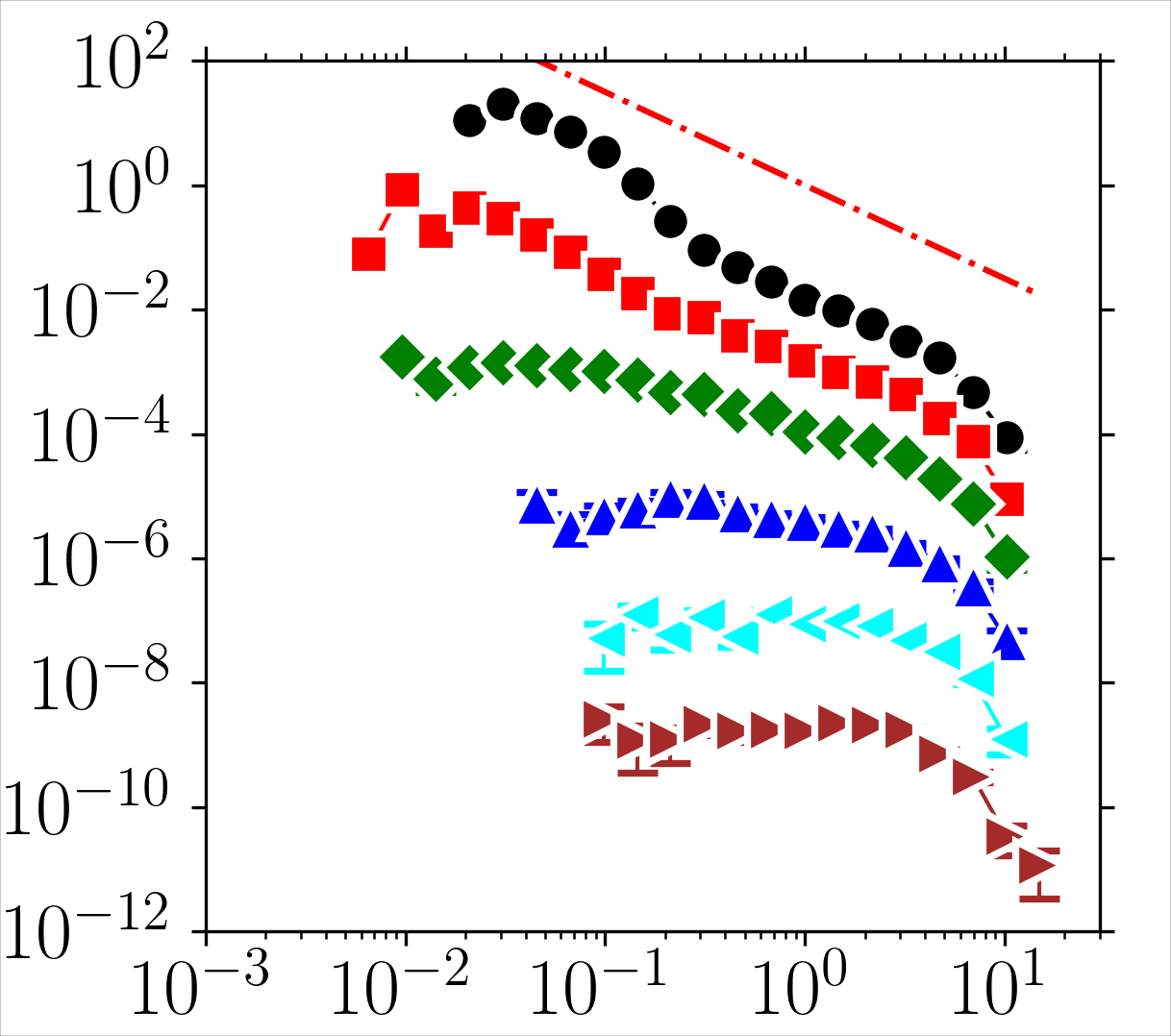}
        \put(96,32){\includegraphics[width=0.05\textwidth]{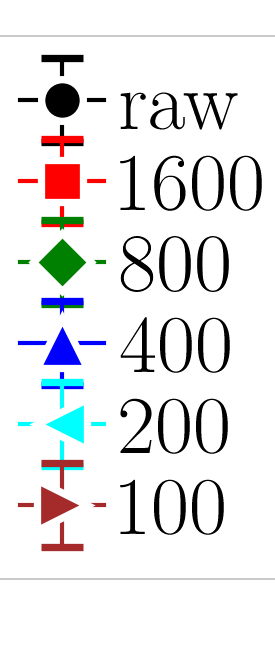}}
        \LabelFig{16}{87}{$c)$ \scriptsize Ni}
        \Labelxy{50}{-6}{0}{$S$~\scriptsize (Gpa)}
        \Labelxy{-6}{44}{90}{$P(S)$}
        \Labelxy{104}{80}{0}{$\omega_c$}
    \end{overpic}
    %
    \caption{Avalanche size distributions $P(S)$ at multiple $\omega_c$ and $\rate=10^8$ s$^{-1}$ corresponding to \textbf{a}) Cantor \textbf{b}) \glsix, and \textbf{c}) Ni. The dashdotted lines denote power laws $P(S)\propto S^{-\tau}$ with $\tau=3/2$.}
    \label{fig:ps_kernels}
\end{figure*}

\begin{figure*}
    \begin{overpic}[width=0.24\textwidth]{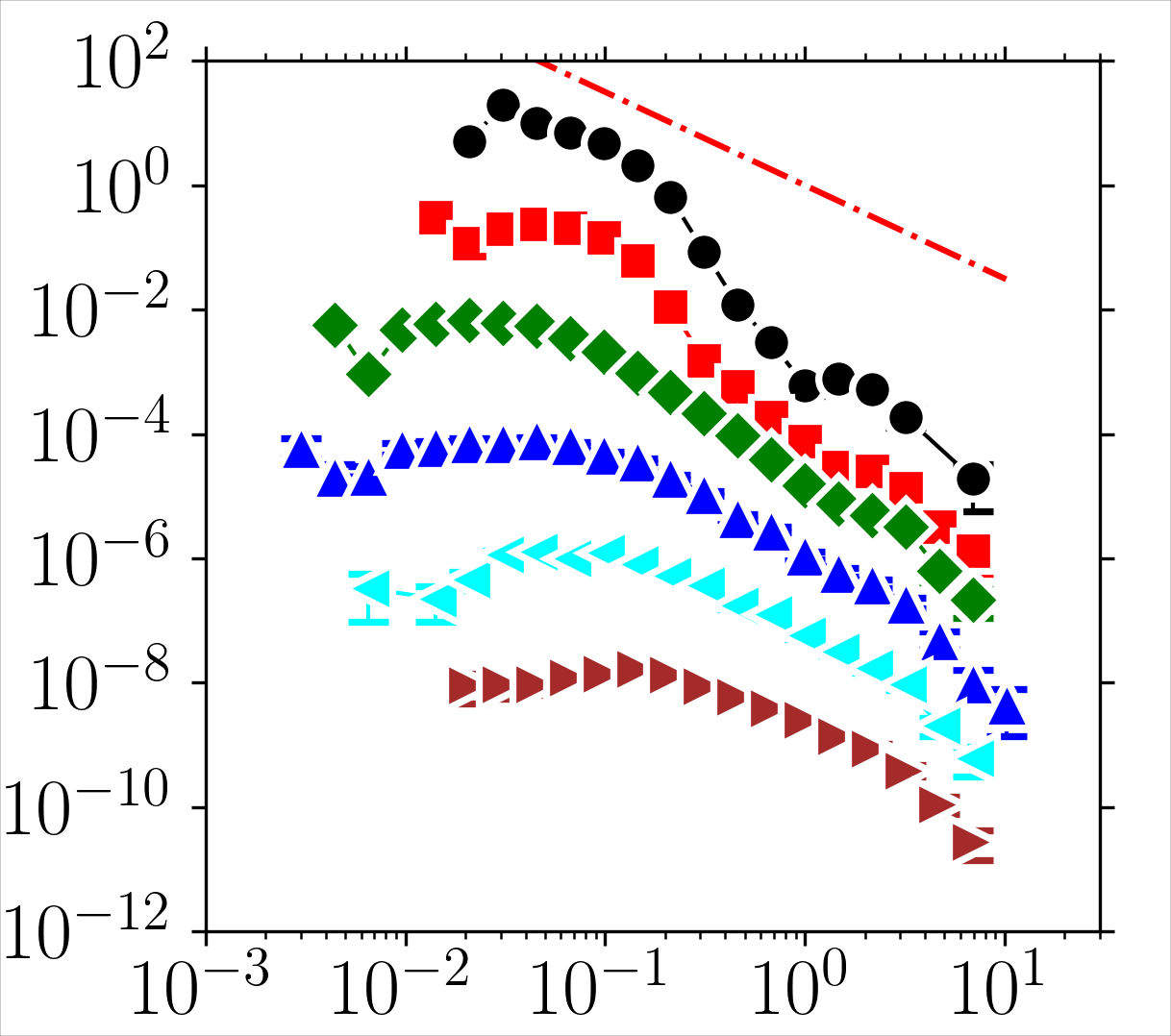}
        \LabelFig{16}{87}{$a)$ \scriptsize\glfour}
        \Labelxy{50}{-6}{0}{$S$~\scriptsize (Gpa)}
        \Labelxy{-6}{44}{90}{$P(S)$}
    \end{overpic}
    \begin{overpic}[width=0.24\textwidth]{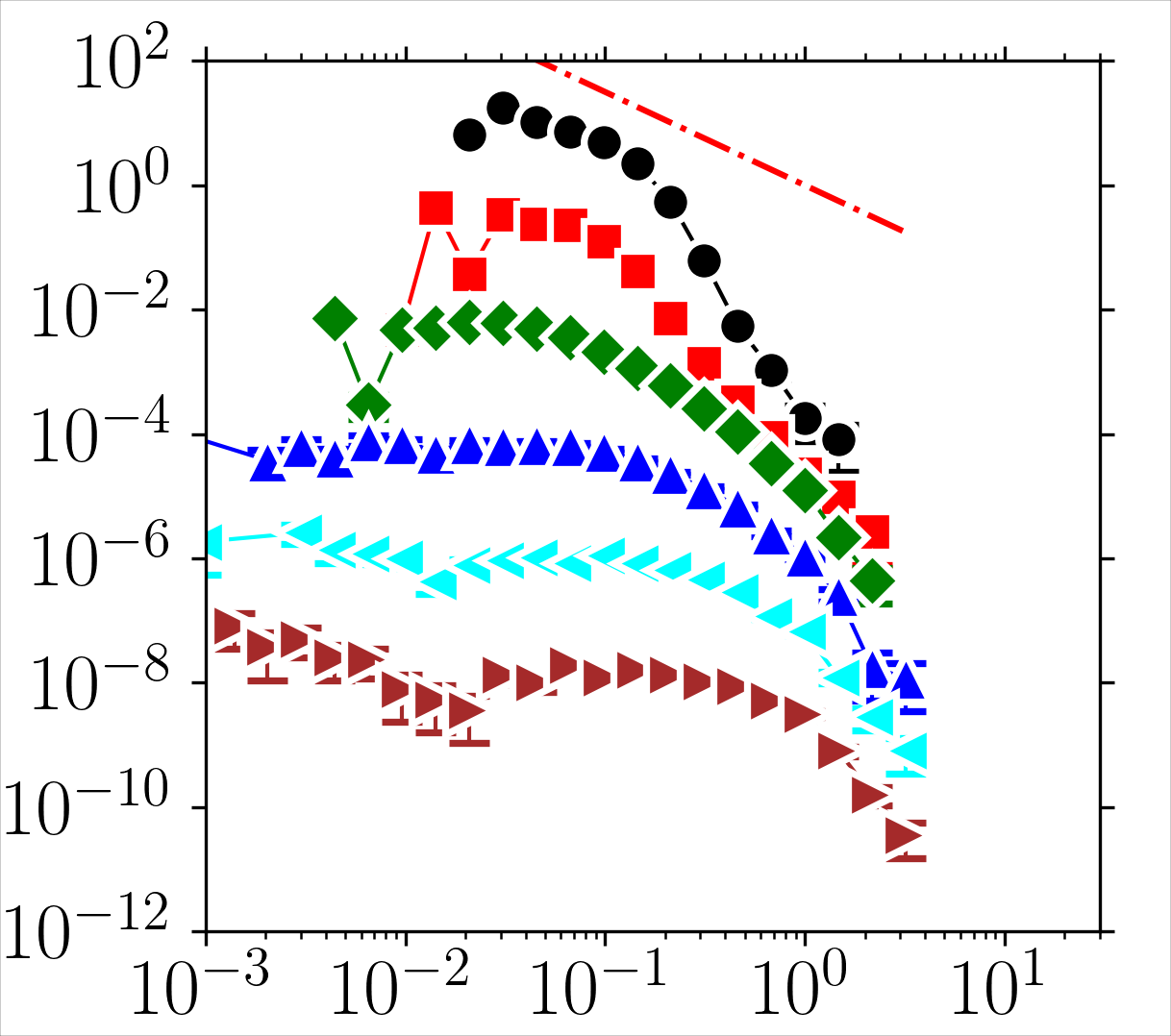}
        \LabelFig{16}{87}{$b)$ \scriptsize\glsix}
        \Labelxy{50}{-6}{0}{$S$~\scriptsize (Gpa)}
        \Labelxy{-6}{44}{90}{$P(S)$}
    \end{overpic}
    \begin{overpic}[width=0.24\textwidth]{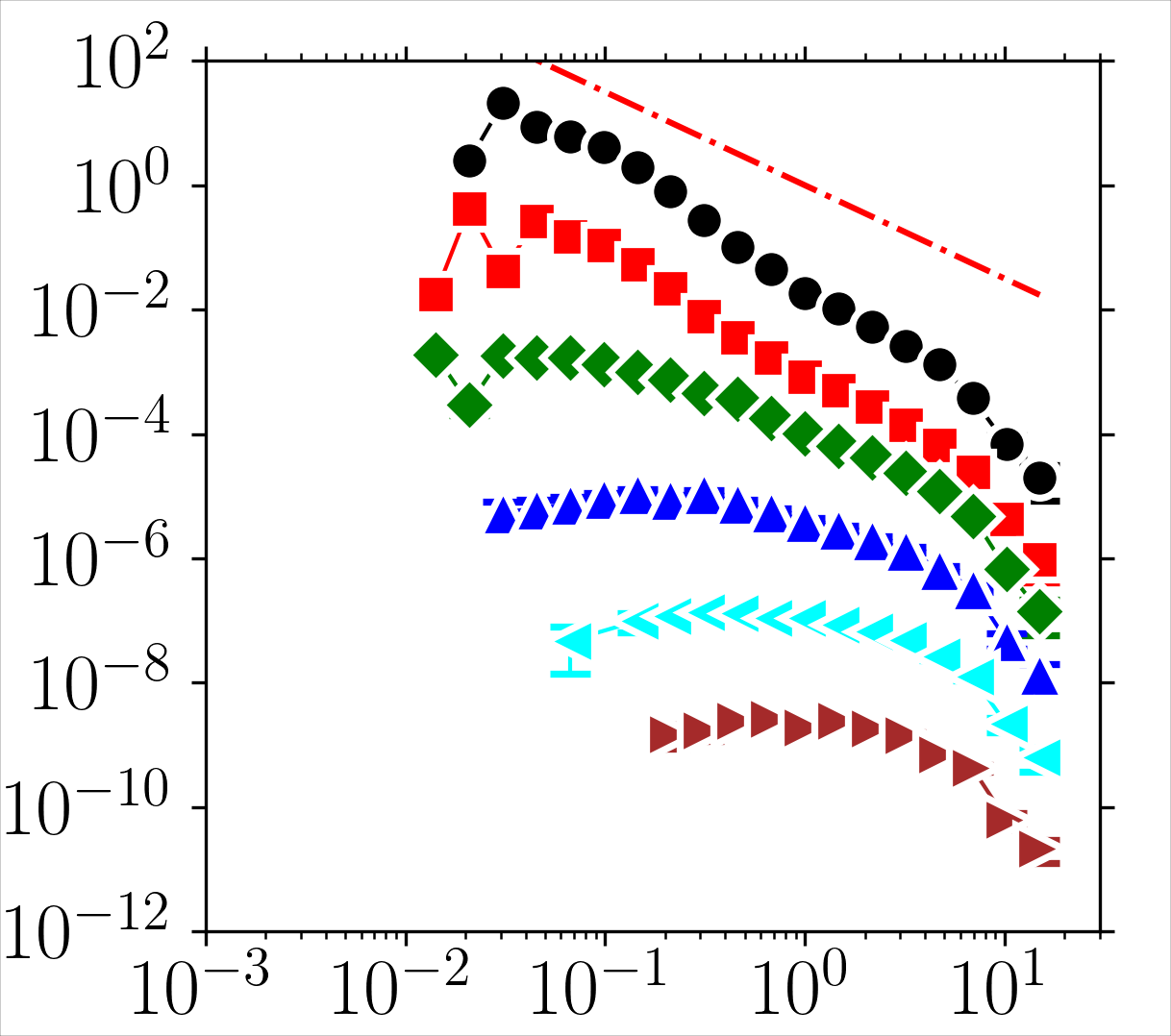}
        \put(96,32){\includegraphics[width=0.05\textwidth]{Figs/pdf_s_cantor_multipleFilters_legend.png}}
        \LabelFig{16}{87}{$c)$ \scriptsize Ni}
        \Labelxy{50}{-6}{0}{$S$~\scriptsize (Gpa)}
        \Labelxy{-6}{44}{90}{$P(S)$}
        \Labelxy{104}{80}{0}{$\omega_c$}
    \end{overpic}
    %
    \caption{Avalanche size distributions $P(S)$ at multiple $\omega_c$ and $\ratethree$ corresponding to \textbf{a}) Cantor \textbf{b}) \glsix, and \textbf{c}) Ni. The dashdotted lines denote power laws $P(S)\propto S^{-\tau}$ with $\tau=3/2$. }
    \label{fig:ps_kernelsRateThree}
\end{figure*}

\begin{figure*}
    \begin{overpic}[width=0.24\textwidth]{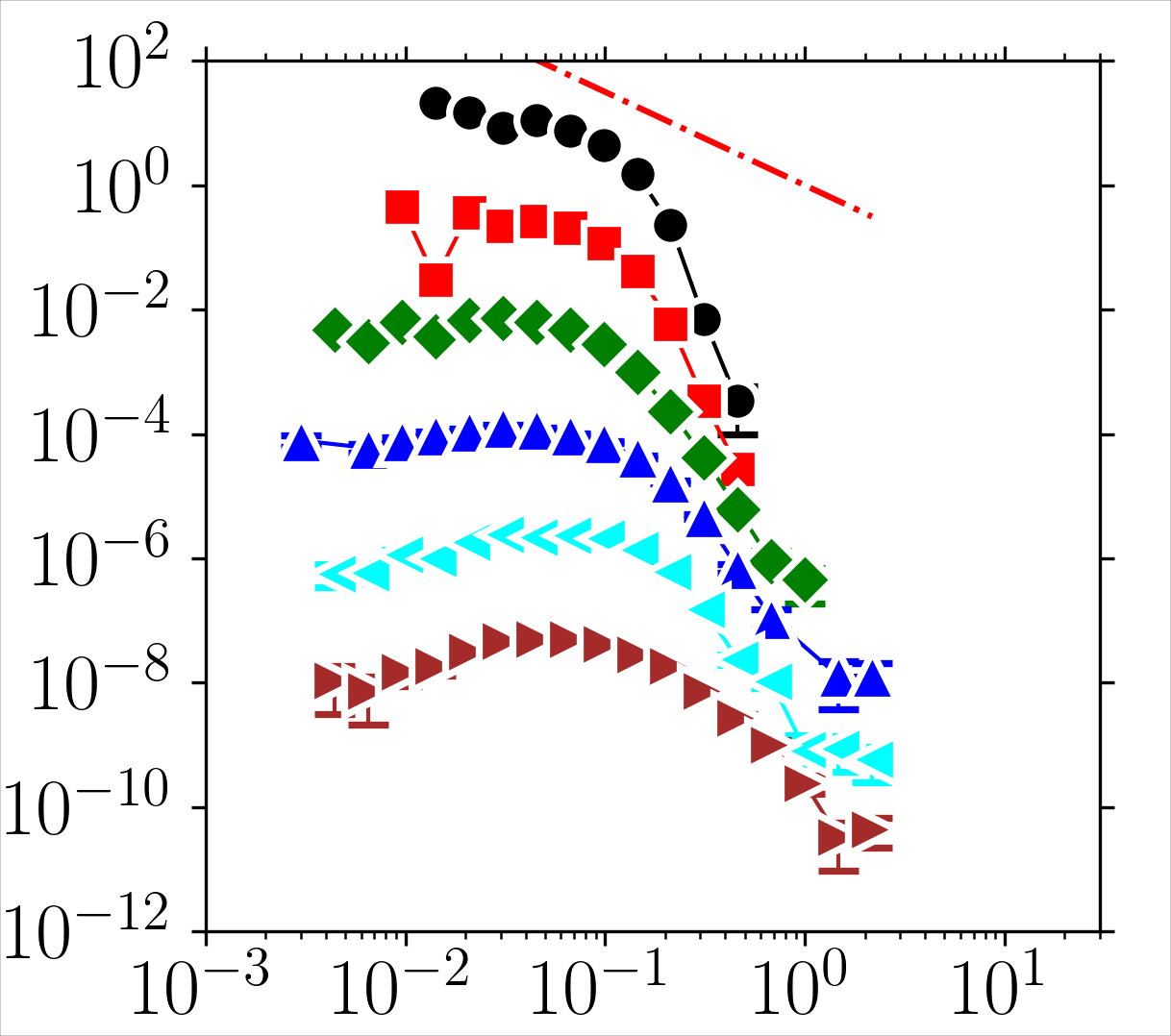}
        \LabelFig{16}{87}{$a)$ \scriptsize\glfour}
        \Labelxy{50}{-6}{0}{$S$~\scriptsize (Gpa)}
        \Labelxy{-6}{44}{90}{$P(S)$}
    \end{overpic}
    \begin{overpic}[width=0.24\textwidth]{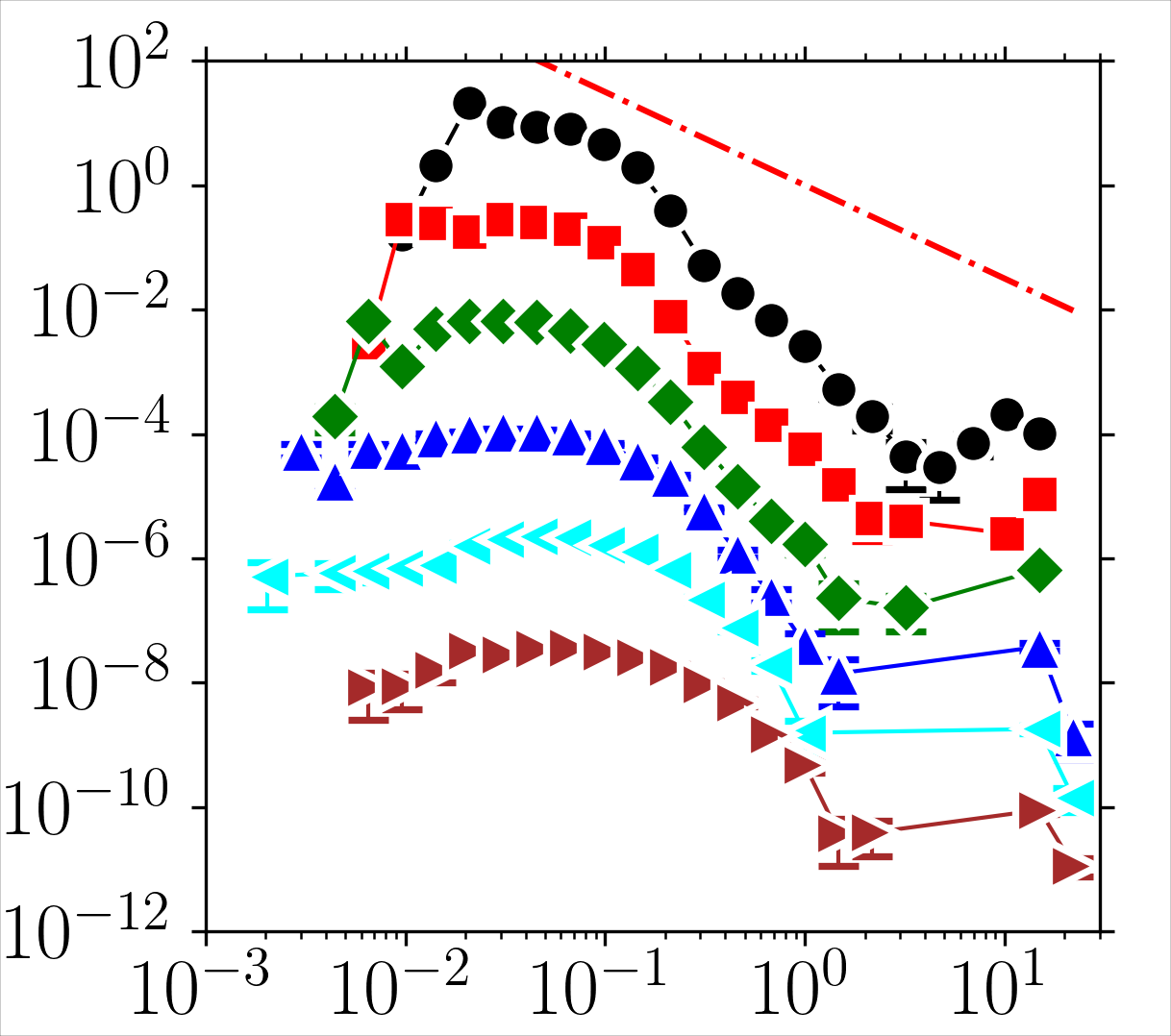}
        \LabelFig{16}{87}{$b)$ \scriptsize\glsix}
        \Labelxy{50}{-6}{0}{$S$~\scriptsize (Gpa)}
        \Labelxy{-6}{44}{90}{$P(S)$}
    \end{overpic}
    \begin{overpic}[width=0.24\textwidth]{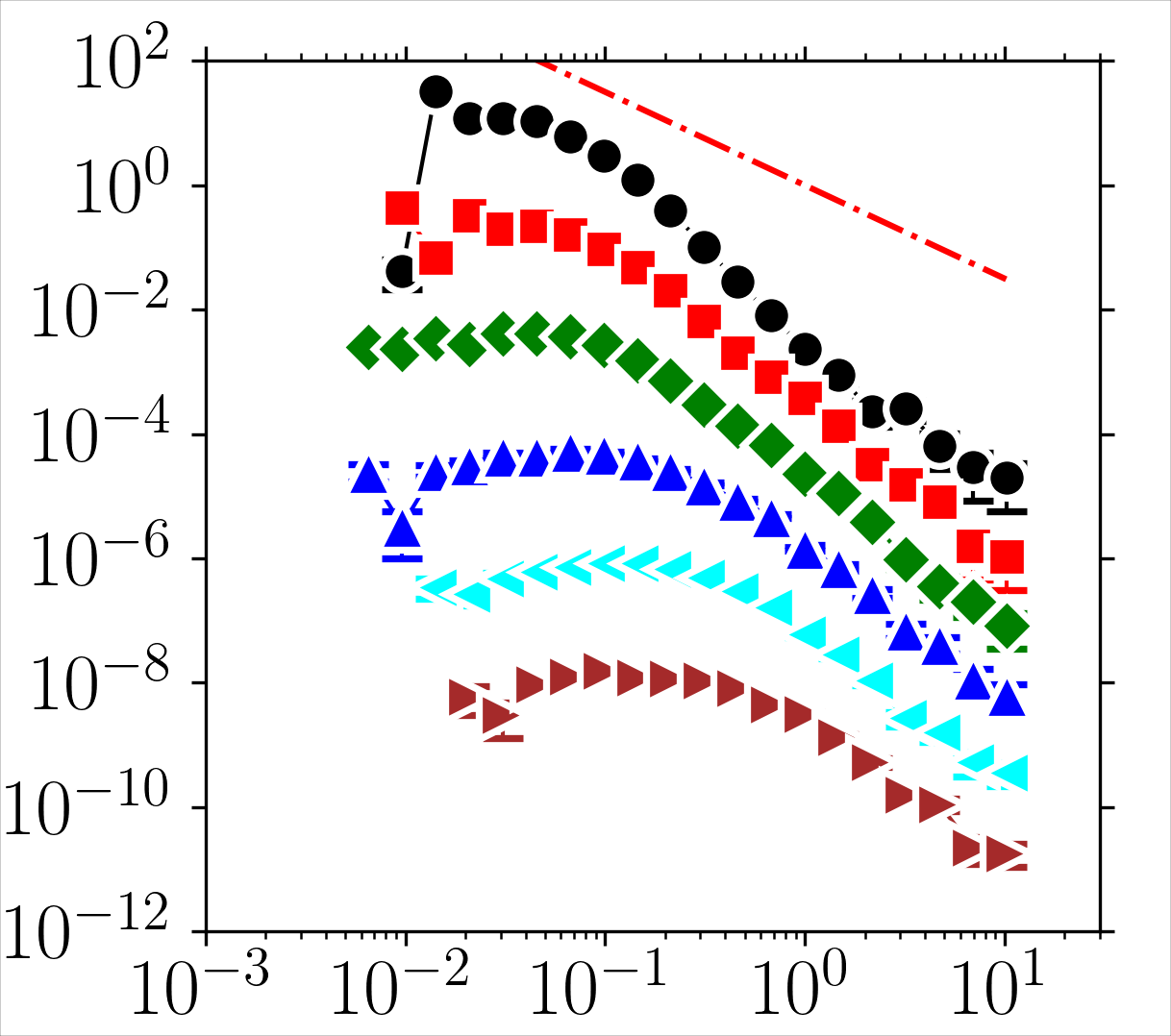}
        \put(96,32){\includegraphics[width=0.05\textwidth]{Figs/pdf_s_cantor_multipleFilters_legend.png}}
        \LabelFig{16}{87}{$c)$ \scriptsize Ni}
        \Labelxy{50}{-6}{0}{$S$~\scriptsize (Gpa)}
        \Labelxy{-6}{44}{90}{$P(S)$}
        \Labelxy{104}{80}{0}{$\omega_c$}
    \end{overpic}
    %
    \caption{Avalanche size distributions $P(S)$ at multiple $\omega_c$ and $\ratefour$ corresponding to \textbf{a}) Cantor \textbf{b}) \glsix, and \textbf{c}) Ni. The dashdotted lines denote power laws $P(S)\propto S^{-\tau}$ with $\tau=3/2$.}
    \label{fig:ps_kernelsRateFour}
\end{figure*}

In Fig.~\ref{fig:ps_kernels}, we plot size distributions corresponding to the smeared signals at $\ratezero$ and various $\omega_c$.
We note that choosing an optimal frequency near the observed upturn in Fig.~\ref{fig:kernels}(e), within the range $\omega_c=800-1600$, results in a fairly robust power-law decay in $P(S)$.
The observed trend seems to be also valid across $\ratethree$ and $\ratefour$ as in Fig.~\ref{fig:ps_kernelsRateThree} and \ref{fig:ps_kernelsRateFour}. 

\bibliography{references}